\documentclass[twocolumn,showpacs,showkeys]{revtex4}
\usepackage{graphicx}
\usepackage{bm}
\usepackage{color}
\usepackage{amsmath}
\usepackage{natbib}

\begin{document}

\title{Extended hydrodynamics of the degenerate partially spin polarized fermions
with the short-range interaction up to the third order by the interaction radius approximation}

\author{Pavel A. Andreev}
\email{andreevpa@physics.msu.ru}
\affiliation{Faculty of physics, Lomonosov Moscow State University, Moscow, Russian Federation, 119991.}

\date{\today}

\begin{abstract}
A possibility of the hydrodynamic description of ultracold fermions via the microscopic derivation of the model is described.
Differently truncated hydrodynamic models are derived and compared.
All models are based on the microscopic many-particle Schr\"{o}dinger equation.
Minimal coupling model based on the continuity and Euler equations are considered.
The extended hydrodynamic model including the independent dynamics of the momentum flux (the pressure evolution) is derived.
Influence of the spin polarization is described.
The short-range interaction is considered in the isotropic limit.
The interaction is considered up to the third order by the interaction radius.
Therefore, the single fluid model of spin-1/2 fermions
and the two fluid model of spin-1/2 fermions
are under consideration in this paper.
Spectra of bulk collective excitations are derived and compared in terms of different models.
\end{abstract}



\maketitle


\section{Introduction}

Currently, the ultracold fermions \cite{Wang PRA 19, Bradlyn PRA 16, Qin PRA 18 b, Qin PRA 18, Wang PRA 18}
has interest equal to the interest to the Bose-Einstein condensates (BECs).
However, there is a simple tool for the theoretical analysis of BEC.
It is the Gross-Pitaevskii equation or equivalent quantum hydrodynamic equations \cite{Dalfovo RMP 99}.
But the application of similar tools for the ultracold fermions is limited \cite{Giorgini RMP 08}.
While the application of kinetic models seems to be more sophisticated.
However, the hydrodynamic models are considerably simpler.
So, it is highly useful for description of static and dynamic behavior of fermions.
A wider possibility of application of the quantum hydrodynamic models to the ultracold fermions is discussed here.
A systematic derivation of mean-field models of fermions starting from the microscopic Schr\"{o}dinger equation is developed.
Let us point out that
spin-1/2 fermions with repulsive short-range interaction between spin-up and spin-down fermions are considered.
Moreover, the interaction between fermions of the same spin projection is included.

Simple look on the hydrodynamic model shows
that the application of the equilibrium Fermi pressure to the dynamical processes
such as the wave propagation
gives partially incorrect results.
This problem can be solved ad hoc by introduction of the advanced equation of state.
However, deeper look on this problem is suggested in Refs. \cite{Tokatly PRB 99, Tokatly PRB 00},
where spectra of the collective excitations of degenerate charged fermions are studied by extended sets of hydrodynamic equations.
The second order hydrodynamics including the pressure tensor evolution
and the fourth order hydrodynamics including evolution of tensors up to the average of product of four momentums are developed and applied to consider properties of degenerate repulsive fermions.
This analysis is based on the kinetic model,
which is a macroscopic method.
While our goal is the microscopic justification based on the Schr\"{o}dinger equation of hydrodynamic model of fermions.

The goal of this paper is the hydrodynamic model of sound waves.
Therefore, extended model is limited by the account of the pressure tensor evolution equation
in addition to the continuity and Euler equations.
Presenting derivation is based on the many-particle quantum hydrodynamic method developed
in Refs. \cite{Maksimov QHM 99, Andreev PRA08, Andreev PRB 11, Andreev IJMPB 13}.
Further development of this method for the derivation of the pressure evolution equation
and calculation of the short-range interaction in this equation is demonstrated.

The application of  simple hydrodynamic model
for description of collective motion of fermions goes back to the first half of the XX century (Bloch's hydrodynamic theory) \cite{Bloch ZP 33}.
The last decades show application of simple hydrodynamic model \cite{Kulkarni PRA 12}
along with the development of new modifications \cite{Zyl PRA 14, Zyl PRA 13}.

We can describe fermions as the single fluid.
Or we can consider spin-s fermions as $2s+1$ different fluids.
Both regimes are studied below.
Description of spin-s bosons as several fluids is widely used being sometimes hidden
as spinor nonlinear Pauli (Schr\"{o}dinger) equation
\cite{Pixley PRL 15, Ho PRL 98, Ohmi JPSJ 98, Szankowski PRL 10, Pogosov PRA 05, Stamper-Kurn RMP 13, Mizushima PRL 02}.
Two-fluid model of the partially spin polarized spin-1/2 fermions
in the first order by the interaction radius is discussed in Ref. \cite{Andreev LPL 18}.
The spin waves are addressed there via the dynamics of the hydrodynamic spin density.
It is presented along with the sound waves.

This paper is organized as follows.
In Sec. II the formulation of basic ideas of the many-particle quantum hydrodynamics method is presented.
In Sec III the feature of the short-range interaction in the momentum balance equation are described.
In Sec. IV the contribution of the short-range interaction in the momentum flux balance equation is presented.
In Sec. V equation for the pressure tensor is discussed.
In Sec. VI features of the minimal coupling model based hydrodynamic equations with no pressure evolution are described
in the first order by the interaction radius.
In Sec. VII the minimal coupling model is demonstrated in the third order by the interaction radius.
In Sec. VIII separate spin evolution extended hydrodynamics is presented.
In Sec. IX the linear collective excitations are considered.
In Sec. X a brief summary of obtained results is presented.

\section{Derivation of hydrodynamic equations}

\subsection{General structure of equations}

Many collective processes,
such as the formation of different structures, clasters, cristals,
formation of wave patterns, solitons, vorticities
reveal patterns in three dimensional physical space.
However, the fundamental microscopic theories (the classical mechanics and the quantum mechanics)
are formulated in multudimensional configurational space.
Proper description of the collective effects requires representation of the mechanics in terms of
the field variables defined in three-dimensional physical space.
So happened that the hydrodynamics is a natural representation of classic and quantum mechanics in physical space in terms of collective observables.
This conclusion does not introduces the concept that
this is one possible representation.
The density functional theory is another example of similar class of models.
Moreover, the hydrodynamics is not structures existing in the momentum space,
where kinetic model have obvious advantage.

It is well-known that the hydrodynamic equations can be derived from the kinetic theory.
However,
proper truncation procedure for degenerate fermions requires the account of the pressure tensor evolution \cite{Tokatly PRB 99, Tokatly PRB 00}.
Although some kinetic models can be derived from the microscopic theories,
so there is a link between the hydrodynamics and the microscopic description,
there is more ambitious problem.
It is the direct derivation of hydrodynamic equations from the microscopic theories.
Being focused on the degenerate fermions
which is a quantum system,
it is necessary to start the derivation from the many-particle Schr\"{o}dinger equation $\imath\hbar\partial_{t}\Psi=\hat{H}\Psi$.
Neutral fermions interact by the short-range interaction which can be presented by the potential of general form $U_{ij}=U(\textbf{r}_{i}-\textbf{r}_{j})$,
where $i$ and $j$ are numbers of particles.
External fields creating traps are also included in the microscopic model via corresponding potential $V_{ext}(\textbf{r}_{i},t)$.
Overall, the fermions can be described by rather simple Hamiltonian
\begin{equation}\label{aSRIff Hamiltonian micro}
\hat{H}=\sum_{i}\biggl(\frac{\hat{\textbf{p}}^{2}_{i}}{2m_{i}}+V_{ext}(\textbf{r}_{i},t)\biggr)
+\frac{1}{2}\sum_{i,j\neq i}U(\textbf{r}_{i}-\textbf{r}_{j}) ,\end{equation}
where $m_{i}$ is the mass of i-th particle, $\hat{\textbf{p}}_{i}=-\imath\hbar\nabla_{i}$ is the momentum of i-th particle.

We consider interacting fermions.
However, we restrict ourselves with the repulsive interaction
since the attraction between fermions with different spin projections leads to the Cooper pair formation and formation of dimers.

Our goal is to create a model of collective motion of degenerate fermions.
To this end, we need to chose a suitable collective variables.
One of simple traditional hydrodynamic variables is the local concentration of particles.
It is a scalar field defined in the following form
\begin{equation}\label{aSRIff concentration def} n(\textbf{r},t)=\int
dR\sum_{i}\delta(\textbf{r}-\textbf{r}_{i})\Psi^{*}(R,t)\Psi(R,t),\end{equation}
where $dR=\prod_{i=1}^{N}d\textbf{r}_{i}$ is the element of volume in $3N$ dimensional configurational space,
with $N$ is the number of particles.
This definition is made in accordance with definition of the quantum observables
as the quantum average of the corresponding operator.
The operator of concentration is chosen as the quantization of the classic microscopic local concentration
which is a sum of $N$ delta functions depicting the point-like objects $\hat{n}=\sum_{i}\delta(\textbf{r}-\textbf{r}_{i})$.

In the definition of the concentration (\ref{aSRIff concentration def}) it is assumed that
the many-particle wave function is normalized on 1.
It would be an incomplete expression for the infinite motion of particles
which happens for the infinite mediums.
Therefore, let us keep in mind expression $n=\int
dR\sum_{i}\delta(\textbf{r}-\textbf{r}_{i})\Psi^{*}(R,t)\Psi(R,t)
/\int dR\Psi^{*}(R,t)\Psi(R,t)$
while explicit calculations are made with expression (\ref{aSRIff concentration def}).
Coefficient $1/\int dR\Psi^{*}(R,t)\Psi(R,t)$ does not depend on $\textbf{r}$ and $t$.
Hence, it can be considered as a constant.
Therefore, the coefficient does not affect the presented calculations.

To obtain an equation governing the evolution of concentration
it is necessary to take derivative of function (\ref{aSRIff concentration def})
with respect to time.
This derivative acts on the wave functions located under the integral.
The time derivative of the wave function is obtained from the Schr\"{o}dinger equation via the Hamiltonian of the system.
After the straightforward calculation find that
the time derivative of concentration is equal to the divergence of the vector function. It gives the continuity equation:
\begin{equation}\label{aSRIff cont eq via j} \partial_{t}n+\nabla\cdot \textbf{j}=0, \end{equation}
where the current $\textbf{j}$ is defined via the many-particle wave function of the system:
$$\textbf{j}(\textbf{r},t)
=\int dR\sum_{i}\delta(\textbf{r}-\textbf{r}_{i})\times$$
\begin{equation}\label{aSRIff j def}
\times\frac{1}{2m_{i}}(\Psi^{*}(R,t)\hat{\textbf{p}}_{i}\Psi(R,t)+c.c.).\end{equation}

The derivation of the continuity equation (\ref{aSRIff cont eq via j})
provides the extra collective variable.

The many-particle wave function is an equivalent of $2N$ independent real scalar functions of three coordinates.
Number $N$ comes from the number of particles
while number $2$ comes from the fact that each single particle wave function (which is a complex function) is equivalent to two real functions.
On this stage
we have four functions:
one scalar function $n$ and
three projections of the current $\textbf{j}$.
Therefore, we can expect appearance of $2N-4$ functions if $N>2$.

If we have a representation of the two particle system
the set of functions $n$ and $\textbf{j}$ looks complete.

For the single particle systems,
the current simplifies to the gradient of the scalar function $\textbf{j}=n\nabla\varphi$,
where $\varphi$ is the potential of the velocity field $\textbf{v}\equiv\textbf{j}/n=\nabla\varphi$.

Anyway,
it is necessary to derive equation for the current evolution.
Let us mention that
$m\textbf{j}$ is the density of momentum.
Hence,
equation for $\textbf{j}$ is the momentum evolution equation.
For the derivation of the momentum balance equation
differentiate the definition of current (\ref{aSRIff cont eq via j}) with respect to time.
Use the Schr\"{o}dinger equation for the time derivatives of the wave function.
During calculation separate two kinds of terms.
The terms containing the interaction
which give the force field
and the terms
which have the kinetic nature.
The last group appears as the divergence of a second rank tensor.
The momentum balance equation has the following structure
\begin{equation} \label{aSRIff Euler eq 1 via j}
\partial_{t}j^{\alpha}+\partial_{\beta}\Pi^{\alpha\beta}
=-\frac{1}{m}n\partial_{\alpha}V_{ext}+\frac{1}{m}F^{\alpha}_{int}, \end{equation}
where
$$\Pi^{\alpha\beta}=\int dR\sum_{i}\delta(\textbf{r}-\textbf{r}_{i}) \frac{1}{4m_{i}^{2}}
[\Psi^{*}(R,t)\hat{p}_{i}^{\alpha}\hat{p}_{i}^{\beta}\Psi(R,t)$$
\begin{equation} \label{aSRIff Pi def} +\hat{p}_{i}^{\alpha *}\Psi^{*}(R,t)\hat{p}_{i}^{\beta}\Psi(R,t)+c.c.] \end{equation}
is the momentum flux (containing the pressure tensor),
and
\begin{equation} \label{aSRIff F alpha def via n2}
F^{\alpha}_{int}=-\int (\partial^{\alpha}U(\textbf{r}-\textbf{r}'))
n_{2}(\textbf{r},\textbf{r}',t)d\textbf{r}', \end{equation}
with the following expression for the two-particle concentration
\begin{equation} \label{aSRIff n2 def} n_{2}(\textbf{r},\textbf{r}',t)=\int
dR\sum_{i,j\neq i}\delta(\textbf{r}-\textbf{r}_{i})\delta(\textbf{r}'-\textbf{r}_{j})\Psi^{*}(R,t)\Psi(R,t) .\end{equation}

The external force field (the density of the force) can be also introduced $F^{\alpha}_{ext}=-n\partial_{\alpha}V_{ext}$.
Two force fields combined together give the full force field $F^{\alpha}=F^{\alpha}_{ext}+F^{\alpha}_{int}$.

There are models of fermions, including the degenerate fermions,
where the truncation is made in the momentum balance equation.
So, the dynamics of fermions is described by two equation.
These hydrodynamic models approximately correspond to the non-linear Schr\"{o}dinger equations.
However, these hydrodynamic models have a fundamental drawback.

Explain it for the degenerate fermions.
The models require an equation of state for the pressure.
An approximate expression is usually taken in the form of the equilibrium ideal quantum gas pressure.
The application of the equilibrium expression to the dynamical processes is questionable.
However, this problem is more complicated
when the choice of the equation of state.
It is demonstrated that
the dynamics of fermions requires evolution equation for the pressure tensor
with the account of the nondiagonal elements \cite{Tokatly PRB 99}.
It is necessary even for the degenerate fermions.

Below, we consider a model based on the continuity and the momentum balance equation,
but now we develop a background for the more appropriate models.

Consider the evolution of the momentum flux tensor (\ref{aSRIff Pi def}).
Similarly to the derivation described above,
take the derivative of tensor (\ref{aSRIff Pi def}) with respect to time and apply the Schr\"{o}dinger equation.

Derivation of the momentum flux evolution (leading to the pressure evolution equation) is more bulging,
but it is similar to the derivation of the momentum balance equation.
The result has the following form
$$\partial_{t}\Pi^{\alpha\beta}+\partial_{\gamma}M^{\alpha\beta\gamma}
=-\frac{1}{m}j^{\beta}\partial_{\alpha}V_{ext}-\frac{1}{m}j^{\alpha}\partial_{\beta}V_{ext}$$
$$-\frac{1}{m}\int[\partial^{\beta}U(\textbf{r}-\textbf{r}')]j_{2}^{\alpha}(\textbf{r},\textbf{r}',t)d\textbf{r}'$$
\begin{equation} \label{aSRIff eq for Pi alpha beta}
-\frac{1}{m}\int[\partial^{\alpha}U(\textbf{r}-\textbf{r}')]j_{2}^{\beta}(\textbf{r},\textbf{r}',t)d\textbf{r}', \end{equation}
where
$$M^{\alpha\beta\gamma}=\int dR\sum_{i}\delta(\textbf{r}-\textbf{r}_{i}) \frac{1}{8m_{i}^{3}}\biggl[\Psi^{*}(R,t)\hat{p}_{i}^{\alpha}\hat{p}_{i}^{\beta}\hat{p}_{i}^{\gamma}\Psi(R,t)$$
$$+\hat{p}_{i}^{\alpha *}\Psi^{*}(R,t)\hat{p}_{i}^{\beta}\hat{p}_{i}^{\gamma}\Psi(R,t)
+\hat{p}_{i}^{\alpha *}\hat{p}_{i}^{\gamma *}\Psi^{*}(R,t)\hat{p}_{i}^{\beta}\Psi(R,t)$$
\begin{equation} \label{aSRIff M alpha beta gamma def}
+\hat{p}_{i}^{\gamma *}\Psi^{*}(R,t)\hat{p}_{i}^{\alpha}\hat{p}_{i}^{\beta}\Psi(R,t)+c.c.\biggr] \end{equation}
is the extra flux function of higher tensor dimension, it can be called the flux of the momentum flux, the trace of tensor $L^{\alpha\beta\gamma}$ on two indexes and find the energy flux $q^{\alpha}=L^{\alpha\beta\beta}$
and
the following expression for the two-particle current-concentration function
$$\textbf{j}_{2}(\textbf{r},\textbf{r}',t)=\int
dR\sum_{i,j\neq i}\delta(\textbf{r}-\textbf{r}_{i})\delta(\textbf{r}'-\textbf{r}_{j})\times$$
\begin{equation} \label{aSRIff j 2 def}
\times\frac{1}{2m_{i}}(\Psi^{*}(R,t)\hat{\textbf{p}}_{i}\Psi(R,t)+c.c.) .\end{equation}
If quantum correlations are dropped function $j_{2}^{\alpha}(\textbf{r},\textbf{r}',t)$ splits on product of the current $j^{\alpha}(\textbf{r},t)$ and the concentration $n(\textbf{r}',t)$.

Extended sets of hydrodynamic equations are used in plasma physics.
An example is discussed in Ref. \cite{Miller PoP 16}.

The application of the kinetic equation suggests that a partial truncation is made
since the kinetic equations form a chain of equations,
hence the single kinetic equation such as the Vlasov or Boltzmann equation is a truncated model.
Deriving the hydrodynamic equations from the microscopic theory
we need to introduce all necessary truncations in terms of hydrodynamic variables or the wave functions constructing the hydrodynamic variables.

\subsection{Velocity field in the hydrodynamic equations}

Traditionally the hydrodynamic equations are written in terms of the velocity field $\textbf{v}$.
The transition to the velocity field also allows to calculate the functions related to the thermal effects
or other mechanisms of the distribution of particles on quantum states with different energies like the Pauli blocking.

The velocity field itself can be defined via the concentration of particles $n(\textbf{r},t)$ and
the particle current $\textbf{j}(\textbf{r},t)$ in the following way: $\textbf{v}=\textbf{j}/n$.

Next, consider the particle current in more details.
To this end,
represent the many-particle wave function $\Psi$ via two real functions $\Psi(R,t)=a(R,t)\exp(\imath S(R,t)/\hbar)$
(the exponential form of the complex function),
where $a(R,t)\in\Re$ and $S(R,t)\in\Re$.
Therefore, the current $\textbf{j}$ (\ref{aSRIff j def}) can be represented as follows
\begin{equation} \label{aSRIff} \textbf{j}=\frac{\hbar}{m}\int dR\sum_{i}\delta(\textbf{r}-\textbf{r}_{i})a^{2}\nabla_{i}S, \end{equation}
where it is assumed that
all particles belongs to the single species,
and, therefore,
have equal mass.

Function $\hbar\nabla_{i}S/m$ can be interpreted as the velocity of $i$-th quantum particle.
It is in agreement with the fact that the current $\textbf{j}$ is proportional to the velocity field
being the average velocity $\textbf{v}$ multiplied by the concentration $n$.
Since $\textbf{v}$ is the average velocity the full velocity of each particle is the superposition of the average velocity and the deviation from the average velocity $\textbf{u}_{i}(\textbf{r},R,t)$.
Mostly this deviation is related to the thermal motion.
Therefore,
it is called the thermal velocity of $i$-th particle.
It leads to a representation of the current
\begin{equation} \label{aSRIff}
\textbf{j}=\int dR\sum_{i}\delta(\textbf{r}-\textbf{r}_{i})a^{2}(\textbf{v}+\textbf{u}_{i})=n\textbf{v}, \end{equation}
which gives the following equation for the thermal part of the current
\begin{equation} \label{aSRIff condition of u goes to zero}
\textbf{j}_{th}=\int dR\sum_{i}\delta(\textbf{r}-\textbf{r}_{i})a^{2}\textbf{u}_{i}=0. \end{equation}

Similar can be written for the two-particle current-concentration function
$$\textbf{j}_{2}(\textbf{r},\textbf{r}',t)=\frac{\hbar}{m}\int dR\sum_{i,j\neq i}\delta(\textbf{r}-\textbf{r}_{i})\delta(\textbf{r}'-\textbf{r}_{j})a^{2}(\textbf{v}(\textbf{r},t)+\textbf{u}_{i})$$
\begin{equation} \label{aSRIff j 2 via v n2} =\textbf{v}(\textbf{r},t)\cdot n_{2}(\textbf{r},\textbf{r}',t)
+\textbf{J}_{2}(\textbf{r},\textbf{r}',t), \end{equation}
with
\begin{equation} \label{aSRIff} \textbf{J}_{2}(\textbf{r},\textbf{r}',t)=
\int dR\sum_{i,j\neq i}\delta(\textbf{r}-\textbf{r}_{i})\delta(\textbf{r}'-\textbf{r}_{j})a^{2}\textbf{u}_{i}, \end{equation}
where the last term is not equal to zero,
since it contains an extra delta function under the integral.
Obtained representation is in agreement with the correlationless form of $\textbf{j}_{2}$ described after equation (\ref{aSRIff j 2 def}).
Consider it in more depth.
Function $n_{2}(\textbf{r},\textbf{r}',t)$ splits as $n_{2}(\textbf{r},\textbf{r}',t)=n(\textbf{r},t) n(\textbf{r}',t)$ and $n\textbf{v}=\textbf{j}$.
The second term in (\ref{aSRIff j 2 via v n2}) splits on $n(\textbf{r}',t)\cdot\int dR\sum_{i}\delta(\textbf{r}-\textbf{r}_{i})a^{2}\textbf{u}_{i}$,
where the last multiplier is equal to zero.

Use the exponential form of the complex function $\Psi(R,t)=a(R,t)\exp(\imath S(R,t)/\hbar)$ for the analysis of the momentum flux $\Pi^{\alpha\beta}$.
Substitute equation $\Psi(R,t)=a(R,t)\exp(\imath S(R,t)/\hbar)$ in the definition of the momentum flux (\ref{aSRIff Pi def}).
After some calculations, find the following microscopic representation
$$\Pi^{\alpha\beta}=\int dR\sum_{i}\delta(\textbf{r}-\textbf{r}_{i})\frac{1}{2m_{i}^{2}}\times$$
\begin{equation} \label{aSRIff Pi alpha beta via S and a}\times[2(\partial_{i}^{\alpha}S)(\partial_{i}^{\beta}S)a^{2} +\hbar^{2}(\partial_{i}^{\alpha}a)(\partial_{i}^{\beta}a)-\hbar^{2}a\partial_{i}^{\alpha}\partial_{i}^{\beta}a].\end{equation}

As it is stated above $\nabla_{i}^{\alpha}S/m$ is the microscopic velocity of $i$-th quantum particle.
Therefore, the first term in equation (\ref{aSRIff Pi alpha beta via S and a}) can be rewritten as
\begin{equation} \label{aSRIff Pi alpha beta via v and a} \Pi^{\alpha\beta}_{cl} =\int dR\sum_{i}\delta(\textbf{r}-\textbf{r}_{i})v_{i}^{\alpha}v_{i}^{\beta}a^{2} .\end{equation}
Next, split the velocity of each part on the local average velocity and the thermal velocity $v_{i}^{\alpha}(R,t)=v^{\alpha}(\textbf{r},t)+u_{i}^{\alpha}(R,\textbf{r},t)$.
It gives four terms.
Two of them are equal to zero due to the condition (\ref{aSRIff condition of u goes to zero}).
Two nonzero terms can be written in the following form
$\Pi^{\alpha\beta}_{cl}=nv^{\alpha}v^{\beta}+p^{\alpha\beta}$,
where
\begin{equation} \label{aSRIff}
p^{\alpha\beta}=\int dR\sum_{i}\delta(\textbf{r}-\textbf{r}_{i})a^{2}u_{i}^{\alpha}u_{i}^{\beta}\end{equation}
is the thermal pressure tensor.
The thermal pressure tensor in a comoving frame
(it is a remnance of the stress tensor in the noninteracting limit)
becomes diagonal $p^{\alpha\beta}=p\cdot\delta^{\alpha\beta}$,
where $p$ is the local pressure.
Tensor $p^{\alpha\beta}=p\cdot\delta^{\alpha\beta}$ is related to the distribution of particles on quantum states with different momentum.
In the degenerate regime, the nondiagonal elements of this tensor describe the Fermi sphere deformation.

Consider the two last terms in equation (\ref{aSRIff Pi alpha beta via S and a}).
They are proportional to the square of the Plank constant $\hbar^{2}$.
Hence,
it is expected that they give some quantum contribution in the momentum flux:
\begin{equation} \label{aSRIff T alpha beta via a}
T^{\alpha\beta}= \int dR\sum_{i}\delta(\textbf{r}-\textbf{r}_{i})
\frac{\hbar^{2}}{2m_{i}^{2}}[\partial^{\alpha}_{i}a\cdot\partial^{\beta}_{i}a-a\partial^{\alpha}_{i}\partial^{\beta}_{i}a]. \end{equation}
Tensor $T^{\alpha\beta}$ does not have any straightforward representation in terms of the hydrodynamic variables.

Start the analysis of $T^{\alpha\beta}$ with the single particle case.
In this case $n=a^{2}$, $\textbf{v}=\nabla S/m$, $\textbf{u}=0$,
and
\begin{equation} \label{aSRIff Bohm tensor single part}
T^{\alpha\beta}=-\frac{\hbar^{2}}{4m^{2}}\biggl[\partial_{\alpha}\partial_{\beta}n
-\frac{\partial_{\alpha}n\cdot\partial_{\beta}n}{n}\biggr].\end{equation}

The momentum balance equation contains the divergence of tensor $T^{\alpha\beta}$:
\begin{equation} \label{aSRIff Bohm potential divergence single part}  \partial_{\beta}T^{\alpha\beta}= -\frac{\hbar^{2}}{4m^{2}}n \partial^{\alpha}\frac{\triangle\sqrt{n}}{\sqrt{n}}. \end{equation}

Let us mention that
tensor $T^{\alpha\beta}$ simplifies for the bosons being in the Bose-Einstein condensate state due to the fact that
all particles are in the same state.
Hence, the calculations almost reduces to the single particle case.

Another example is the ideal gas of fermions at the arbitrary temperature.
The single particle wave function of all fermions are well-known.
Hence, use them to calculate (\ref{aSRIff T alpha beta via a}).

The plane waves $\varphi_{\textbf{k}}=A\cdot e^{\imath \textbf{k}\textbf{r}}$ have constant amplitudes, so $\partial_{i}^{\alpha}a=0$.
It gives the quantum Bohm potential equal to zero.

This example of the explicit calculation of the quantum Bohm potential for particular cases.
However, a part of tensor $T^{\alpha\beta}$ can be calculated for the arbitrary single particle wave functions.
Consider
$\partial^{\alpha}_{i}\partial^{\beta}_{i}a^{2}$
$=2\partial^{\alpha}_{i}a\cdot\partial^{\beta}_{i}a+2\partial^{\alpha}_{i}\partial^{\beta}_{i}a$,
then
\begin{equation} \label{aSRIff T partially via n} T^{\alpha\beta}=-\frac{\hbar^{2}}{4m^{2}}\partial^{\alpha}\partial^{\beta}n
+\frac{\hbar^{2}}{m^{2}}\int dR\sum_{i}\delta(\textbf{r}-\textbf{r}_{i})\partial^{\alpha}_{i}a\cdot\partial^{\beta}_{i}a.\end{equation}
The first term in equation (\ref{aSRIff T partially via n}) corresponds to the linear part of $T^{\alpha\beta}$ at the analysis of the small amplitude perturbations.
Hence, it can be used to study the wave phenomena.
However, the second term requires an equation of state.
As a rough approximation, allowing an estimation of the nonlinear term contribution,
consider the nonlinear term existing in the single particle case (\ref{aSRIff Bohm tensor single part}).

It provides the structure of the momentum flux tensor:
\begin{equation} \label{aSRIff}\Pi^{\alpha\beta}=nv^{\alpha}v^{\beta} +p^{\alpha\beta}+T^{\alpha\beta}.\end{equation}

Similar calculations which are rather more bulky gives the representation of flux of the momentum flux:
$$M^{\alpha\beta\gamma}= nv^{\alpha}v^{\beta}v^{\gamma} +v^{\alpha} p^{\beta\gamma} +v^{\beta} p^{\alpha\gamma} $$
\begin{equation} \label{aSRIff} +v^{\gamma} p^{\alpha\beta}+Q^{\alpha\beta\gamma}+T^{\alpha\beta\gamma}+L^{\alpha\beta\gamma}, \end{equation}
where
\begin{equation} \label{aSRIff Q definition} Q^{\alpha\beta\gamma}=\int dR\sum_{i}\delta(\textbf{r}-\textbf{r}_{i})a^{2}u_{i}^{\alpha}u_{i}^{\beta}u_{i}^{\gamma} \end{equation}
presents the purely thermal part of tensor $M^{\alpha\beta\gamma}$,
$$T^{\alpha\beta\gamma}
=\frac{\hbar^{2}}{2m^{2}}\biggl[-\frac{1}{6}n(\partial^{\alpha}\partial^{\beta} v^{\gamma}
+\partial^{\alpha}\partial^{\gamma} v^{\beta}
+\partial^{\beta}\partial^{\gamma} v^{\alpha})$$
$$-\sqrt{n}\partial^{\beta}\partial^{\gamma}\sqrt{n}\cdot v^{\alpha}
-\sqrt{n}\partial^{\alpha}\partial^{\beta}\sqrt{n}\cdot v^{\gamma}
-\sqrt{n}\partial^{\alpha}\partial^{\gamma}\sqrt{n}\cdot v^{\beta}$$
\begin{equation} \label{aSRIff T alpha beta gamma explicit}
+\partial^{\beta}\sqrt{n}\cdot\partial^{\gamma}\sqrt{n}\cdot v^{\alpha}
+\partial^{\alpha}\sqrt{n}\cdot\partial^{\beta}\sqrt{n}\cdot v^{\gamma}
+\partial^{\alpha}\sqrt{n}\cdot\partial^{\gamma}\sqrt{n}\cdot v^{\beta}\biggr] \end{equation}
gives the purely quantum part of tensor $M^{\alpha\beta\gamma}$
(equation (\ref{aSRIff T alpha beta gamma explicit}) is a simplified form of tensor $T^{\alpha\beta\gamma}$
analogous to equation (\ref{aSRIff Bohm tensor single part}),
the general form of $T^{\alpha\beta\gamma}$ similar to equation (\ref{aSRIff T alpha beta via a}) is not demonstrated),
and
$L^{\alpha\beta\gamma}$ presents quantum-thermal terms
$$L^{\alpha\beta\gamma}=\int dR\sum_{i}\frac{\hbar^{2}}{2m_{i}^{2}}\delta(\textbf{r}-\textbf{r}_{i}) \times$$ $$\times\biggl[-\frac{1}{6}a^{2}(\partial_{i}^{\alpha}\partial_{i}^{\beta}u_{i}^{\gamma} +\partial_{i}^{\beta}\partial_{i}^{\gamma}u_{i}^{\alpha}+\partial_{i}^{\alpha}\partial_{i}^{\gamma}u_{i}^{\beta})$$
$$-a(\partial_{i}^{\beta}\partial_{i}^{\gamma}a)u_{i}^{\alpha} -a(\partial_{i}^{\alpha}\partial_{i}^{\gamma}a)u_{i}^{\beta}-(\partial_{i}^{\alpha}\partial_{i}^{\beta}a)u_{i}^{\gamma}$$
\begin{equation} \label{aSRIff L alpha beta gamma via thermal velocity}  +(\partial_{i}^{\beta}a)(\partial_{i}^{\gamma}a)u_{i}^{\alpha} +(\partial_{i}^{\alpha}a)(\partial_{i}^{\gamma}a)u_{i}^{\beta}+(\partial_{i}^{\alpha}a)(\partial_{i}^{\beta}a)u_{i}^{\gamma}\biggr]. \end{equation}

Next consider equation of state for tensor $L^{\alpha\beta\gamma}$ in the weakly interacting limit.
To this end, the plane function approximation for the single particle states can be used.
For the plane waves the amplitude $a$ is a constant.
Therefore, the first group of terms in equation (\ref{aSRIff L alpha beta gamma via thermal velocity}) can  be nonzero,
while other terms containing the derivatives of the amplitude are equal to zero.
Consider the first group of terms.
The space derivative can be taken out of the integral since the amplitude is constant.
Remaining integral is proportional to the thermal current (\ref{aSRIff condition of u goes to zero}).
Hence, it is equal to zero.
Therefore, it is obtained that tensor $L^{\alpha\beta\gamma}=0$ in the plane wave approximation.
This result is used below for the truncation of the chain of equations.

Tensor $T^{\alpha\beta\gamma}$ presented by equation (\ref{aSRIff T alpha beta gamma explicit}) can be presented in a form similar to representation (\ref{aSRIff T alpha beta via a}) for tensor $T^{\alpha\beta}$.

The first group of terms in equation
(\ref{aSRIff T alpha beta gamma explicit})
which is proportional to $n/3$ is the exact result for the arbitrary distribution of particles on quantum states.
Other terms in equation
(\ref{aSRIff T alpha beta gamma explicit})
are written in the single particle approximation as an approximate equation of state for the system of degenerate fermions.

Substitute tensors $\Pi^{\alpha\beta}$ and $M^{\alpha\beta\gamma}$
presented via the velocity field into the left-hand side of equation (\ref{aSRIff eq for Pi alpha beta})
find the following representation of the left-hand side
$$\partial_{t}\Pi^{\alpha\beta}+\partial_{\gamma}M^{\alpha\beta\gamma}= \partial_{t}p^{\alpha\beta}+\partial_{\gamma}T^{\alpha\beta\gamma}$$
$$+\partial_{\gamma}L^{\alpha\beta\gamma}+\partial_{\gamma}Q^{\alpha\beta\gamma} +p^{\alpha\gamma}\partial_{\gamma}v^{\beta}+p^{\beta\gamma}\partial_{\gamma}v^{\alpha}$$
\begin{equation} \label{aSRIff eq for Pi alpha beta lhs via vel}
+\partial_{\gamma}(p^{\alpha\beta}v^{\gamma}) +\frac{1}{m}v^{\alpha}F^{\beta}+\frac{1}{m}v^{\beta}F^{\alpha}+(\partial_{t}+v^{\gamma}\partial_{\gamma})T^{\alpha\beta},\end{equation}
where the continuity and Euler equations are used to eliminate $\partial_{t}n$ and $\partial_{t}\textbf{v}$.

Next, consider the representation of the right-hand side of the momentum flux evolution equation (\ref{aSRIff eq for Pi alpha beta})
at the introduction of the velocity field
$$\partial_{t}\Pi^{\alpha\beta}+\partial_{\gamma}M^{\alpha\beta\gamma}=\frac{1}{m}v^{\alpha}F^{\beta}+\frac{1}{m}v^{\beta}F^{\alpha}$$
$$-\frac{1}{m}\int[\partial^{\alpha}U(\textbf{r}-\textbf{r}')]J_{2}^{\beta}(\textbf{r},\textbf{r}',t)d\textbf{r}'$$
\begin{equation} \label{aSRIff eq for Pi alpha beta rhs via vel}
-\frac{1}{m}\int[\partial^{\beta}U(\textbf{r}-\textbf{r}')]J_{2}^{\alpha}(\textbf{r},\textbf{r}',t)d\textbf{r}'. \end{equation}

Combine equations (\ref{aSRIff eq for Pi alpha beta lhs via vel}) and (\ref{aSRIff eq for Pi alpha beta rhs via vel})
\emph{and} find
$$\partial_{t}p^{\alpha\beta} +v^{\gamma}\partial_{\gamma}p^{\alpha\beta} +p^{\alpha\gamma}\partial_{\gamma}v^{\beta} +p^{\beta\gamma}\partial_{\gamma}v^{\alpha}$$
$$+p^{\alpha\beta}\partial_{\gamma}v^{\gamma} +\partial_{\gamma}Q^{\alpha\beta\gamma} +\partial_{\gamma}T^{\alpha\beta\gamma}
+\partial_{\gamma}L^{\alpha\beta\gamma}$$
$$+\frac{\hbar^{2}}{4m^{2}}\biggl[\partial_{\alpha}\partial_{\beta}\partial_{\gamma}(nv^{\gamma})
-v^{\gamma}\partial_{\alpha}\partial_{\beta}\partial_{\gamma}n
-\frac{1}{n}(\partial_{\gamma}v^{\gamma})\partial_{\alpha}n\cdot\partial_{\beta}n$$
$$ -\frac{\partial_{\beta}n}{n}\cdot \partial_{\gamma}(n\cdot\partial_{\alpha}v^{\gamma})
-\frac{\partial_{\alpha}n}{n}\cdot \partial_{\gamma}(n\cdot\partial_{\beta}v^{\gamma}) \biggr]$$
$$=-\frac{1}{m}\int[\partial^{\alpha}U(\textbf{r}-\textbf{r}')]J_{2}^{\beta}(\textbf{r},\textbf{r}',t)d\textbf{r}'$$
\begin{equation} \label{aSRIff eq for p alpha beta}
-\frac{1}{m}\int[\partial^{\beta}U(\textbf{r}-\textbf{r}')]J_{2}^{\alpha}(\textbf{r},\textbf{r}',t)d\textbf{r}'. \end{equation}

The pressure $p^{\alpha\beta}$ is an isotropic tensor $p^{\alpha\beta}=p^{\beta\alpha}$.
In the simple isotropic equilibrium states described by the Maxwellian or Fermi-Dirac distribution
the thermal pressure tensor is proportional to the unit tensor $p^{\alpha\beta}_{eq}=p\cdot\delta^{\alpha\beta}$.
The scalar pressure $p$ is a traditional physical quantity.
Moreover, the existence of the scalar pressure illustrates the isotropy of the system.
If there is an anisotropy direction
the diagonal pressure includes different elements $diag(p^{\alpha\beta})=\{p_{\perp},p_{\perp},p_{\parallel}\}$.
Hence, it is useful to use the scalar pressure $p$ for the isotropic systems.
However, the deviation from the equilibrium state existing in waves and other phenomena leads to nonzero nondiagonal elements
which can be introduced as an independent variable $\pi^{\alpha\beta}$.
Hence, the pressure tensor has the following structure:
$p^{\alpha\beta}=p\cdot\delta^{\alpha\beta}+\pi^{\alpha\beta}$,
where $Tr\pi^{\alpha\beta}=\pi^{\alpha\alpha}=0$,
and $p=Tr p^{\alpha\beta}/3$.

All terms (except $\partial_{\gamma}T^{\alpha\beta\gamma}$) in equation (\ref{aSRIff eq for p alpha beta}) are proportional to the thermal velocity.

The left-hand side of equation (\ref{aSRIff eq for p alpha beta}) contains the kinematic terms, while the right-hand side contains interaction.
It is interesting to point out that there is no action of the external potential in this equation.

Consider equation for the scalar pressure $p$.
Multiply equation (\ref{aSRIff eq for p alpha beta}) by $\delta^{\alpha\beta}/3$
$$\partial_{t}p +v^{\alpha}\partial_{\alpha}p +\frac{5}{3}p\partial_{\alpha}v^{\alpha}+\frac{2}{3}\pi^{\alpha\gamma}\partial_{\gamma}v^{\alpha}$$
$$+\frac{1}{3}\partial_{\gamma}Q^{\alpha\alpha\gamma} +\frac{1}{3}\partial_{\gamma}T^{\alpha\alpha\gamma} +\frac{1}{3}\partial_{\gamma}L^{\alpha\alpha\gamma}$$
$$+\frac{\hbar^{2}}{4m^{2}}\biggl[\triangle\nabla(n\textbf{v})-\textbf{v}\triangle\nabla n -\frac{1}{n}(\nabla\textbf{v}) (\nabla n)^{2}$$
$$-2\partial^{\beta}n\cdot\partial^{\alpha}\partial^{\beta}v^{\alpha} -\frac{2}{n}\partial^{\beta}n\cdot\partial^{\alpha}n\cdot\partial^{\beta}v^{\alpha}\biggr]$$
\begin{equation} \label{aSRIff eq for scalar p} =-\frac{2}{3m}\int\partial^{\alpha}U(\textbf{r}-\textbf{r}') J_{2}^{\alpha}(\textbf{r},\textbf{r}',t)d\textbf{r}'.\end{equation}

Consider the equilibrium case for the noninteracting classic particles
then we can drop the right-hand side of equation (\ref{aSRIff eq for scalar p}),
drop $L^{\alpha\alpha\gamma}$ and $T^{\alpha\alpha\gamma}$ (for classic particles),
and drop $\pi^{\alpha\gamma}$ since $\pi^{\alpha\gamma}$ corresponds to the nonequilibrium states.

Introduce the full derivative $d/dt\equiv \partial_{t} +\textbf{v}\cdot\nabla$
and apply the continuity equation for $\partial_{\alpha}v^{\alpha}=(1/n)dn/dt$.

After described simplifications and manipulations equation (\ref{aSRIff eq for scalar p}) simplifies to
\begin{equation} \label{aSRIff} \frac{dp}{dt} +\frac{5}{3}\frac{p}{n}\frac{dn}{dt}+\frac{1}{3}\partial_{\alpha}q^{\alpha}=0, \end{equation}
where $q^{\alpha}\equiv Q^{\alpha\beta\beta}=Q^{\beta\beta\alpha}$.

Let us make few comments on the truncation of the derived chain of equations for degenerate fermions.
Account of the next order equations
(equation for the average of higher degree of the momentum operator $j^{\alpha}$, $\Pi^{\alpha\beta}$, $M^{\alpha\beta\gamma}$, etc) and
calculation of appearing corrections gives a possibility to understand validity of the application of the hydrodynamic equation set restricted
by the lower order \cite{Tokatly PRB 99, Tokatly PRB 00}.
Functions $\pi^{\alpha\beta}$ and $Q^{\alpha\beta\gamma}$ vanish for the locally equilibrium distribution functions.

\section{Short range interaction in the momentum balance equation}

The short range interaction is presented in two equations.
They are the momentum balance equation and the momentum flux evolution equation.
Consider them separately.

The analysis is different for the identical particles and the interaction of particles of different species.
Start our consideration with the identical particles.

\subsection{Identical particles}

The hydrodynamic equations can be truncated after obtaining  of the momentum balance equation or the after the account of the higher rank tensors evolution like the momentum flux $\Pi^{\alpha\beta}$.
Anyway, it requires approximate calculation of the force field.
An equation for the two-particle concentration
or for the quantum stress tensor can give more detailed picture of interaction,
but these generalizations will be consider elsewhere.
Present the force field in terms of the many-particle wave function with no application of two-particle concentration:
$$\textbf{F}_{int}(\textbf{r},t)$$
\begin{equation} \label{aSRIff ForceField} =-\int dR\sum_{i,j\neq i}\delta(\textbf{r}-\textbf{r}_{i})(\nabla_{i}U(\textbf{r}_{ij}))\Psi^{*}(R,t)\Psi(R,t).\end{equation}

At the description of the identical particles
it can be symmetrized relatively pair of interacting particles
$$\textbf{F}_{int}(\textbf{r},t)=-\frac{1}{2}\int dR\sum_{i,j\neq i}[\delta(\textbf{r}-\textbf{r}_{i})-\delta(\textbf{r}-\textbf{r}_{j})]\times$$
\begin{equation} \label{aSRIff ForceField symm} \times\nabla_{i}U(\textbf{r}_{ij})\cdot\Psi^{*}(R,t)\Psi(R,t) .\end{equation}
Next, introducing the coordinates of relative motion and center of mass for each pair of particles $\textbf{R}_{ij}=\frac{1}{2}(\textbf{r}_{i}+\textbf{r}_{j})$, $\textbf{r}_{ij}=\textbf{r}_{i}-\textbf{r}_{j}$, represent coordinates of $i$-th and $j$-th particles
via $\textbf{r}_{ij}$ and $\textbf{R}_{ij}$.

Here, the symmetry of the wave function $\Psi(R,t)$ relatively the permutation of particles is used.
Function $\Psi(..., \textbf{r}_{i}, ..., \textbf{r}_{j}, ..., t)$ is presented in equation (\ref{aSRIff ForceField})
and in the first term in equation (\ref{aSRIff ForceField symm}).
Function $\Psi(..., \textbf{r}_{j}, ..., \textbf{r}_{i}, ..., t)$ is used in the second term in equation (\ref{aSRIff ForceField symm}),
where notation $j$ is used instead of $i$,
but it is represented via $(-1)\Psi(..., \textbf{r}_{i}, ..., \textbf{r}_{j}, ..., t)$.
However,
the sign does not affect the square of the wave function module $\Psi^{*}(R,t)\Psi(R,t)$.

The wave functions $\Psi(R,t)$ entering the equation (\ref{aSRIff ForceField symm}) have the following explicit structure of arguments:
\begin{equation} \label{aSRIff Psi with arguments} \Psi(\textbf{r}_{1},\textbf{r}_{2}, ..., \textbf{R}_{ij}+\textbf{r}_{ij}/2,... ,\textbf{R}_{ij}-\textbf{r}_{ij}/2, ..., \textbf{r}_{N},t)\end{equation}

After the expansion in the Teylor series up to the third order of the small parameter $\textbf{r}_{ij}$
obtain the following expression:\begin{widetext}
$$\textbf{F}_{int}=\frac{1}{2}\int dR\sum_{i,j\neq i}\nabla_{i}U(\textbf{r}_{ij})
\Biggl[r_{ij}^{\alpha}\partial^{\alpha}\delta_{i}\Psi^{*}(R',t)\Psi(R',t)
+\frac{1}{2}r_{ij}^{\alpha}r_{ij}^{\beta}\partial^{\alpha}\delta_{i}
\Biggl(\Psi^{*}(R',t)\biggl(\partial_{R1}^{\beta}\Psi(R',t)-\partial_{R2}^{\beta}\Psi(R',t)\biggr)+c.c.\Biggr)$$
$$+\frac{1}{4}r_{ij}^{\alpha}r_{ij}^{\beta}r_{ij}^{\gamma}\partial^{\alpha}\delta_{i} \Biggl(\Psi^{*}(R',t)\biggl(\frac{1}{2}\partial_{R1}^{\beta}\partial_{R1}^{\gamma}\Psi(R',t)
+\frac{1}{2}\partial_{R2}^{\beta}\partial_{R2}^{\gamma}\Psi(R',t)
-\partial_{R1}^{\beta}\partial_{R2}^{\gamma}\Psi(R',t)\biggr)$$
$$+\biggl(\frac{1}{2}\partial_{R1}^{\beta}\partial_{R1}^{\gamma}\Psi^{*}(R',t)
+\frac{1}{2}\partial_{R2}^{\beta}\partial_{R2}^{\gamma}\Psi^{*}(R',t)
-\partial_{R1}^{\beta}\partial_{R2}^{\gamma}\Psi^{*}(R',t)\biggr)\Psi(R',t)$$
\begin{equation} \label{aSRIff F expansion up to the third order}
+\biggl(\partial_{R1}^{\beta}\Psi^{*}(R',t)-\partial_{R2}^{\beta}\Psi^{*}(R',t)\biggr) \biggl(\partial_{R1}^{\gamma}\Psi(R',t)-\partial_{R2}^{\gamma}\Psi(R',t)\biggr)\Biggr)
+\frac{1}{24}r_{ij}^{\alpha}r_{ij}^{\beta}r_{ij}^{\gamma} \partial^{\alpha}\partial^{\beta}\partial^{\gamma}\delta_{i} \cdot\Psi^{*}(R',t)\Psi(R',t)\Biggr],\end{equation}\end{widetext}
where
$\delta_{i}=\delta(\textbf{r}-\textbf{R}_{ij})$,
$R'=\{\textbf{r}_{1}, ..., \textbf{R}_{ij}, ..., \textbf{R}_{ij}, ..., \textbf{r}_{N}\}$,
$\partial_{R1}=\frac{\partial}{\partial \textbf{R}_{ij}}$ for $\textbf{R}_{ij}$ located at the $i$-th place,
$\partial_{R2}=\frac{\partial}{\partial \textbf{R}_{ij}}$ for $\textbf{R}_{ij}$ located at the $j$-th place,
and $c.c.$ is the complex conjugation.
In equation (\ref{aSRIff F expansion up to the third order}) and similar equations below we have $dR=dR_{N-2} dR_{ij} dr_{ij}$.
Einstein rule of summation on the repeating index is assumed here and below.

The expression (\ref{aSRIff F expansion up to the third order}) can be rewritten in a more compact form
via the derivatives of the product of the wave functions $\Psi^{*}(R',t)\Psi(R',t)$.
However, the explicit form (\ref{aSRIff F expansion up to the third order}) is more useful for the further analysis.

Consider the spin-polarized fermions with the full spin polarization.
It means that the spin part of the wave function is symmetric relatively the permutations of particles
while the coordinate part of the wave function is antisymmetric.
The force field (\ref{aSRIff F expansion up to the third order}) explicitly contains the wave function $\Psi(R',t)$,
where two arguments coincides.
Hence, the terms containing $\Psi(R',t)$ are equal to zero since the function $\Psi(R',t)=0$ due to the antisymmetry.

Consequently,
equation (\ref{aSRIff F expansion up to the third order}) simplifies to the following force field
$$\textbf{F}_{int}=
\frac{1}{8}\int dR\sum_{i,j\neq i}\nabla_{i}U(\textbf{r}_{ij})
\Biggl[r_{ij}^{\alpha}r_{ij}^{\beta}r_{ij}^{\gamma}\partial^{\alpha}\delta_{i} \biggl(\partial_{R1}^{\beta}\Psi^{*}(R',t)$$
\begin{equation} \label{aSRIff F expansion up to the third order Simplified}
-\partial_{R2}^{\beta}\Psi^{*}(R',t)\biggr)
\biggl(\partial_{R1}^{\gamma}\Psi(R',t)-\partial_{R2}^{\gamma}\Psi(R',t)\biggr)\Biggr].\end{equation}

The force field (\ref{aSRIff F expansion up to the third order}) can be rewritten via the stress tensor of quantum gas: $F^{\alpha}_{int}=-\partial_{\beta}\sigma^{\alpha\beta}$.
It happens that this representation of the force field of the short-range interaction appears for the interspecies interaction.
The delta function in equation is the single part of the equation
which depends on the coordinate $\textbf{r}$.
Therefore, the derivative $\partial^{\alpha}$ can be placed in front of the integral.
Recalling the summation index from $\alpha$ to $\beta$
while free index placed in $\nabla_{i}U(\textbf{r}_{ij})$ call $\alpha$.
It provides the final form of the quantum stress tensor
$$\sigma^{\alpha\beta}(\textbf{r},t)=
-\frac{1}{8}\int dR\sum_{i,j\neq i}\frac{\partial U(\textbf{r}_{ij})}{\partial r_{ij}^{\alpha}}\times$$
$$\times\Biggl[r_{ij}^{\beta}r_{ij}^{\gamma}r_{ij}^{\delta}
\delta(\textbf{r}-\textbf{R}_{ij}) \biggl(\partial_{R1}^{\gamma}\Psi^{*}(R',t)$$
\begin{equation} \label{aSRIff F expansion up to the third order Simplified}
-\partial_{R2}^{\gamma}\Psi^{*}(R',t)\biggr)
\biggl(\partial_{R1}^{\delta}\Psi(R',t)-\partial_{R2}^{\delta}\Psi(R',t)\biggr)\Biggr].\end{equation}

The quantum stress tensor allows to rewrite the Euler equation in the following form:
\begin{equation}\label{aSRIff Euler via vel} mn(\partial_{t}+\textbf{v}\cdot\nabla)v^{\alpha}+ \partial_{\beta}(p^{\alpha\beta}+\sigma^{\alpha\beta}+T^{\alpha\beta}) =-n\partial^{\alpha}V_{ext} ,\end{equation}
where the current is represented via the velocity field $\textbf{j}(\textbf{r},t)=n(\textbf{r},t)\textbf{v}(\textbf{r},t)$.

Integral over $\textbf{r}_{ij}$ and integral over $\textbf{R}_{ij}$ are independent.
Separate them explicitly.
Include that the sum over $i$ and $j$ gives $N(N-1)$ identical pairs of the interacting particles.
Integral over $\textbf{R}_{ij}$ is illiminated via the delta function $\delta(\textbf{r}-\textbf{R}_{ij})$.
The result for the quantum stress tensor can be presented in the following form
\begin{equation} \label{aSRIff sigma via Upsilon and Tr g alpha beta}
\sigma^{\alpha\beta}=-\frac{1}{8}\Upsilon_{2}^{\alpha\beta\gamma\delta} Tr g^{\gamma\delta}(\textbf{r},\textbf{r}',t), \end{equation}
where
\begin{equation} \label{aSRIff Upsilon 2 tensor DEF} \Upsilon_{2}^{\alpha\beta\gamma\delta}=\int
r^{\beta}r^{\gamma}r^{\delta}\frac{\partial U(\textbf{r})}{\partial r^{\alpha}}d\textbf{r}, \end{equation}
and
$$g^{\alpha\beta}(\textbf{r},\textbf{r}',t)=N(N-1)\int dR_{N-2}
\biggl(\partial_{1}^{\alpha}\Psi^{*}(R'',t)$$
\begin{equation} \label{aSRIff g alpha beta r r'}
-\partial_{2}^{\alpha}\Psi^{*}(R'',t)\biggr)
\biggl(\partial_{1}^{\beta}\Psi(R'',t)-\partial_{2}^{\beta}\Psi(R'',t)\biggr),\end{equation}
with
$R''=\{\textbf{r}_{1}, ..., \textbf{r}, ..., \textbf{r}', ..., \textbf{r}_{N}\}$,
where
$\textbf{r}$ and $\textbf{r}'$ are placed in $i$-th and $j$-th places, correspondingly.

Particularly, the calculation of the tensor $\Upsilon_{2}^{\alpha\beta\gamma\delta}$ (\ref{aSRIff Upsilon 2 tensor DEF}) for the isotropic potential of the interatomic interaction gives one scalar interaction constant in the following way:
\begin{equation} \label{aSRIff tensor structure in 3 order} \Upsilon_{2}^{\alpha\beta\gamma\delta}=
-g_{2}I_{0}^{\alpha\beta\gamma\delta},\end{equation}
where
$g_{2}=\frac{1}{3}\int r^{2}U(r)d\textbf{r}$,
and
\begin{equation} \label{aSRIff tensor I 0}
I_{0}^{\alpha\beta\gamma\delta}=
\delta^{\alpha\beta}\delta^{\gamma\delta}+\delta^{\alpha\gamma}\delta^{\beta\delta}+\delta^{\alpha\delta}\delta^{\beta\gamma}.\end{equation}

Consider the interaction of fermions with the same spin projection.
It describes the fermions if the system of fermions has the full spin polarization.
Or it gives the partial description for the partial spin polarization or for the zero spin polarization.

Let us present further calculation of equation (\ref{aSRIff g alpha beta r r'}).
To this end, write the many-particle wave function in the representation of the occupation numbers
\begin{equation}\label{aSRIff} \Psi(R'',t)=\langle \textbf{r}, \textbf{r}', R_{N-2},t |n_1, n_2 \ldots\rangle \end{equation}

Next, present an expansion of the wave function in the weakly interacting limit as the Slater determinant \cite{Shveber}:
$$\langle \textbf{r}, \textbf{r}', R_{N-2},t |n_1, n_2 \ldots\rangle$$
$$=\sum_f \sum_{f'<f}
\sqrt{\frac{n_f}{N}} \sqrt{\frac{n_{f'}}{N-1}} (-1)^{\sum\limits_{f'\le q<f} n_q }$$
$$\: \left(\:
\langle \textbf{r},t | f\rangle \: \langle\textbf{r}',t | f'\rangle-
\langle\textbf{r}',t | f\rangle \: \langle\textbf{r},t | f'\rangle
\:\right)\times$$
\begin{equation}\times
\langle R_{N-2},t |n_1, \ldots (n_{f'}-1),\ldots (n_f-1), \ldots \rangle.
\label{aSRIff fermi determinant expansion}
\end{equation}

The following normalization integral is used for the $N-2$ particle wave function from equation (\ref{aSRIff fermi determinant expansion})
$$\langle n_1, \ldots (n_{f'}-1), \ldots (n_f-1), \ldots |
n_1, \ldots (n_{g'}-1),\ldots (n_g-1),\ldots \rangle$$
\begin{equation}=\delta (f-g) \delta (f'-g')-\delta
(f-g')\delta (f'-g).
\end{equation}

For the product of the wave functions we find
$$\Psi^{*}(R'',t)\Psi(R'',t)=\frac{1}{4}\sum_f \sum_{f'\neq f}\sum_g \sum_{g'\neq g}
\sqrt{\frac{n_f}{N}} \sqrt{\frac{n_{f'}}{N-1}}\times$$
$$\times(-1)^{\sum\limits_{f'\le q<f} n_q }
\sqrt{\frac{n_g}{N}} \sqrt{\frac{n_{g'}}{N-1}} (-1)^{\sum\limits_{g'\le q<g} n_q }\times$$
$$\times\left(\:
\langle \textbf{r},t | f\rangle \: \langle\textbf{r}',t | f'\rangle-
\langle\textbf{r}',t | f\rangle \: \langle\textbf{r},t | f'\rangle \:\right)\times$$
$$\times\left(\:
\langle g | \textbf{r},t\rangle \: \langle g'|\textbf{r}',t \rangle-
\langle g|\textbf{r}',t \rangle \: \langle g'|\textbf{r},t \rangle \:\right)\times$$
\begin{equation}\label{aSRIff} \times
(\delta (f-g) \delta (f'-g')-\delta(f-g')\delta (f'-g)). \end{equation}

After simplification the product of the wave functions can be represented as follows
$$\Psi^{*}(R'',t)\Psi(R'',t)=\frac{1}{2}\sum_f \sum_{f'\neq f}
\frac{n_f}{N} \frac{n_{f'}}{N-1}\times$$
$$\times\left(\:
\langle \textbf{r},t | f\rangle \: \langle\textbf{r}',t | f'\rangle-
\langle\textbf{r}',t | f\rangle \: \langle\textbf{r},t | f'\rangle \:\right)\times$$
\begin{equation}\label{aSRIff} \times\left(\:
\langle f| \textbf{r},t\rangle \: \langle f'|\textbf{r}',t \rangle-
\langle f|\textbf{r}',t \rangle \: \langle f'|\textbf{r},t \rangle \:\right).\end{equation}

Similar representation can be done for the derivative of the wave function presented in the expression under the integral
in equations (\ref{aSRIff F expansion up to the third order Simplified}) and (\ref{aSRIff g alpha beta r r'}) (see Appendix A)

Consider the trace of the function find superposition of four identical terms
$$Tr g^{\alpha\beta}(\textbf{r},\textbf{r}',t)
=4\biggl[\sum_f (n_{f}\partial_{\alpha}\varphi_{f}^{*}\cdot\partial_{\beta}\varphi_{f}) \sum_{f'}n_{f'}\varphi_{f'}^{*}\varphi_{f'}$$
\begin{equation} \label{aSRIff Tr g alpha beta explicit}  -\sum_f (n_{f}\partial_{\alpha}\varphi_{f}^{*}\cdot\varphi_{f})
\sum_{f'}n_{f'}\varphi_{f'}^{*}\partial_{\beta}\varphi_{f'}\biggr],\end{equation}
where $n=\sum_{f'}n_{f'}\varphi_{f'}^{*}\varphi_{f'}$.

Equation (\ref{aSRIff Tr g alpha beta explicit}) together with equations (\ref{aSRIff sigma via Upsilon and Tr g alpha beta})
and (\ref{aSRIff Euler via vel}) give an intermediate representation of the interaction in the Euler equation.
Further interpretation and combination of our results for the Euler and the momentum flux evolution equations are presented below.
Adaptation of the obtained results for the fermions with different spin polarization are described below either.

Substituting equations (\ref{aSRIff tensor structure in 3 order}), (\ref{aSRIff tensor I 0})
and equation (\ref{aSRIff Tr g alpha beta explicit}) in equation (\ref{aSRIff sigma via Upsilon and Tr g alpha beta})
\emph{and} obtain the quantum stress tensor in terms of the single particle wave functions as follows:
$$ \sigma^{\alpha\beta}=\frac{1}{2}g_{2}\biggl[\delta^{\alpha\beta}\biggl(n\sum_{g}n_{g}|\nabla\varphi_{g}|^{2}$$
$$-|\sum_{g}n_{g}\varphi^{*}_{g}\nabla\varphi_{g}|^{2}\biggr) +\biggl[n\sum_{g}n_{g}(\partial^{\alpha}\varphi^{*}_{g})\partial^{\beta}\varphi_{g}$$
\begin{equation}\label{sigma fer in 2 or} -\sum_{g}n_{g}\varphi_{g}^{*}(\partial^{\alpha}\varphi_{g}) \sum_{g'}n_{g'}(\partial^{\beta}\varphi_{g'}^{*})\varphi_{g'}+c.c.\biggr)\biggr]\end{equation}
where
$\varphi_{g}(\textbf{r},t)$ are the arbitrary single-particle wave functions.

The quantity
$\sum_{g}n_{g}(\partial^{\alpha}\varphi^{*}_{g})\partial^{\beta}\varphi_{g}$,
in the plane wave approximation
is similar to $m^{2}/\hbar^{2}\Pi^{\alpha\beta}$, but they are not equal to each other.
The tensor $\sum_{g}n_{g}(\partial^{\alpha}\varphi^{*}_{g})\partial^{\beta}\varphi_{g}$
equals to the sum $m^{2}nv^{\alpha}v^{\beta}+m^{2}p^{\alpha\beta}+\hbar^{2}\partial^{\alpha}\sqrt{n}\cdot\partial^{\beta}\sqrt{n}$.
Next, consider $\sum_f (n_{f}\partial_{\alpha}\varphi_{f}^{*}\cdot\varphi_{f})$ and
$\sum_{f'}n_{f'}\varphi_{f'}^{*}\partial_{\beta}\varphi_{f'}$.
Find expressions for the plane waves
$\sum_f (n_{f}\partial_{\alpha}\varphi_{f}^{*}\cdot\varphi_{f})=\frac{\imath\hbar}{2}\partial^{\beta}n +mnv^{\beta}$
and
$\sum_{f'}n_{f'}\varphi_{f'}^{*}\partial_{\beta}\varphi_{f'}=-\frac{\imath\hbar}{2}\partial^{\beta}n +mnv^{\beta}$.
Their product has the following form
$\sum_f (n_{f}\partial_{\alpha}\varphi_{f}^{*}\cdot\varphi_{f})$
$\sum_{f'}n_{f'}\varphi_{f'}^{*}\partial_{\beta}\varphi_{f'}$
$=\frac{\hbar^{2}}{4}\partial^{\alpha}n\cdot\partial^{\beta}n +m^{2}n^{2}v^{\alpha}v^{\beta}$
$+mn\imath\hbar(\partial^{\alpha}n\cdot v^{\beta}-\partial^{\beta}n\cdot v^{\alpha})/2$.
Include that tensor $g^{\gamma\delta}$ is multiplied by the symmetric tensor $\Upsilon_{2}^{\alpha\beta\gamma\delta}$
therefore the last (imaginary) term gives no contribution in the quantum stress tensor $\sigma^{\alpha\beta}$.

If we do not consider evolution of the pressure we need to use an equation of state to close the set of equations.

The kinetic pressure tensor is diagonal $p^{\alpha\beta}=p_{\uparrow\uparrow}\delta^{\alpha\beta}$,
and that $p_{\uparrow\uparrow}=(6\pi^{2})^{2/3}\hbar^{2}n^{5/3}/5m^{2}$ \cite{Landau 5}.
It is explicitly seen that
physics dimension of pressure is changed by multiplier $1/m$.
It is done to give symmetric presentation for equations given through the paper.
The multiplier $1/m$ can be replaced in front of pressure tensor $p^{\alpha\beta}$ to restore the traditional physical dimension.

Finally, we derive the expression for the stress tensor in the form
\begin{equation}\label{aSRIff sigma fer in 2 or pl w short form}
\sigma^{\alpha\beta}=\frac{1}{2}g_{2}
\delta^{\alpha\beta}(6\pi^{2})^{\frac{2}{3}}n^{\frac{8}{3}}.
\end{equation}

In Ref. \cite{Andreev PRA08} term $\sum_{g}n_{g}\varphi_{g}^{*}\partial^{\alpha}\varphi_{g}$ was interpreted via the current $j^{\alpha}$.
However,  more accurate analysis presented above shows that $\sum_{g}n_{g}\varphi_{g}^{*}\partial^{\alpha}\varphi_{g}$
has an imaginary part proportional to the gradient of the concentration.
The account of this part compensate derivatives of concentration coming from
$\sum_{g}n_{g}\partial^{\alpha}\varphi_{g}^{*}\cdot\partial^{\beta}\varphi_{g}$,
as it is presented before equation (\ref{aSRIff sigma fer in 2 or pl w short form}).
Hence, the final equation for the quantum stress tensor (\ref{aSRIff sigma fer in 2 or pl w short form}) becomes relatively simple.
Moreover, it is a local term since it is proportional to the first derivative of the concentration.
It make fundamental difference from the older results for fermions
or the third order by the interaction radius model of bosons \cite{Andreev PRA08}, \cite{Andreev MPL B 12}, \cite{Andreev LP 19},
where the third order by the interaction radius approximation is presented via nonlocal terms containing hither derivatives of concentration or the products of the first derivatives.

In this paper we are focused on the extended hydrodynamics of the degenerate fermions.
Therefore, we consider the evolution of the pressure.
Hence, we find a more general expression for the quantum stress tensor:
\begin{equation}\label{aSRIff sigma fer TOIR via pressure and I 0}
\sigma^{\alpha\beta}=\frac{m^{2}}{2\hbar^{2}}g_{2} I_{0}^{\alpha\beta\gamma\delta}np^{\gamma\delta}. \end{equation}

There is an association with the p-wave
\cite{Roth PRA 02, Roth PRA 01}.
Moreover, d-waves are also consider in context of quantum gases
\cite{Idziaszek PRA 09, Derevianko PRA 05, Yao PRA 19}.

\subsection{Different species}

Generally speaking,
all fermions of the single species are equivalent.
Hence,
all permutations of arguments and symmetries used in equations (\ref{aSRIff ForceField})-(\ref{aSRIff Psi with arguments}) are applicable
for fermions in the same spin state and fermions in different spin states.
However, approximate analysis of noninteracting fermions shows that
the best approximation for the many-particle wave function in terms of the single-particle wave functions
is the product of two (for spin-1/2) Slater determinants \cite{Landau 3, Fock book quantum mech, Comparin PRA 19} (see eq. 5).
The first determinant for the spin-up fermions and the second determinant for the spin-down fermions.
Each determinant ensures the antisymmetry of the coordinate part of the wave function relatively permutations of fermions with the same spin projection.
However,
there is no symmetry for the pair of fermions with different spin projections.
Therefore,
they are considered as two different species.
Consequently,
we present an independent analysis of the interaction of fermions with different spin projections.
Start with equation (\ref{aSRIff ForceField}).

Substitute the coordinates of the i-th and j-th particles via their center of mass
and the coordinate of their relative motion
$$\textbf{F}_{int}(\textbf{r},t)=-\int dR\sum_{i,j\neq i}\delta\biggl(\textbf{r}-\textbf{R}_{ij}-\frac{1}{2}\textbf{r}_{ij}\biggr)(\nabla_{i}U(\textbf{r}_{ij}))\times$$
$$\times\Psi^{*}\biggl(..., \textbf{R}_{ij}+\frac{1}{2}\textbf{r}_{ij}, ..., \textbf{R}_{ij}-\frac{1}{2}\textbf{r}_{ij}, ...,t\biggr)\times$$
\begin{equation} \label{aSRIff ForceField nonsymmetric via R,r} \times  \Psi\biggl(..., \textbf{R}_{ij}+\frac{1}{2}\textbf{r}_{ij}, ..., \textbf{R}_{ij}-\frac{1}{2}\textbf{r}_{ij}, ...,t\biggr).\end{equation}

Expand the obtained equation (\ref{aSRIff ForceField nonsymmetric via R,r}) on the coordinate of the relative motion.
Let us present the expansion for the delta function limited by the third order
$$\delta\biggl(\textbf{r}-\textbf{R}_{ij}-\frac{1}{2}\textbf{r}_{ij}\biggr)
=\delta(\textbf{r}-\textbf{R}_{ij})-\frac{1}{2}r_{ij}^{\alpha}\frac{\partial\delta}{\partial r^{\alpha}}$$
\begin{equation} \label{aSRIff}
+\frac{1}{2!}\frac{1}{2^{2}}r_{ij}^{\alpha}r_{ij}^{\beta} \frac{\partial^{2}\delta}{\partial r^{\alpha}\partial r^{\beta}} -\frac{1}{3!}\frac{1}{2^{3}}r_{ij}^{\alpha}r_{ij}^{\beta}r_{ij}^{\gamma}
\frac{\partial^{2}\delta}{\partial r^{\alpha}\partial r^{\beta}\partial r^{\gamma}}.\end{equation}

Next, let us present the expansion for the wave function
$$\Psi\biggl(..., \textbf{R}_{ij}+\frac{1}{2}\textbf{r}_{ij}, ..., \textbf{R}_{ij}-\frac{1}{2}\textbf{r}_{ij}, ...,t\biggr)$$
$$=\Psi(..., \textbf{R}_{ij}, ..., \textbf{R}_{ij}, ...,t)
+\frac{1}{2}r_{ij}^{\alpha}\biggl(\partial_{R1}^{\alpha}\Psi-\partial_{R2}^{\alpha}\Psi\biggr)$$
$$+\frac{1}{2!}\frac{1}{2^{2}}r_{ij}^{\alpha}r_{ij}^{\beta} \biggl(\partial_{R1}^{\alpha}\partial_{R1}^{\beta}\Psi
-2\partial_{R1}^{\alpha}\partial_{R2}^{\beta}\Psi
+\partial_{R2}^{\alpha}\partial _{R 2}^{\beta}\Psi\biggr)$$
$$+\frac{1}{3!}\frac{1}{2^{3}}r_{ij}^{\alpha}r_{ij}^{\beta}r_{ij}^{\gamma}
\biggl(\partial _{R1}^{\alpha}\partial _{R1}^{\beta}\partial _{R1}^{\gamma}\Psi
-3\partial_{R1}^{\alpha}\partial_{R1}^{\beta}\partial_{R2}^{\gamma}\Psi$$
\begin{equation} \label{aSRIff}
+3\partial _{R1}^{\alpha}\partial _{R2}^{\beta}\partial _{R2}^{\gamma}\Psi
-\partial_{R2}^{\alpha}\partial_{R2}^{\beta}\partial_{R2}^{\gamma}\Psi \biggr). \end{equation}

In order to study the interaction up to the third order by the interaction radius
we make the expansion up to the third order on the coordinate of the relative motion.

Substituting these expansions into the force field (\ref{aSRIff ForceField nonsymmetric via R,r})
find a rather large expression,
where we have four kinds of terms.
These terms differ from each other by the degree of the relative distance, having terms from the zeroth and up to the third degree.
After integration over the relative distance find that terms of the zeroth and second orders go to zero.
Hence, we have expressions in the first and third order by the interaction radius.

Consider the terms existing in the first order by the interaction radius:
$$F^{\alpha}_{FOIR}=\frac{1}{2}\int dR\sum_{i,j\neq i}r^{\alpha}_{ij}r^{\beta}_{ij}\frac{1}{r_{ij}}\frac{\partial U}{\partial r_{ij}} \frac{\partial\delta_{i}}{\partial r^{\beta}}\Psi^{*}(R',t)\Psi(R',t)$$
\begin{equation} \label{aSRIff F FOIR via Psi} -\frac{1}{2}\int dR\sum_{i,j\neq i}r^{\alpha}_{ij}r^{\beta}_{ij}\frac{1}{r_{ij}}\frac{\partial U}{\partial r_{ij}} \delta_{i}\Biggl(\Psi^{*}\biggl(\partial_{R1}^{\beta}\Psi-\partial_{R2}^{\beta}\Psi\biggr)+c.c.\Biggr), \end{equation}
where FOIR stands for the first order by the interaction radius.
Let us repeat the following notations
$\delta_{i}=\delta(\textbf{r}-\textbf{R}_{ij})$,
$R'=\{\textbf{r}_{1}, ..., \textbf{R}_{ij}, ..., \textbf{R}_{ij}, ..., \textbf{r}_{N}\}$,
$dR=dR_{N-2} dR_{ij} dr_{ij}$.

Integral over the relative motion can be separated from other in the force field (\ref{aSRIff F FOIR via Psi}) (in both terms).
It can be presented as the following second-rank tensor
$\Upsilon^{\alpha\beta}=\int d^{3}r_{ij}r^{\alpha}_{ij}r^{\beta}_{ij}\frac{1}{r_{ij}}\frac{\partial U}{\partial r_{ij}}$
$=\delta^{\alpha\beta}\frac{4\pi}{3}\int dr_{ij} r^{3}_{ij}\frac{\partial U}{\partial r_{ij}}=-\delta^{\alpha\beta}g$.

Consider the force field in the regime of weakly interacting particles.
To the start consider
$I_{1}=\int d^{3}R_{ij}dR_{N-2}\delta_{i}\Psi^{*}(\partial_{R1}^{\alpha}\Psi-\partial_{R2}^{\alpha}\Psi)$
presented in the force field (\ref{aSRIff F FOIR via Psi}).
Absence of the symmetry allows to write the wave function as the product of partial wave functions without (anti-)symmetrization.
Since we have $N_{1}N_{2}$ equivalent pairs of interacting particles we can consider $N_{1}N_{2}$ identical integrals instead of summation on $i$ and $j$ under the integral.

After expansion of the wave function integral $I_{1}$ simplifies to the following form
$I_{1}=\int d^{3}R_{ij}\delta(\textbf{r}-\textbf{R}_{ij})
\psi_{1}^{*}(R_{ij})\psi_{2}^{*}(R_{ij})[\psi_{1}(R_{ij}) \psi_{2}(R_{ij}) -\psi_{1}(R_{ij}) \psi_{2}(R_{ij})]$,
where it is used that $\int dR_{N-2}\Psi^{*}(R_{N-2},t)\Psi(R_{N-2},t)=1$.

Next, find the representation of integral $I_{1}$ via the concentrations
$N_{1}N_{2}\cdot I_{1}=N_{1} n_{(2)}\int d^{3}R_{ij}\delta_{i}\psi_{1}^{*}(R_{ij})\partial_{R}^{\beta}\psi_{1}(R_{ij})
- N_{2} n_{(1)}\int d^{3}R_{ij}\delta_{i}\psi_{2}^{*}(R_{ij})\partial_{R}^{\beta}\psi_{2}(R_{ij})$.
The final form of the representation appears after adding of the complex conjugate part:
$N_{1}N_{2}(I_{1}+c.c.)=n_{(2)}\partial^{\beta}n_{(1)} -n_{(1)}\partial^{\beta}n_{(2)}$.

The derivative acting on the delta function in the first term in the force field (\ref{aSRIff F FOIR via Psi})
can be placed outside of the integral.
Remaining integral contains the product of the wave functions
$\psi_{1}^{*}(R_{ij})\psi_{2}^{*}(R_{ij})\psi_{1}(R_{ij})\psi_{2}(R_{ij})$,
so it gives the product of concentrations
$n_{(1)}n_{(2)}=N_{1}N_{2}\int d^{3}R_{ij}\delta_{i}\psi_{1}^{*}(R_{ij})\psi_{2}^{*}(R_{ij})\psi_{1}(R_{ij})\psi_{2}(R_{ij})$.

Combining all described in the equation (\ref{aSRIff F FOIR via Psi}) find the following result for the force field:
$$F^{\alpha}_{FOIR}= -\frac{1}{2}g\partial^{\alpha}(n_{(1)}n_{(2)})
-\frac{1}{2}gn_{(1)}\partial^{\alpha}n_{(2)} $$
\begin{equation} \label{aSRIff F TOIR via n1 n2}
+\frac{1}{2}gn_{(2)}\partial^{\alpha}n_{(1)}
=-gn_{(1)}\partial^{\alpha}n_{(2)}. \end{equation}
Let us to point out that species "2" is the source of the force acting on species "1".

Consider the force field existing in the third order by the interaction radius in the weakly-interacting limit:\begin{widetext}
$$F^{\alpha}_{TOIR}=-\Upsilon_{2}^{\alpha\beta\gamma\delta} N_1 N_2 \int dR_{N-1}
\Biggl[-\frac{1}{3!}\frac{1}{2^{3}}\partial_{\beta}\partial_{\gamma}\partial_{\delta}\delta_{i}
\psi_{1}^{*}\psi_{2}^{*}\psi_{1}\psi_{2}
+\frac{1}{2!}\frac{1}{2^{3}}\partial^{\beta}\partial^{\gamma}\delta_{i}
\cdot\biggl(\psi_{1}^{*}\psi_{2}^{*}(\partial^{\delta}_{(1)}-\partial^{\delta}_{(2)})\psi_{1}\psi_{2}+c.c.\biggr)$$
$$-\frac{1}{2}\frac{1}{2^{3}}\partial^{\beta}\delta_{i} \cdot\biggl(\psi_{1}^{*}\psi_{2}^{*}(\partial^{\gamma}_{(1)}\partial^{\delta}_{(1)} -2\partial^{\gamma}_{(1)}\partial^{\delta}_{(2)}+\partial^{\gamma}_{(2)}\partial^{\delta}_{(2)})\psi_{1}\psi_{2}
+(\partial^{\gamma}_{(1)}\partial^{\delta}_{(1)}-2\partial^{\gamma}_{(1)}\partial^{\delta}_{(2)}+\partial^{\gamma}_{(2)}\partial^{\delta}_{(2)})\psi_{1}^{*}\psi_{2}^{*}\cdot\psi_{1}\psi_{2}$$
$$+2(\partial^{\gamma}_{(1)}-\partial^{\gamma}_{(2)})\psi_{1}^{*}\psi_{2}^{*}
\cdot(\partial^{\delta}_{(1)}-\partial^{\delta}_{(2)})\psi_{1}\psi_{2}\biggr)
+\frac{1}{3!}\frac{1}{2^{3}}\delta_{i}\biggl(\psi_{1}^{*}\psi_{2}^{*} (\partial^{\beta}_{(1)}\partial^{\gamma}_{(1)}\partial^{\delta}_{(1)}
-3\partial^{\beta}_{(1)}\partial^{\gamma}_{(1)}\partial^{\delta}_{(2)} $$
\begin{equation} \label{aSRIff F TOIR via Psi} +3\partial^{\beta}_{(1)}\partial^{\gamma}_{(2)}\partial^{\delta}_{(2)} -\partial^{\beta}_{(2)}\partial^{\gamma}_{(2)}\partial^{\delta}_{(2)})\psi_{1}\psi_{2}
+3(\partial^{\beta}_{(1)}-\partial^{\beta}_{(2)})\psi_{1}^{*}\psi_{2}^{*} \cdot(\partial^{\gamma}_{(1)}\partial^{\delta}_{(1)}
-2\partial^{\gamma}_{(1)}\partial^{\delta}_{(2)}+\partial^{\gamma}_{(2)}\partial^{\delta}_{(2)})\psi_{1}\psi_{2}+c.c.\biggr)\Biggr],
\end{equation}\end{widetext}
where $\psi_{l}=\psi_{l}(R_{ij})$, $l=1,2$, and TOIR stands for third order by the interaction radius.

Equation (\ref{aSRIff F TOIR via Psi}) consists of four group of terms
which are proportional to the third, second,  first and zeroth derivatives of the delta function correspondingly.
Calculating each group find that the force field is a function of concentrations of each species of particles
$$F^{\alpha}_{TOIR}=-\Upsilon_{2}^{\alpha\beta\gamma\delta}
\Biggl[-\frac{1}{3!}\frac{1}{2^{3}}\partial_{\beta}\partial_{\gamma}\partial_{\delta}[n_{(1)}n_{(2)}]$$
$$+\frac{1}{2!}\frac{1}{2^{3}}\partial_{\beta}\partial_{\gamma}[\partial_{\delta}n_{(1)}\cdot n_{(2)}-n_{(1)}\partial_{\delta}n_{(2)}] $$
$$-\frac{1}{2}\frac{1}{2^{3}}\partial_{\beta}[n_{(2)}\partial_{\gamma}\partial_{\delta}n_{(1)}
+n_{(1)}\partial_{\gamma}\partial_{\delta}n_{(2)}-\partial_{\gamma}n_{(1)}\cdot\partial_{\delta}n_{(2)}]$$
$$ +\frac{1}{3!}\frac{1}{2^{3}} [n_{(2)}\partial_{\beta}\partial_{\gamma}\partial_{\delta}n_{(1)} -n_{(1)}\partial_{\beta}\partial_{\gamma}\partial_{\delta}n_{(2)}$$
\begin{equation} \label{aSRIff F TOIR via n1 n2 intermediate}
+3\partial_{\beta}n_{(1)}\cdot \partial_{\gamma}\partial_{\delta}n_{(2)}
-3\partial_{\gamma}\partial_{\delta}n_{(1)}\cdot \partial_{\beta} n_{(2)}]\Biggr], \end{equation}
where
terms are grouped with accordance with equation (\ref{aSRIff F TOIR via Psi}).

After simplification the force field in the third order by the interaction radius approximation appears in a rather simple form:
\begin{equation} \label{aSRIff F TOIR via n1 n2 final}
F^{\alpha}_{TOIR}=
\frac{1}{3!}\Upsilon_{2}^{\alpha\beta\gamma\delta} n_{(1)}\partial_{\beta}\partial_{\gamma}\partial_{\delta}n_{(2)}. \end{equation}

Simplified and combined force field appears in the following form:
\begin{equation} \label{aSRIff F up to TOIR via n1 n2}
F^{\alpha}=-gn_{(1)}\partial^{\alpha}n_{(2)}-\frac{1}{2}g_{2}n_{(1)}\partial^{\alpha}\triangle n_{(2)}.\end{equation}
This result corresponds to equation 54 in Ref. \cite{Andreev PRA08} (where parameter $\Upsilon=-g$).
The nonlocal interactions in BECs are described in Refs.
\cite{Andreev PRA08, Andreev MPL B 12, Andreev LP 19, Braaten PRA 01, Rosanov PLA 02}.

A p-wave scattering model for boson fermion interaction is described in Ref. \cite{Parker PRA 12}.

\section{SRI in the momentum flux evolution equation}

\subsection{Identical particles}

The short-range interaction enters the momentum flux evolution equation
via the force tensor field $F^{\alpha\beta}_{int}$
which contains the gradient of the potential $\partial^{\alpha}U[\textbf{r}-\textbf{r}']$
(similarly to the traditional force field)
and
the two-particle current-concentration function
$j_{2}^{\alpha}(\textbf{r},\textbf{r}',t)$ (instead of the two-particle concentration).

Consider the force tensor field
\begin{equation} \label{aSRIff}
F^{\alpha\beta}_{int}=-\int (\partial^{\alpha}U[\textbf{r}-\textbf{r}']) j_{2}^{\beta}(\textbf{r},\textbf{r}',t)d\textbf{r}' \end{equation}
which is not a symmetric tensor.

Neglecting the thermal part of function $j_{2}^{\beta}$ obtain a relation between the force tensor field and the traditional force vector field:
\begin{equation} \label{aSRIff}
F^{\alpha\beta}_{int}=-v^{\beta}\int (\partial^{\alpha}U[\textbf{r}-\textbf{r}']) n_{2}(\textbf{r},\textbf{r}',t)d\textbf{r}'
=v^{\beta}F^{\alpha}_{int}. \end{equation}
Therefore, tensor $F^{\alpha\beta}_{int}$ can be considered as the flux of the force field.

The account of the thermal part does not change this interpretation.
Consider, for instance, the flux of the momentum $\Pi^{\alpha\beta}$.
It contains the macroscopic part of the flux $nv^{\alpha}v^{\beta}=j^{\alpha}v^{\beta}$,
but tensor $\Pi^{\alpha\beta}$ contains the thermal part of the flux $p^{\alpha\beta}$
which is the traditional thermal pressure.

To the short-range interaction analysis present the force tensor field via the microscopic many-particle wave function
\begin{equation} \label{aSRIff F alpha beta original} F^{\alpha\beta}_{int}=-\int dR \sum_{i,j \neq i}\delta(\textbf{r}-\textbf{r}_{i})(\partial_{i}^{\alpha}U(\textbf{r}_{ij}))\frac{1}{2m_{i}}(\Psi^{*}\hat{p}_{i}^{\beta}\Psi+c.c.) , \end{equation}
where $\textbf{r}_{ij}=\textbf{r}_{i}-\textbf{r}_{j}$.

Analysis of the identical particles allows to make a partial symmetrization:
$$F^{\alpha\beta}_{int}
=-\frac{1}{4m}\int dR \sum_{i,j \neq i} (\partial_{i}^{\alpha}U(\textbf{r}_{ij}))
\biggl(\delta(\textbf{r}-\textbf{r}_{i})\cdot$$
\begin{equation} \label{aSRIff}  \cdot(\Psi^{*}\hat{p}_{i}^{\beta}\Psi+c.c.)
-\delta(\textbf{r}-\textbf{r}_{j})(\Psi^{*}\hat{p}_{j}^{\beta}\Psi+c.c.)\biggr),\end{equation}
where
$\partial_{j}^{\alpha}U[\textbf{r}_{i}-\textbf{r}_{j}]=-\partial_{i}^{\alpha}U[\textbf{r}_{i}-\textbf{r}_{j}]$ and $m_{i}=m_{j}=m$ are used.

To continue we introduce the interparticle distance and the center of mass coordinates for the pair of interacting particles,
make the expansion on the interparticle distance up to the third order,
include that integral on the interparticle distance is equal to zero for the zeroth and second orders.
Moreover, include that in the first and third orders of the expansion of the wave function $\Psi(R',t)$ is equal to zero
if it is not affected by other operators.
As the result, we obtain the expressions for $F^{\alpha\beta}_{int}$ in the first order by the interaction radius and third order by the interaction radius approximations.

Present the force tensor field in the first order by the interaction radius:
$$F^{\alpha\beta}_{int,FOIR}=-\frac{1}{8m}\int dR  \sum_{i,j\neq i}(r_{ij}^{\gamma}\partial_{i}^{\alpha}U(\textbf{r}_{ij}))\delta_{i}\times$$
\begin{equation} \label{aSRIff F int FOIR via Psi} \times\biggl((\partial_{R1}^{\gamma}-\partial_{R2}^{\gamma})\Psi^{*}\cdot(-\imath\hbar)(\partial_{R1}^{\beta}-\partial_{R2}^{\beta})\Psi+c.c.\biggr). \end{equation}

Represent the expression for the force tensor field $F^{\alpha\beta}_{int}$ in the first order by the interaction radius in the following form:
\begin{equation} \label{aSRIff}
F^{\alpha\beta}_{int,FOIR}=-\frac{1}{8m}g[-\imath\hbar Tr g^{\alpha\beta}(\textbf{r},\textbf{r}',t) +c.c.].
\end{equation}

Here tensor $Tr g^{\alpha\beta}(\textbf{r},\textbf{r}',t)$ defined by equation (\ref{aSRIff g alpha beta r r'}) is used.
Let us to point out that $(Tr g^{\alpha\beta})^{*}=Tr g^{\beta\alpha}\neq Tr g^{\alpha\beta}$ (in general case).
Next, consider a simplified expression (\ref{aSRIff Tr g alpha beta explicit}) for tensor $g^{\alpha\beta}$.
The plane waves can be used as the single particle wave function for the weakly interacting particles
$\varphi_{f=\textbf{k}}=Ae^{\imath \textbf{k}\textbf{r}}$.
It gives
$Tr g^{\alpha\beta}=4[n\sum_{\textbf{k}}(n_{\textbf{k}}k^{\alpha}k^{\beta}\varphi_{\textbf{k}}^{*}\varphi_{\textbf{k}})
-\sum_{\textbf{k}}(n_{\textbf{k}}k^{\alpha}\varphi_{\textbf{k}}^{*}\varphi_{\textbf{k}})
\sum_{\textbf{k}'}(n_{\textbf{k}'}k'^{\beta}\varphi_{\textbf{k}'}^{*}\varphi_{\textbf{k}'})]$,
where $Tr g^{\alpha\beta}\in Re$.
Consequently, the force tensor field in the first order by the interaction radius is equal to zero: $F^{\alpha\beta}_{int,FOIR}=0$.

More general analysis can be applied to $F^{\alpha\beta}_{int,FOIR}$.
To this end, use the Madelung decomposition in equation (\ref{aSRIff F int FOIR via Psi}) and find the following representation
\begin{equation} \label{aSRIff F int FOIR final}
F^{\alpha\beta}_{int,FOIR}=-\frac{1}{2}g n\sum_{f}n_{f}(\partial_{\delta}a^{2}\cdot u^{\beta}-\partial_{\delta}a^{2}\cdot u^{\delta}).\end{equation}

The expression (\ref{aSRIff F int FOIR final}) requires no further analysis.
Since the pressure evolution equation contains the symmetric combination of the force tensor fields $F^{\alpha\beta}$: $F^{\alpha\beta}_{int,FOIR}+F^{\beta\alpha}_{int,FOIR}$.
The antisymmetry of expression (\ref{aSRIff F int FOIR final}) leads to the zero value of the symmetric combination
$F^{\alpha\beta}_{int,FOIR}+F^{\beta\alpha}_{int,FOIR}=0$.

Next step is the calculation of the force tensor field appearing in the third order by the interaction radius approximation $F^{\alpha\beta}_{int,TOIR}$.
It has a huge expression.
Therefore, it is splitted on four groups of terms $F^{\alpha\beta}_{int,i}$,
with $i=I,II,III,IV$:
$F^{\alpha\beta}_{int,TOIR}=\sum_{i=I}^{IV}F^{\alpha\beta}_{int,i}$.
They have the following forms \begin{widetext}
\begin{equation} \label{aSRIff}
F^{\alpha\beta}_{int,I}=\frac{1}{4m}\frac{1}{2!}\frac{1}{2^{3}}\int dR  \sum_{i,j\neq i}(r_{ij}^{\gamma}r_{ij}^{\delta}r_{ij}^{\mu}\partial_{i}^{\alpha}U(\textbf{r}_{ij}))
\partial^{\gamma}\partial^{\delta}\delta_{i}\cdot\biggl((\partial_{R1}^{\gamma}-\partial_{R2}^{\gamma})\Psi^{*} \cdot(\imath\hbar)(\partial_{R1}^{\beta}-\partial_{R2}^{\beta})\Psi +c.c.\biggr),\end{equation}
\begin{equation} \label{aSRIff}
F^{\alpha\beta}_{int,II}=-\frac{1}{4m}\frac{1}{2!}\frac{1}{2^{3}}\int dR  \sum_{i,j\neq i}(r_{ij}^{\gamma}r_{ij}^{\delta}r_{ij}^{\mu}\partial_{i}^{\alpha}U(\textbf{r}_{ij}))\partial^{\gamma}\delta_{i}
\biggl[\biggl((\partial_{R1}^{\delta}-\partial_{R2}^{\delta})(\partial_{R1}^{\mu}-\partial_{R2}^{\mu})\Psi^{*} \cdot(\imath\hbar)(\partial_{R1}^{\beta}+\partial_{R2}^{\beta})\Psi\biggr)+c.c.\biggr],\end{equation}
\begin{equation} \label{aSRIff}
F^{\alpha\beta}_{int,III}=-\frac{1}{4m}\frac{1}{2^{3}}\int dR  \sum_{i,j\neq i}(r_{ij}^{\gamma}r_{ij}^{\delta}r_{ij}^{\mu}\partial_{i}^{\alpha}U(\textbf{r}_{ij}))
\biggl[\partial^{\gamma}\delta_{i}\cdot\biggl((\partial_{R1}^{\delta}-\partial_{R2}^{\delta})\Psi^{*} \cdot(\imath\hbar)(\partial_{R1}^{\beta}+\partial_{R2}^{\beta})(\partial_{R1}^{\mu}-\partial_{R2}^{\mu})\Psi\biggr)+c.c.\biggr],\end{equation}
and
$$F^{\alpha\beta}_{int,IV}=\frac{(\imath\hbar)}{4m}\frac{1}{3!}\frac{1}{2^{3}}\int dR  \sum_{i,j\neq i}(r_{ij}^{\gamma}r_{ij}^{\delta}r_{ij}^{\mu}\partial_{i}^{\alpha}U(\textbf{r}_{ij}))\delta_{i}
\biggl[3\biggl((\partial_{R1}^{\delta}-\partial_{R2}^{\delta})(\partial_{R1}^{\mu}-\partial_{R2}^{\mu})\Psi^{*} \cdot(\partial_{R1}^{\beta}-\partial_{R2}^{\beta})(\partial_{R1}^{\gamma}-\partial_{R2}^{\gamma})\Psi\biggr)$$
\begin{equation} \label{aSRIff F alpha beta int,TOIR,IV via Single part WF}
+3 \biggl((\partial_{R1}^{\gamma}-\partial_{R2}^{\gamma})\Psi^{*} \cdot(\partial_{R1}^{\beta}-\partial_{R2}^{\beta})(\partial_{R1}^{\delta}-\partial_{R2}^{\delta}) (\partial_{R1}^{\mu}-\partial_{R2}^{\mu})\Psi\biggr)
+\biggl((\partial_{R1}^{\gamma}-\partial_{R2}^{\gamma})(\partial_{R1}^{\delta}-\partial_{R2}^{\delta}) (\partial_{R1}^{\mu}-\partial_{R2}^{\mu})\Psi^{*}
\cdot(\partial_{R1}^{\beta}-\partial_{R2}^{\beta})\Psi\biggr)\biggr]+c.c..\end{equation}
\end{widetext}

After further calculation in the approximation of the weakly interacting particles find the following expressions for the partial force tensor fields.
Start our list with expression for $F^{\alpha\beta}_{int,I}$:
$$F^{\alpha\beta}_{int,I}= \frac{\imath\hbar}{2!m}\frac{1}{2^{3}}\Upsilon_{2}^{\alpha\delta\mu\nu}\partial_{\mu}\partial_{\nu}
\sum_{f,f'\neq f}n_{f}n_{f'} (\partial_{\beta}\varphi_{f}^{*}\cdot\partial_{\delta}\varphi_{f}\cdot\varphi_{f'}^{*}\varphi_{f'}$$
\begin{equation} \label{aSRIff F alpha beta int,TOIR,I via Single part WF}
-\partial_{\delta}\varphi_{f}^{*}\cdot\varphi_{f} \varphi_{f'}^{*}\partial_{\beta}\varphi_{f'})+c.c..\end{equation}

Using the Madelung decomposition and introducing the velocity field in $F^{\alpha\beta}_{int,TOIR,I}$ find the following representation
\begin{equation} \label{aSRIff F alpha beta int,TOIR,I via a u}
F^{\alpha\beta}_{int,I}=-\frac{1}{2!}\frac{1}{2^{3}} \Upsilon_{2}^{\alpha\delta\mu\nu} \partial_{\mu}\partial_{\nu}[n\sum_{f}n_{f}(\partial_{\delta}a^{2}\cdot u^{\beta}-\partial_{\delta}a^{2}\cdot u^{\delta})].\end{equation}
In the plane wave single particle wave function approximation it becomes equal to zero
$F^{\alpha\beta}_{int,I}=0$
similarly to $F^{\alpha\beta}_{int,FOIR}$ discussed above.

Expression $F^{\alpha\beta}_{int,II}$ equals to zero for arbitrary single particle wave functions
$\varphi_{f}$: $F^{\alpha\beta}_{int,II}=0$.

We also have\begin{widetext}
$$F^{\alpha\beta}_{int,III}=\frac{1}{m\hbar^{2}}\frac{1}{2^{3}}\Upsilon_{2}^{\alpha\gamma\delta\mu}
\cdot\partial^{\gamma}\biggl[n
\sum_{f}n_{f}\biggl((p^{\mu}\varphi_{f})^{*}(p^{\beta}p^{\delta}\varphi_{f})+(p^{\beta}p^{\delta}\varphi_{f})^{*}(p^{\mu}\varphi_{f})\biggr)
+2m j^{\beta}\sum_{f}n_{f}(p^{\mu}\varphi_{f})^{*}(p^{\delta}\varphi_{f})$$
\begin{equation} \label{aSRIff F alpha beta int,TOIR,III via Single part WF} -\sum_{f,f'}n_{f}n_{f'}\biggl((p^{\mu}\varphi_{f})^{*}(p^{\beta}\varphi_{f})+(p^{\mu}p^{\delta}\varphi_{f})^{*}\varphi_{f}\biggr) \varphi_{f}^{*}(p^{\delta}\varphi_{f}) -\sum_{f,f'}n_{f}n_{f'}\biggl(\biggl((p^{\beta}\varphi_{f})^{*}(p^{\mu}\varphi_{f})+(\varphi_{f})^{*}p^{\mu}p^{\delta}\varphi_{f}\biggr) (p^{\delta}\varphi_{f})^{*}\varphi_{f}\biggr)\biggr],\end{equation}
and
$$F^{\alpha\beta}_{int,IV}=
\frac{-\imath}{3! \cdot 2^{3}}\frac{1}{m\hbar^{3}}\Upsilon_{2}^{\alpha\delta\mu\nu}
\times\Biggl\{n\sum_{f}n_{f}(p^{\delta}p^{\mu}p^{\nu}\varphi_{f})^{*} p^{\beta}\varphi_{f}
-3\sum_{f}n_{f}(p^{\nu}\varphi_{f})^{*} \varphi_{f} \sum_{f'}n_{f'}(p^{\delta}p^{\mu}\varphi_{f})^{*} p^{\beta}\varphi_{f}$$
$$+3\sum_{f}n_{f}(p^{\mu}p^{\nu}\varphi_{f})^{*} \varphi_{f} \sum_{f'}n_{f'}(p^{\delta}\varphi_{f'})^{*} p^{\beta}\varphi_{f'}
-\sum_{f}n_{f}(p^{\delta}p^{\mu}p^{\nu}\varphi_{f})^{*} \varphi_{f} \sum_{f'}n_{f'}(\varphi_{f'})^{*} p^{\beta}\varphi_{f'}
+3n \sum_{f}n_{f}(p^{\delta}\varphi_{f})^{*} p^{\mu}p^{\nu}p^{\beta}\varphi_{f}$$
$$-6\sum_{f}n_{f}(\varphi_{f})^{*} p^{\nu}\varphi_{f} \sum_{f'}n_{f'}(p^{\delta}\varphi_{f'})^{*} p^{\mu}p^{\beta}\varphi_{f'}
+3\sum_{f}n_{f}(\varphi_{f})^{*} p^{\mu}p^{\nu}\varphi_{f} \sum_{f'}n_{f'}(p^{\delta}\varphi_{f'})^{*} p^{\beta}\varphi_{f'}
-3\sum_{f}n_{f}(\varphi_{f})^{*} p^{\beta}\varphi_{f} \times$$
\begin{equation} \label{aSRIff F alpha beta int,TOIR,IV via Single part WF}
\times\sum_{f'}n_{f'}(p^{\delta}\varphi_{f'})^{*} p^{\mu}p^{\nu}\varphi_{f}
+6\sum_{f}n_{f}(\varphi_{f})^{*} p^{\nu}p^{\beta}\varphi_{f} \sum_{f'}n_{f'}(p^{\delta}\varphi_{f})^{*} p^{\mu}\varphi_{f'}
-3\sum_{f}n_{f}(\varphi_{f})^{*} p^{\mu}p^{\nu}p^{\beta}\varphi_{f} \sum_{f'}n_{f'}(p^{\delta}\varphi_{f'})^{*} \varphi_{f'}
\Biggr\}+c.c. .
\end{equation}\end{widetext}
Two parts have nonzero values.
They are found in terms of the single particle wave functions.
They have no immediate expressions of the partial force tensor fields via the hydrodynamic functions.
Therefore, find some approximate relations between the partial force tensor fields and the hydrodynamic functions.

Let us describe the transformation of equations (\ref{aSRIff F alpha beta int,TOIR,III via Single part WF}) and
(\ref{aSRIff F alpha beta int,TOIR,IV via Single part WF}) to get their representation in the hydrodynamic functions.
Equation (\ref{aSRIff F alpha beta int,TOIR,IV via Single part WF}) is a part of equation (\ref{aSRIff F alpha beta int,TOIR,III via Single part WF}).
Hence, focus on equation (\ref{aSRIff F alpha beta int,TOIR,III via Single part WF})
and use obtained results to represent equation (\ref{aSRIff F alpha beta int,TOIR,IV via Single part WF}) either.

Present equation (\ref{aSRIff F alpha beta int,TOIR,III via Single part WF}) via functions having intermediate meaning.
It reappear as follows
$$F^{\alpha\beta}_{int,III}
=\frac{1}{2^{3}m\hbar^{2}}\Upsilon_{2}^{\alpha\delta\mu\nu}\times$$
\begin{equation} \label{aSRIff F alpha beta int,TOIR,III via Single part WF simplified}
\times\partial^{\nu}\biggl[n \Xi^{\delta,\beta\mu}+\Lambda^{\beta} P^{\delta\mu}
-\Lambda^{\delta *}R^{\mu\beta}-\Lambda^{\mu}P^{\delta\beta}\biggr]+c.c.,\end{equation}
where
\begin{equation} \label{aSRIff Lambda via varphi} \Lambda^{\alpha}=\sum_{g}n_{g} \varphi_{g}^{*}(p^{\alpha}\varphi_{g}), \end{equation}
\begin{equation} \label{aSRIff P via varphi} P^{\alpha\beta}=\sum_{g}n_{g}(p^{\alpha}\varphi_{g})^{*}(p^{\beta}\varphi_{g}), \end{equation}
\begin{equation} \label{aSRIff R via varphi} 
R^{\alpha\beta}=\sum_{g}n_{g}(\varphi_{g})^{*}p^{\alpha}p^{\beta}\varphi_{g}, \end{equation}
and
\begin{equation} \label{aSRIff A via varphi} \Xi^{\alpha,\beta\gamma}=
\sum_{g}n_{g}(p^{\alpha}\varphi_{g})^{*}(p^{\beta}p^{\gamma}\varphi_{g}). \end{equation}

Functions $\Lambda^{\alpha}$, $P^{\alpha\beta}$, $R^{\alpha\beta}$, $\Xi^{\alpha,\beta\gamma}$ are written in terms of the occupation numbers of the single-particle states.
Present the single-particle wave functions via the amplitudes and phases
$\varphi_{g}=a_{g}e^{\imath S_{g}/\hbar}$.
Next, calculate the described functions including corresponding forms for the hydrodynamic variables:
$n=\sum_{g}n_{g} \varphi_{g}^{*}\varphi_{g}=\sum_{g}n_{g}a_{g}^{2}$,
$j^{\delta}=(1/2m)\sum_{g}n_{g} [\varphi_{g}^{*}(p^{\delta}\varphi_{g})+c.c.]
=(\hbar/m)\sum_{g}n_{g}a_{g}^{2}\partial^{\delta}S_{g}=nv^{\delta}$,
$\Pi^{\alpha\beta}=(1/4m^{2})\sum_{g}n_{g} [\varphi_{g}^{*}(p^{\alpha}p^{\beta}\varphi_{g})+(p^{\alpha}\varphi_{g})^{*}(p^{\beta}\varphi_{g})+c.c.]$
$=nv^{\alpha}v^{\beta}+p^{\alpha\beta}+T^{\alpha\beta}$.

Consider vector function (\ref{aSRIff Lambda via varphi}):
$\Lambda^{\alpha}=\sum_{g}n_{g}[-\imath\hbar a_{g}\partial^{\alpha}a_{g}+\hbar a_{g}^{2}\partial^{\alpha}S_{g}]$
$=\frac{-\imath\hbar}{2}\partial^{\alpha}n+mnv^{\alpha}+m j_{th}^{\alpha}$.
Including that the average thermal velocity is equal to zero (\ref{aSRIff condition of u goes to zero}) $j_{th}^{\alpha}=0$
find
\begin{equation} \label{aSRIff}
\Lambda^{\alpha}
=\frac{-\imath\hbar}{2}\partial^{\alpha}n+mnv^{\alpha}.\end{equation}

Present result for tensor $P^{\mu\delta}$ (\ref{aSRIff P via varphi}) after the segregation of the amplitude and phase of the wave function:
$$P^{\alpha\beta}=\sum_{g}n_{g}[\hbar^{2}\partial^{\alpha}a_{g}\cdot\partial^{\beta}a_{g}
+\hbar^{2}a_{g}^{2}\partial^{\alpha}S_{g}\cdot\partial^{\beta}S_{g}$$
\begin{equation} \label{aSRIff} +\imath\hbar^{2}a_{g}(\partial^{\alpha}a_{g} \cdot\partial^{\beta}S_{g}-\partial^{\alpha}S_{g}\cdot\partial^{\beta}a_{g})].\end{equation}
The last (imaginary) term disappears if it is multiplied by the symmetric tensor.
Hence, drop the last term and present tensor $P^{\alpha\beta}$ via hydrodynamic functions:
$$P^{\alpha\beta}=m^{2}\biggl(nv^{\alpha}v^{\beta} +p^{\alpha\beta}
+\frac{\hbar^{2}}{m^{2}}\sum_{f}n_{f}\partial^{\alpha}a_{f}\cdot\partial^{\beta}a_{f}\biggr)
+\frac{\imath}{2}m\hbar\cdot$$
\begin{equation} \label{aSRIff}
\cdot\biggl(\partial^{\alpha}n\cdot v^{\beta}-\partial^{\beta}n\cdot v^{\alpha}+\sum_{f}n_{f}(\partial^{\alpha}a_{f}^{2}\cdot u_{f}^{\beta}-\partial^{\beta}a_{f}^{2}\cdot u_{f}^{\alpha})\biggr),\end{equation}
where the single-particle approximation is used for the first term similarly to equation (\ref{aSRIff Bohm tensor single part}).

Next, consider tensor $R^{\alpha\beta}$ (\ref{aSRIff R via varphi}).
In the considering case it has the following form:
$$R^{\alpha\beta}=\sum_{g}n_{g}[2\hbar^{2}a_{g}^{2}\partial^{\mu}S_{g}\cdot\partial^{\beta}S_{g}
-\hbar^{2}a_{g}\partial^{\mu}\partial^{\beta}a_{g}+\hbar^{2}\partial^{\mu}a_{g}\cdot\partial^{\beta}a_{g}$$
\begin{equation} \label{aSRIff}-\imath\hbar^{2}a_{g}\partial^{\mu}a_{g}\cdot\partial^{\beta}S_{g} -\imath\hbar^{2}a_{g}^{2}\partial^{\mu}\partial^{\beta}S_{g}].\end{equation}
It allows to get a representation via hydrodynamic functions:
$$R^{\alpha\beta}=m^{2}\biggl(nv^{\alpha}v^{\beta}+p^{\alpha\beta}
-\frac{\hbar^{2}}{m^{2}}\sum_{f}n_{g}a_{g}\partial^{\alpha}\partial^{\beta}a_{g}\biggr)$$
\begin{equation} \label{aSRIff}
-\imath\frac{1}{2}\hbar m\biggl(\partial^{\alpha}(nv^{\beta})
+\partial^{\beta}(nv^{\alpha})\biggr),\end{equation}
where it is included that
$\partial^{\alpha}\sum_{g}n_{g}[a_{g}^{2}u_{g}^{\beta}]=\partial^{\alpha}j_{th}^{\beta}=0$.

Present similar result for the real part of the third order tensor (\ref{aSRIff A via varphi})\begin{widetext}
$$Re \Xi^{\alpha,\beta\gamma}=
m^{3}\biggl[n v^{\alpha}v^{\beta}v^{\gamma}+Q^{\alpha\beta\gamma}
+v^{\alpha}p^{\beta\gamma}+v^{\beta}p^{\alpha\gamma}
+v^{\gamma}p^{\alpha\beta}
+\frac{\hbar^{2}}{4m^{2}}\partial^{\alpha}n\cdot(\partial^{\beta}v^{\gamma} +\partial^{\gamma}v^{\beta})$$
$$+\frac{\hbar^{2}}{m^{2}}\biggl(v^{\beta}\sum_{g}n_{g}[\partial_{g}^{\gamma}a_{g}\cdot\partial_{g}^{\alpha}a_{g}]
+v^{\gamma}\sum_{g}n_{g}[\partial_{g}^{\beta}a_{g}\cdot\partial_{g}^{\alpha}a_{g}]
-v^{\alpha}\sum_{g}n_{g}[a_{g}\partial_{g}^{\gamma}\partial_{g}^{\beta}a_{g}]
+\frac{\hbar^{2}}{m^{2}}\biggl(\sum_{g}n_{g}[u_{g}^{\gamma}\partial_{g}^{\beta}a_{g}\cdot\partial_{g}^{\alpha}a_{g}]$$
\begin{equation} \label{aSRIff}
+\sum_{g}n_{g}[u_{g}^{\beta}\partial_{g}^{\gamma}a_{g}\cdot\partial_{g}^{\alpha}a_{g}]
-\sum_{g}n_{g}[u_{g}^{\alpha}a_{g}\partial_{g}^{\gamma}\partial_{g}^{\beta}a_{g}]
+\frac{1}{2} \sum_{g}n_{g}[a_{g}\partial_{g}^{\alpha}a_{g}\cdot(\partial_{g}^{\beta}u_{g}^{\gamma}+\partial_{g}^{\gamma}u_{g}^{\beta})]\biggr)\biggr],\end{equation}
where tensor $Q^{\beta\delta\mu}=\sum_{g}n_{g}[a_{g}^{2} u_{g}^{\beta}u_{g}^{\delta}u_{g}^{\mu}]$ equivalent to tensor (\ref{aSRIff Q definition}).

Next present result for the imaginary part of the third order tensor (\ref{aSRIff A via varphi}):
$$Im \Xi^{\alpha,\beta\gamma}=m^{2}\hbar\Biggl[\frac{1}{2}
\biggl(v^{\gamma}\partial^{\alpha}v^{\beta}-v^{\gamma}\partial^{\beta}v^{\alpha}-v^{\beta}\partial^{\gamma}v^{\alpha}
-nv^{\alpha}(\partial^{\beta}v^{\gamma}
+\partial^{\gamma}v^{\beta})\biggr)
-\frac{1}{2}\partial^{\beta}p^{\alpha\gamma} -\frac{1}{2}\partial^{\gamma}p^{\alpha\beta}
+\sum_{f}n_{f}\biggl(a_{f}\partial^{\alpha}a_{f}\cdot u_{f}^{\beta}u_{f}^{\gamma}$$
\begin{equation} \label{aSRIff}
+\frac{1}{2}a_{f}^{2}(u_{f}^{\beta}\partial^{\gamma}u_{f}^{\alpha}+u_{f}^{\gamma}\partial^{\beta}u_{f}^{\alpha})\biggr)
+v^{\beta}\sum_{f}n_{f}a_{f}(u_{f}^{\gamma}\partial^{\alpha}a_{f}- u_{f}^{\alpha}\partial^{\gamma}a_{f})
+v^{\gamma}\sum_{f}n_{f}a_{f}(u_{f}^{\beta}\partial^{\alpha}a_{f}- u_{f}^{\alpha}\partial^{\beta}a_{f})
-\frac{\hbar^{2}}{m^{2}}\sum_{f}n_{f}\partial^{\alpha}a_{f}\partial^{\beta}\partial^{\gamma}a_{f}\Biggr] \end{equation}

Combine all described results to get the force tensor field $F^{\alpha\beta}_{int,III}$
presented by equations (\ref{aSRIff F alpha beta int,TOIR,III via Single part WF})
and (\ref{aSRIff F alpha beta int,TOIR,III via Single part WF simplified})
$$F^{\alpha\beta}_{int,III}=\frac{m^{2}}{4\hbar^{2}}\Upsilon_{2}^{\alpha\delta\mu\nu}\partial^{\nu}
\Biggl\{n Q^{\delta\mu\beta}+2nv^{\beta}p^{\delta\mu}
+\frac{\hbar^{2}}{m^{2}}\Biggl[2nv^{\beta}\sum_{f}n_{f}\partial^{\delta}a_{f}\cdot\partial^{\mu}a_{f}
+\frac{1}{2}n\sum_{f}n_{f}a_{f}\partial^{\delta}a_{f}\cdot(\partial^{\beta}u_{f}^{\mu}+\partial^{\mu}u_{f}^{\delta})$$
\begin{equation} \label{aSRIff F alpha beta int,TOIR,III via hydrodynamics}
+n\sum_{f}n_{f}\biggl(u^{\beta}\partial^{\delta}a_{f}\cdot\partial^{\mu}a_{f}
+u^{\mu}\partial^{\delta}a_{f}\cdot\partial^{\beta}a_{f}
-u^{\delta}a_{f}\partial^{\beta}\partial^{\mu}a_{f}\biggr)
-\frac{1}{2}v^{\beta}\partial^{\delta}n\cdot\partial^{\mu}n
-\frac{1}{4}\partial^{\mu}n\cdot\sum_{f}n_{f}(u_{f}^{\beta}\partial^{\delta}a_{f}^{2}-u_{f}^{\delta}\partial^{\beta}a_{f}^{2})\Biggr]\Biggr\}.\end{equation}
\end{widetext}

Similarly, obtain the force tensor field $F^{\alpha\beta}_{int,IV}$ given by equation
(\ref{aSRIff F alpha beta int,TOIR,IV via Single part WF})
$$F^{\alpha\beta}_{int,IV}=-\frac{\imath}{m\hbar^{3}}\frac{1}{3!\cdot 2^{3}}\Upsilon^{\alpha\delta\mu\nu}
\Biggl(n D^{\beta,\delta\mu\nu} -3\Lambda^{\nu *}\Xi^{\beta,\delta\mu *}$$
$$+P^{\delta\beta} R^{\mu\nu *}
-\Lambda^{\beta}B^{\delta\mu\nu *}
+3n D^{\delta,\beta\mu\nu}
-6\Lambda^{\nu} \Xi^{\delta,\beta\mu}
+3R^{\mu\nu}P^{\delta\beta}$$
\begin{equation} \label{aSRIff F alpha beta int,TOIR,IV via Single part WF simplified}
-3\Lambda^{\beta}\Xi^{\delta,\mu\nu}
+6R^{\beta\nu}P^{\delta\mu} -3\Lambda^{\delta *}B^{\beta\mu\nu}
\Biggr)+c.c. .\end{equation}

The force field $F^{\alpha\beta}_{int,IV}$ contains extra functions
which are not introduced above.
Their definitions have the following form:
\begin{equation} \label{aSRIff}
B^{\alpha\beta\gamma}=\sum_{f}n_{f}\varphi_{f}^{*}p^{\alpha}p^{\beta}p^{\gamma}\varphi_{f}, \end{equation}
and
\begin{equation} \label{aSRIff}
D^{\alpha,\beta\gamma\delta}=\sum_{f}n_{f}(p^{\alpha}\varphi_{f})^{*}p^{\beta}p^{\gamma}p^{\delta}\varphi_{f}. \end{equation}
Expressions for $B^{\alpha\beta\gamma}$ and $D^{\alpha,\beta\gamma\delta}$ in terms of hydrodynamic functions are huge.
They are placed in Appendix B.

Substitution of all necessary expressions in equation (\ref{aSRIff F alpha beta int,TOIR,IV via Single part WF simplified})
leads to the explicit form of $F^{\alpha\beta}_{int,IV}$:
\begin{widetext}
$$F^{\alpha\beta}_{int,IV}=-\frac{1}{4}\frac{1}{3!}\frac{m^{2}}{\hbar^{2}}\Upsilon_{2}^{\alpha\delta\mu\nu}
\Biggl\{-9n^{2} v^{\beta}v^{\nu}\partial^{\delta}v^{\mu}+\partial^{\beta}n\cdot Q^{\delta\mu\nu} +6np^{\mu\nu}(\partial^{\beta}v^{\delta}+\partial^{\delta}v^{\beta})
+6nv^{\beta}\sum_{f}n_{f}\partial^{\mu}a_{f}^{2}\cdot u_{f}^{\nu}u_{f}^{\delta}$$
$$+3n\sum_{f}n_{f}a_{f}^{2}u_{f}^{\nu}u_{f}^{\delta}\partial^{\mu}u_{f}^{\beta}
+3p^{\mu\nu}\sum_{f}n_{f}\partial^{\beta}a_{f}^{2}\cdot u_{f}^{\delta}
-3p^{\mu\nu}\sum_{f}n_{f}u_{f}^{\beta}\partial^{\delta}a_{f}^{2}
-3nv^{\delta}\sum_{f}n_{f}a_{f}^{2}u_{f}^{\mu}\partial^{\beta}u_{f}^{\nu}
-3nv^{\delta}\sum_{f}n_{f}a_{f}^{2}u_{f}^{\beta}\partial^{\mu}u_{f}^{\nu}$$
$$+\frac{\hbar^{2}}{m^{2}}\Biggl[
n\partial^{\delta}n\cdot\partial^{\mu}\partial^{\nu}v^{\beta}
+2n\partial^{\delta}n\cdot\partial^{\beta}\partial^{\mu}v^{\nu}
+\frac{3}{2}\partial^{\nu} n\cdot\partial^{\delta} n\cdot(\partial^{\beta}v^{\mu}+\partial^{\mu}v^{\beta})
+\frac{3}{2}\partial^{\beta}n\cdot\partial^{\mu}\partial^{\delta}n\cdot v^{\nu}$$
$$+\frac{3}{2}n\partial^{\mu}\partial^{\delta}n\cdot(\partial^{\beta}v^{\nu}+\partial^{\nu}v^{\beta})
+2nv^{\beta}\sum_{f}n_{f}\partial^{\delta}a_{f}\cdot\partial^{\mu}\partial^{\nu} a_{f} +n\sum_{f}n_{f}u_{f}^{\beta}a_{f}\partial^{\mu}\partial^{\nu}\partial^{\delta}a_{f}
-3n\sum_{f}n_{f}u_{f}^{\delta}a_{f}\partial^{\mu}\partial^{\nu}\partial^{\beta}a_{f}$$
$$-3n\sum_{f}n_{f}u_{f}^{\mu}\partial^{\beta}a_{f}\cdot\partial^{\nu}\partial^{\delta}a_{f}
+3n\sum_{f}n_{f}u_{f}^{\beta}\partial^{\delta}a_{f}\cdot\partial^{\mu}\partial^{\nu} a_{f}
+6n\sum_{f}n_{f}u_{f}^{\nu}\partial^{\delta}a_{f}\cdot\partial^{\mu}\partial^{\beta}a_{f}
+3n\sum_{f}n_{f}\partial^{\nu}a_{f}\cdot\partial^{\delta}a_{f}\cdot(\partial^{\beta}u_{f}^{\mu}+\partial^{\mu}u_{f}^{\beta})$$
$$-\frac{1}{2}n\sum_{f}n_{f}\partial^{\beta}a_{f}^{2}\partial^{\mu}\partial^{\nu}u_{f}^{\delta}
+\frac{1}{2}n\sum_{f}n_{f}\partial^{\delta}a_{f}^{2}\partial^{\mu}\partial^{\nu}u_{f}^{\beta}
+n\sum_{f}n_{f}\partial^{\delta}a_{f}^{2}\partial^{\mu}\partial^{\beta}u_{f}^{\nu}
+3n(\partial^{\beta}v^{\mu}+\partial^{\mu}v^{\beta})\sum_{f}n_{f}(\partial^{\nu}a_{f}\cdot\partial^{\delta}a_{f}-a_{f}\partial^{\nu}\partial^{\delta}a_{f})$$
$$+6\partial^{\nu}n\cdot v^{\delta}\sum_{f}n_{f}a_{f}\partial^{\beta}\partial^{\mu}a_{f}
+3\partial^{\nu}n\cdot\sum_{f}n_{f} a_{f}\partial^{\delta}a_{f}\cdot(\partial^{\beta}u_{f}^{\mu}+\partial^{\mu}u_{f}^{\beta})
+\frac{1}{2}\partial^{\nu}n\cdot\sum_{f}n_{f}a_{f}^{2}\partial^{\delta}\partial^{\mu}u_{f}^{\beta}$$
$$+\partial^{\nu}n\cdot\sum_{f}n_{f}a_{f}^{2}\partial^{\delta}\partial^{\beta}u_{f}^{\mu}
-3\partial^{\beta}n\cdot v^{\delta}\sum_{f}n_{f}(a_{f}\partial^{\mu}\partial^{\nu}a_{f}+\partial^{\mu}a_{f}\cdot\partial^{\nu}a_{f})
-3\partial^{\beta}n\cdot\sum_{f}n_{f}u_{f}^{\delta}(a_{f}\partial^{\mu}\partial^{\nu}a_{f}+\partial^{\mu}a_{f}\cdot\partial^{\nu}a_{f})$$
\begin{equation} \label{aSRIff F alpha beta int,TOIR,IV via hydrodynamics}
+3\partial^{\nu}n\cdot\sum_{f}n_{f}u_{f}^{\beta}(a_{f}\partial^{\delta}\partial^{\mu}a_{f}+\partial^{\delta}a_{f}\cdot\partial^{\mu}a_{f})
+\frac{1}{2}\partial^{\beta}n\cdot\sum_{f}n_{f}a_{f}^{2}\partial^{\mu}\partial^{\nu}u_{f}^{\delta}
+\sum_{f}n_{f}a_{f}\partial^{\mu}\partial^{\nu} a_{f}\cdot
\sum_{f'}n_{f'}(u_{f}^{\beta}\partial^{\delta}a_{f}^{2}-u_{f}^{\delta}\partial^{\beta}a_{f}^{2}) \Biggr]\Biggr\}.\end{equation}

\end{widetext}

The quantum terms in equations (\ref{aSRIff F alpha beta int,TOIR,III via hydrodynamics})
and (\ref{aSRIff F alpha beta int,TOIR,IV via hydrodynamics})
contain higher space derivatives.
It means that their role is small in the long-wavelength limit.
Hence, the analysis of the long-wavelength limit allows a simplification of the force tensor fields
(\ref{aSRIff F alpha beta int,TOIR,III via hydrodynamics})
and (\ref{aSRIff F alpha beta int,TOIR,IV via hydrodynamics})
while it does not affect the structure of the of the quantum stress tensor
(\ref{aSRIff sigma fer TOIR via pressure and I 0}).
Present the force tensor field in the long-wavelength limit

$$F^{\alpha\beta}_{int,TOIR}=\frac{m^{2}}{4\hbar^{2}}g_{2} I_{0}^{\alpha\delta\mu\nu}
\Biggl\{\partial^{\nu} (nQ^{\delta\mu\beta})
+\frac{1}{6}\partial^{\beta}n\cdot Q^{\delta\mu\nu}$$
\begin{equation} \label{aSRIff F alpha beta int,TOIR LWLL}
+\frac{3}{2}n^{2}v^{\beta}v^{\nu}\partial^{\delta}v^{\mu}
+np^{\mu\nu}(\partial^{\delta}v^{\beta}-\partial^{\beta}v^{\delta})\Biggl\}.\end{equation}

\subsection{The force tensor field for the interaction of different species}

Contribution of the short-range interaction in the momentum flux evolution equation can be found
by the analysis of equation (\ref{aSRIff F alpha beta original}).
Introduce the interparticle distance and the coordinate of center of mass for pair of particles
($i$-th and $j$-th particles) belonging to the different species.
Next, make the expansion in the series on the interparticle distance $r_{ij}^{\alpha}$ and keep terms up to the third order on $r_{ij}^{\alpha}$.
Similar to the analysis described above,
we find that the zeroth and the second orders contributions are equal to zero.
Hence, we need to study the contributions in the first order by the interaction radius and third order by the interaction radius approximations.
Consider these groups of terms separately.

Start our calculation with terms existing in the first order by the interaction radius approximation.
After the expansion we include that $\Psi(R')=0$
while $\partial^{\alpha}_{Ri}\Psi(R')$ and other derivatives of the expanded wave function are not zero.
It gives us the following expression for the force tensor field:  
$$F^{\alpha\beta}_{int,FOIR}=-\frac{\imath\hbar}{4m}\int dR\sum_{i,j\neq i}r_{ij}^{\gamma}(\partial_{i}^{\alpha}U_{ij})\times$$
$$\times\biggl[\partial_{\gamma}\delta_{i}\cdot\Psi^{*}(R',t)\partial^{\beta}_{R1}\Psi(R',t)$$
$$-\delta_{i}\biggl((\partial^{\gamma}_{R1}-\partial^{\gamma}_{R2})\Psi^{*}(R',t)\cdot\partial^{\beta}_{R1}\Psi(R',t)$$
\begin{equation} \label{aSRIff}
+\Psi^{*}(R',t)\partial^{\beta}_{R1}(\partial^{\gamma}_{R1}-\partial^{\gamma}_{R2})\Psi(R',t)\biggr)\biggr]+c.c..\end{equation}
Here there are two groups of terms: one with the derivative of the delta function, and another one without derivative of the delta function.

Calculation gives the following expression containing contributions of both groups
$F^{\alpha\beta}_{int,FOIR}=
-\frac{1}{2}g\partial^{\alpha} (n_{(2)} \cdot j^{\beta}_{(1)})
-\frac{1}{2}g(j^{\beta}_{(1)}\partial^{\alpha} n_{(2)} - n_{(2)}\partial^{\alpha} j^{\beta}_{(1)})$
which provides the following combination
\begin{equation} \label{aSRIff F alpha beta FOIR diff species} F^{\alpha\beta}_{int,FOIR}=
-g\partial^{\alpha} n_{(2)} \cdot j^{\beta}_{(1)}\end{equation}

The expression for the force tensor field in the third order by the interaction radius approximation is rather large in this regime.
Hence, it is presented as a combination of the partial force tensor fields.
It is splitted on three parts presented below:\begin{widetext}
$$F^{\alpha\beta}_{TOIR,0}=-\frac{1}{2m}\frac{1}{2^{3}}\int dR\sum_{i,j\neq i}r_{ij}^{\gamma}r_{ij}^{\delta}r_{ij}^{\mu}\partial_{i}^{\alpha}U_{ij}\times$$
$$\times\biggl[\frac{1}{3!}\partial^{\gamma}\partial^{\delta}\partial^{\mu}\delta_{i}\cdot\Psi^{*}(R',t)\imath\hbar\partial^{\beta}_{R1}\Psi(R',t)
+\frac{1}{2!}\partial^{\gamma}\delta_{i}\cdot\biggl((\partial^{\mu}_{1}\partial^{\delta}_{1}-2\partial^{\mu}_{1}\partial^{\delta}_{2}+\partial^{\mu}_{2}\partial^{\delta}_{2})\Psi^{*}(R',t) \cdot\imath\hbar\partial^{\beta}_{R1}\Psi(R',t)$$
\begin{equation} \label{aSRIff}
-\Psi^{*}(R',t)\imath\hbar\partial^{\beta}_{R1} (\partial^{\mu}_{1}\partial^{\delta}_{1}-2\partial^{\mu}_{1}\partial^{\delta}_{2}+\partial^{\mu}_{2}\partial^{\delta}_{2})\Psi(R',t)
-(\partial^{\mu}_{R1}-\partial^{\mu}_{R2})\Psi^{*}(R',t) \cdot\imath\hbar\partial^{\beta}_{R1}(\partial^{\delta}_{R1}-\partial^{\delta}_{R2})\Psi(R',t)\biggr)\biggr], \end{equation}
$$F^{\alpha\beta}_{TOIR,1}=\frac{1}{2m}\frac{1}{2^{3}}\int dR\sum_{i,j\neq i}r_{ij}^{\gamma}r_{ij}^{\delta}r_{ij}^{\mu}\partial_{i}^{\alpha}U_{ij}\times$$
\begin{equation} \label{aSRIff}
\times\frac{1}{2!}\partial^{\gamma}\partial^{\delta}\delta_{i}\cdot\biggl(
(\partial^{\mu}_{R1}-\partial^{\mu}_{R2})\Psi^{*}(R',t)\cdot\imath\hbar\partial^{\beta}_{R1}\Psi(R',t) +\Psi^{*}(R',t)\cdot\imath\hbar\partial^{\beta}_{R1}(\partial^{\mu}_{R1}-\partial^{\mu}_{R2})\Psi(R',t)\biggr),\end{equation}
and
$$F^{\alpha\beta}_{TOIR,2}=\frac{1}{2m}\frac{1}{2^{3}}\int dR\sum_{i,j\neq i}r_{ij}^{\gamma}r_{ij}^{\delta}r_{ij}^{\mu}\partial_{i}^{\alpha}U_{ij}\delta_{i}\times$$
$$\times\Biggl[\frac{1}{2!}  (\partial^{\gamma}_{R1}-\partial^{\gamma}_{R2})(\partial^{\delta}_{R1}-\partial^{\delta}_{R2})\Psi^{*} \cdot\imath\hbar\partial^{\beta}_{R1}(\partial^{\mu}_{R1}-\partial^{\mu}_{R2})\Psi
+\frac{1}{2!}  (\partial^{\gamma}_{R1}-\partial^{\gamma}_{R2})\Psi^{*} \cdot\imath\hbar\partial^{\beta}_{R1}(\partial^{\delta}_{R1}-\partial^{\delta}_{R2})(\partial^{\mu}_{R1}-\partial^{\mu}_{R2})\Psi$$
\begin{equation} \label{aSRIff}
+\frac{1}{3!}\Psi^{*} \cdot\imath\hbar\partial^{\beta}_{R1}(\partial^{\gamma}_{R1}-\partial^{\gamma}_{R2}) (\partial^{\delta}_{R1}-\partial^{\delta}_{R2})(\partial^{\mu}_{R1}-\partial^{\mu}_{R2})\Psi
+\frac{1}{3!}
(\partial^{\gamma}_{R1}-\partial^{\gamma}_{R2})(\partial^{\delta}_{R1}-\partial^{\delta}_{R2})(\partial^{\mu}_{R1}-\partial^{\mu}_{R2})\Psi^{*} \cdot\imath\hbar\partial^{\beta}_{R1}\Psi  +c.c.\Biggr].\end{equation}
\end{widetext}

After calculation find the following representations of the corresponding partial force tensor fields:
$$F^{\alpha\beta}_{TOIR,0}=
\frac{1}{2}\frac{1}{3!}\Upsilon_{2}^{\alpha\gamma\delta\mu}\partial^{\gamma}\biggl[\partial^{\delta}\partial^{\mu}n_{(2)}\cdot j^{\beta}_{(1)}$$
\begin{equation} \label{aSRIff}
+n_{(2)}\partial^{\delta}\partial^{\mu}j^{\beta}_{(1)} -\partial^{\delta}n_{(2)}\cdot\partial^{\mu}j^{\beta}_{(1)}\biggr],\end{equation}
\begin{equation} \label{aSRIff} F^{\alpha\beta}_{TOIR,1}=\frac{1}{2^{3}}\frac{1}{2!}\Upsilon_{2}^{\alpha\gamma\delta\mu}\partial^{\gamma}\partial^{\delta}
\biggl[\partial^{\mu}n_{(2)}\cdot j^{\beta}_{(1)}-n_{(2)}\partial^{\mu}j^{\beta}_{(1)}\biggr],\end{equation}
and
$$F^{\alpha\beta}_{TOIR,2}=\frac{1}{2}\frac{1}{3!}\Upsilon_{2}^{\alpha\gamma\delta\mu}
\biggl[n_{(2)}\partial^{\gamma}\partial^{\delta}\partial^{\mu}j^{\beta}_{(1)}
-3\partial^{\gamma}n_{(2)}\cdot \partial^{\delta}\partial^{\mu}j^{\beta}_{(1)}$$
\begin{equation} \label{aSRIff}
+3\partial^{\gamma}\partial^{\delta}n_{(2)}\cdot \partial^{\mu}j^{\beta}_{(1)}
-\partial^{\gamma}\partial^{\delta}\partial^{\mu}n_{(2)}\cdot j^{\beta}_{(1)}\biggr].\end{equation}

Their combination gives the following force tensor field
\begin{equation} \label{aSRIff F alpha beta TOIR diff species}
F^{\alpha\beta}_{int,TOIR}=
\frac{1}{3!}\Upsilon_{2}^{\alpha\gamma\delta\mu}\partial^{\gamma}\partial^{\delta}\partial^{\mu} n_{(2)}\cdot j^{\beta}_{(1)}, \end{equation}
where subindex $(1)$ describes the system under study, subindex $(2)$ represents the species acting on our system.

The following relations exist between the tensor and scalar interaction constants for the isotropic interaction:
$\Upsilon_{2}^{\alpha\beta\gamma\delta}=-g_{2}I_{0}^{\alpha\beta\gamma\delta}$.

\section{Contribution of the SRI in the pressure evolution equation}

The contribution of the short-range interaction in the momentum flux $\Pi^{\alpha\beta}$ evolution equation is calculated above.
Next, it is necessary to consider contribution of the short-range interaction in the pressure $p^{\alpha\beta}$ evolution equation
following our analysis near equation (\ref{aSRIff eq for p alpha beta}).

Equation (\ref{aSRIff eq for p alpha beta}) can be rewritten via general force field $F^{\alpha}$ and the force tensor field $F^{\alpha\beta}$:
$$\partial_{t}p^{\alpha\beta} +v^{\gamma}\partial_{\gamma}p^{\alpha\beta} +p^{\alpha\gamma}\partial_{\gamma}v^{\beta} +p^{\beta\gamma}\partial_{\gamma}v^{\alpha}$$
$$+p^{\alpha\beta}\partial_{\gamma}v^{\gamma} +\partial_{\gamma}Q^{\alpha\beta\gamma} +\partial_{\gamma}T^{\alpha\beta\gamma}
+\partial_{\gamma}L^{\alpha\beta\gamma}$$
$$+\frac{\hbar^{2}}{4m^{2}}\biggl[\partial_{\alpha}\partial_{\beta}\partial_{\gamma}(nv^{\gamma})
-v^{\gamma}\partial_{\alpha}\partial_{\beta}\partial_{\gamma}n
-\frac{1}{n}(\partial_{\gamma}v^{\gamma})\partial_{\alpha}n\cdot\partial_{\beta}n$$
$$-\frac{\partial_{\beta}n}{n}\cdot \partial_{\gamma}(n\cdot\partial_{\alpha}v^{\gamma})
-\frac{\partial_{\alpha}n}{n}\cdot \partial_{\gamma}(n\cdot\partial_{\beta}v^{\gamma}) \biggr]$$
\begin{equation} \label{aSRIff eq for p alpha beta II}
=\frac{1}{m}(F^{\alpha\beta}+F^{\beta\alpha}-F^{\alpha}v^{\beta}-F^{\beta}v^{\alpha}). \end{equation}

Analysis of tensor $L^{\alpha\beta\gamma}$ presented after equation (\ref{aSRIff L alpha beta gamma via thermal velocity}) shows
that $L^{\alpha\beta\gamma}=0$ can be used as an equation of state.

Consider $F^{\alpha\beta}+F^{\beta\alpha}-F^{\alpha}v^{\beta}-F^{\beta}v^{\alpha}$ for interaction between particles of the same species and for interaction between species separately.

Start with regime of different species and find
$F^{\alpha\beta}+F^{\beta\alpha}-F^{\alpha}v^{\beta}-F^{\beta}v^{\alpha}=0$.
Consider of each pair like $F^{\alpha\beta}-F^{\alpha}v^{\beta}$
by substituting equations (\ref{aSRIff F up to TOIR via n1 n2}), (\ref{aSRIff F alpha beta FOIR diff species}), and (\ref{aSRIff F alpha beta TOIR diff species}) \emph{and}
find
$F^{\alpha\beta}-F^{\alpha}v^{\beta}=0$.

Next, consider combination $F^{\alpha\beta}-F^{\alpha}v^{\beta}$ for the interaction of the particles of the same species.
First, mention that $F^{\alpha\beta}$ and $F^{\alpha}$ are equal to zero in the first order by the interaction radius approximation.

The force field $F^{\alpha}=-\partial_{\beta}\sigma^{\alpha\beta}$ is presented
via the quantum stress tensor (\ref{aSRIff sigma fer TOIR via pressure and I 0}).
Combination $F^{\alpha\beta}-F^{\alpha}v^{\beta}$ entering the right-hand side
of the pressure tensor evolution equation (\ref{aSRIff eq for p alpha beta II}) has the following form:
$F^{\alpha\beta}-F^{\alpha}v^{\beta}=
F^{\alpha\beta}
+(m^{2}/2\hbar^{2})g_{2}I_{0}^{\alpha\gamma\delta\mu} v^{\beta}\partial_{\gamma}(np^{\delta\mu})$,
where $F^{\alpha\beta}$ is given by equation
(\ref{aSRIff F alpha beta int,TOIR LWLL}).

Full expression for the right-hand side of the pressure evolution equation (\ref{aSRIff eq for p alpha beta II}) has the following form
$$F^{\alpha\beta}+F^{\beta\alpha}-F^{\alpha}v^{\beta}-F^{\beta}v^{\alpha}$$
$$=-\frac{m}{4\hbar^{2}}g_{2} I_{0}^{\alpha\delta\mu\nu}\biggl[\partial^{\nu}(nQ^{\delta\mu\beta})+\frac{1}{6}Q^{\delta\mu\nu}\partial^{\beta}n\biggr]$$
$$-\frac{m}{8\hbar^{2}}g_{2} I_{0}^{\alpha\gamma\delta\mu}
[3 n_{s}^{2} v_{s}^{\beta}v_{s}^{\delta}\partial^{\gamma}v_{s}^{\mu}
+2n_{s}p_{s}^{\mu\delta}(\partial^{\gamma}v_{s}^{\beta}-\partial^{\beta}v_{s}^{\gamma})]$$
$$-\frac{m}{4\hbar^{2}}g_{2} I_{0}^{\beta\delta\mu\nu}\biggl[\partial^{\nu}(nQ^{\delta\mu\alpha})+\frac{1}{6}Q^{\delta\mu\nu}\partial^{\alpha}n\biggr]$$
\begin{equation} \label{aSRIff eq for FF-Fv}
-\frac{m}{8\hbar^{2}}g_{2} I_{0}^{\beta\gamma\delta\mu}
[3 n_{s}^{2} v_{s}^{\alpha}v_{s}^{\delta}\partial^{\gamma}v_{s}^{\mu}
+2n_{s}p_{s}^{\mu\delta}(\partial^{\gamma}v_{s}^{\alpha}-\partial^{\alpha}v_{s}^{\gamma})]. \end{equation}

The left-hand side of the pressure evolution equation (\ref{aSRIff eq for p alpha beta II})
requires an equation of state for tensor $Q^{\alpha\beta\gamma}$.
Moreover, this tensor is a part of the force tensor field $F^{\alpha\beta}$.
Calculate it for the equilibrium distribution function for the degenerate fermions:
$Q^{\alpha\beta\gamma}=(2/(2\pi\hbar)^{3})\int p^{\alpha} p^{\beta} p^{\gamma} \Theta(p-p_{Fe})d^{3}p$,
where $\Theta(x)$ is the theta function (the function of Heaviside), and $p_{Fe}$ is the Fermi momentum.
Presented integral splits on product of two integrals on the module of the momentum and the angle part:
$(2/(2\pi\hbar)^{3})\int_{0}^{p_{Fe}} p^{5} dp=(3/8)\pi\hbar^{3}n_{0}^{2}$
and
$\int n_{\textbf{p}}^{\alpha} n_{\textbf{p}}^{\beta} n_{\textbf{p}}^{\gamma} d\Omega=0$,
where $\textbf{n}_{\textbf{p}}=\textbf{p}/p$ is the unit vector in the momentum space,
and $\Omega$ is the solid angle.
It gives the zero value of the considered tensor $Q^{\alpha\beta\gamma}=0$ as an equation of state.

\section{Hydrodynamic equations in the first order by the interaction radius}

Nonzero contribution of the interaction in the first order by the interaction radius exists for nonpolarized or the partially polarized systems
while the fully polarized systems have zero contribution in this case
as it is described above.

\subsection{A minimal coupling model: separate spin evolution}

Summarize the results obtained in Secs. II and III including the interaction in the first order by the interaction radius approximation.
Consider the evolution of fermions with different spin projections separately.

In this regime we have two continuity equations:
\begin{equation}\label{aSRIff cont eq via vel spin up} \partial_{t}n_{\uparrow}+\nabla\cdot (n_{\uparrow}\textbf{v}_{\uparrow})=0, \end{equation}
and
\begin{equation}\label{aSRIff cont eq via vel spin down} \partial_{t}n_{\downarrow}+\nabla\cdot (n_{\downarrow}\textbf{v}_{\downarrow})=0. \end{equation}

We also have two Euler (momentum balance) equations
$$mn_{\uparrow}(\partial_{t} +\textbf{v}_{\uparrow}\cdot\nabla)v^{\alpha}_{\uparrow}
-\frac{\hbar^{2}}{2m}n_{\uparrow}\partial^{\alpha}\frac{\triangle\sqrt{n_{\uparrow}}}{\sqrt{n_{\uparrow}}}$$
\begin{equation}\label{aSRIff Euler FOIR spin up}
+\frac{(6\pi^{2})^{\frac{2}{3}}\hbar^{2}}{3m}n_{\uparrow}^{\frac{2}{3}}\partial^{\alpha}n_{\uparrow}
+g_{\uparrow\downarrow} n_{\uparrow}\partial^{\alpha}n_{\downarrow} =-n_{\uparrow}\partial^{\alpha}V_{ext},\end{equation}
and
$$mn_{\downarrow}(\partial_{t} +\textbf{v}_{\downarrow}\cdot\nabla)v^{\alpha}_{\downarrow}
-\frac{\hbar^{2}}{2m}n_{\downarrow}\partial^{\alpha}\frac{\triangle\sqrt{n_{\downarrow}}}{\sqrt{n_{\downarrow}}}$$
\begin{equation}\label{aSRIff Euler FOIR spin down}
+\frac{(6\pi^{2})^{\frac{2}{3}}\hbar^{2}}{3m}n_{\downarrow}^{\frac{2}{3}}\partial^{\alpha}n_{\downarrow}
+g_{\uparrow\downarrow} n_{\downarrow}\partial^{\alpha}n_{\uparrow} =-n_{\downarrow}\partial^{\alpha}V_{ext}.\end{equation}

The short-range interaction does not change the partial concentrations.

Similar approach and notations are used in Ref. \cite{Comparin PRA 19}
studying two-dimensional dipolar fermions.

The spin-spin interaction is not included in equations (\ref{aSRIff cont eq via vel spin up})-(\ref{aSRIff Euler FOIR spin down}),
but it gives the change of the partial concentrations and partial currents (see Ref. \cite{Andreev LPL 18}).

Minimal coupling assumes the application of the continuity and Euler equation with no account of the pressure evolution,
but application of the equation of state for the reduction of the pressure evolution to the concentration evolution.

Equations (\ref{aSRIff cont eq via vel spin up})-(\ref{aSRIff Euler FOIR spin down})
correspond to the non-linear Schr\"{o}dinger equation for fermions or more precisely non-linear Pauli equation:
\begin{equation}\label{aSRIff NLSE Pauli first appearence} \imath\hbar\partial_{t}\Phi
=\biggl(-\frac{\hbar^{2}\nabla^{2}}{2m}+\hat{\pi}+V_{ext} +\left(
                                           \begin{array}{cc}
                                             g_{\uparrow\downarrow} n_{\downarrow} & 0 \\
                                             0 & g_{\uparrow\downarrow} n_{\uparrow} \\
                                           \end{array}
                                         \right)\biggr)\Phi,\end{equation}
which is in a way similar to traditional Gross-Pitaevskii equation for BEC \cite{Dalfovo RMP 99},
with
\begin{equation}\label{aSRIff} \hat{\pi}=\left(
                                           \begin{array}{cc}
                                             \pi_{\uparrow} & 0 \\
                                             0 & \pi_{\downarrow} \\
                                           \end{array}
                                         \right),
\end{equation}
where $\pi_{s}\equiv (6\pi^{2}n_{s})^{\frac{2}{3}}\hbar^{2}/2m$,
and
\begin{equation}\label{aSRIff def Phi} \Phi(\textbf{r},t)=\left(
                                                            \begin{array}{c}
                                                              \sqrt{n_{\uparrow}} e^{\imath m\phi_{\uparrow}/\hbar} \\
                                                              \sqrt{n_{\downarrow}} e^{\imath m\phi_{\downarrow}/\hbar} \\
                                                            \end{array}
                                                          \right)
,\end{equation}
where $\textbf{v}_{\uparrow}=\nabla\phi_{\uparrow}$ and $\textbf{v}_{\downarrow}=\nabla\phi_{\downarrow}$.

The last term in equation (\ref{aSRIff NLSE Pauli first appearence}) describes the interparticle interaction.
This interaction happens between fermions of different spin projections.
It contains the cubic nonlinearity.
The second term on the right-hand side of non-linear Pauli equation (\ref{aSRIff NLSE Pauli first appearence}) is cased by the Fermi pressure.
It is a non-linear term with the fractional nonlinearity.

\subsection{A minimal coupling model: single fluid approach}

Full concentration is the sum of the partial concentrations $n=n_{\uparrow}+n_{\downarrow}$.
Same correct for the current $\textbf{j}=\textbf{j}_{\uparrow}+\textbf{j}_{\downarrow}$ and the momentum flux $\Pi^{\alpha\beta}=\Pi^{\alpha\beta}_{\uparrow}+\Pi^{\alpha\beta}_{\downarrow}$.
When, equations for the spin-up and spin-down fermions combines in the following equations:
\begin{equation}\label{aSRIff cont eq via vel Single Fluid} \partial_{t}n+\nabla\cdot (n\textbf{v})=0, \end{equation}
and
$$ mn(\partial_{t} +\textbf{v}\cdot\nabla)v^{\alpha}
-\frac{\hbar^{2}}{2m}n\partial^{\alpha}\frac{\triangle\sqrt{n}}{\sqrt{n}}$$
\begin{equation}\label{aSRIff Euler equation Single Fluid}
+\vartheta\frac{(3\pi^{2})^{\frac{2}{3}}\hbar^{2}}{3m}n^{\frac{2}{3}}\partial^{\alpha}n +g_{\uparrow\downarrow} \partial^{\alpha}(n_{\uparrow}n_{\downarrow}) =-n\partial^{\alpha}V_{ext},\end{equation}
where
\begin{equation}\label{aSRIff vartheta def}\vartheta=\frac{1}{2}\biggl[(1+\eta)^{\frac{5}{3}}+(1-\eta)^{\frac{5}{3}}\biggr].\end{equation}

After all, obtained result can be rewritten via the quantum stress tensor in the first order by the interaction radius:
\begin{equation}\label{aSRIff sigma fermions FOIR via up down}
\sigma_{1}^{\alpha\beta}=g_{\uparrow\downarrow}\delta^{\alpha\beta} n_{\uparrow}n_{\downarrow}.\end{equation}
However, the quantum stress is not written in terms of the single fluid model.
Use the representation of the partial concentrations via the full concentration and the spin polarization
$\eta=\mid n_{\uparrow}-n_{\downarrow}\mid/(n_{\uparrow}+n_{\downarrow})$.
Consequently, we obtain the quantum stress tensor via the concentration of all fermions
\begin{equation}\label{aSRIff sigma fermions FOIR via n eta}
\sigma_{1}^{\alpha\beta}=\frac{1-\eta^{2}}{4}g_{\uparrow\downarrow}\delta^{\alpha\beta} n^{2}.\end{equation}

Equations (\ref{aSRIff cont eq via vel Single Fluid}), (\ref{aSRIff Euler equation Single Fluid}) correspond
to the non-linear Schr\"{o}dinger equation for fermions:
$$\imath\hbar\partial_{t}\Phi
=\biggl(-\frac{\hbar^{2}\nabla^{2}}{2m}+V_{ext} $$
\begin{equation}\label{aSRIff NLSE first appearence after Pauli}
+\vartheta(3\pi^{2})^{\frac{2}{3}}\frac{\hbar^{2}}{2m}\mid\Phi\mid^{4/3}
+\frac{1-\eta^{2}}{4}g_{\uparrow\downarrow}\mid\Phi\mid^{2}\biggr)\Phi,\end{equation}
which is in a way similar to traditional Gross-Pitaevskii equation for BEC \cite{Dalfovo RMP 99},
with
\begin{equation}\label{aSRIff def Phi} \Phi(\textbf{r},t)=\sqrt{n} e^{\imath m\phi/\hbar},\end{equation}
where $\textbf{v}=\nabla\phi$.

Equation (\ref{aSRIff NLSE first appearence after Pauli}) includes two nonlinear terms.
One of them has fractional nonlinearity and caused by the Fermi pressure (the third term on the right-hand side).
Another nonlinear term is related to the interaction between fermions with different spin projections in the first order by the interaction radius.
Equation (\ref{aSRIff NLSE first appearence after Pauli}) is a single fluid reduction of equation (\ref{aSRIff NLSE Pauli first appearence}).

Equation similar to non-linear Schr\"{o}dinger equation (\ref{aSRIff NLSE first appearence after Pauli}) are used in literature
\cite{Butts PRA97, Belemuk PRA 07, Adhikari PRA05, Bludov PRA06, Rizzi PRA08, Maruyama PRA08, Karpiuk PRA06},
but the partial spin polarization is not included there.

\section{Hydrodynamic minimal coupling model in third order by the interaction radius}

\subsection{Regime of the full spin polarization}

The quantum stress tensor given by equation (\ref{aSRIff sigma fer in 2 or pl w short form})
is a part of the following Euler equation written up to the third order by the interaction radius
$$mn(\partial_{t} +\textbf{v}\cdot\nabla)v^{\alpha} +n\partial^{\alpha}V_{ext} -\frac{\hbar^{2}}{2m}n\partial^{\alpha}\frac{\triangle\sqrt{n}}{\sqrt{n}}$$
\begin{equation}\label{aSRIff Euler TOIR full spin polarization}
+\frac{(6\pi^{2})^{\frac{2}{3}}\hbar^{2}}{3m}n^{\frac{2}{3}}\partial^{\alpha}n =-g_{2}\frac{4}{3}(6\pi^{2})^{\frac{2}{3}}n^{\frac{5}{3}}\partial^{\alpha}n,\end{equation}
where the right-hand side of equation (\ref{aSRIff Euler TOIR full spin polarization})
shows interaction between fermions of the single spin polarization.

The Euler equation (\ref{aSRIff Euler TOIR full spin polarization}) together with the continuity equation
of the traditional form (\ref{aSRIff cont eq via vel Single Fluid}) can be represented as non-linear Schr\"{o}dinger equation:
$$\imath\hbar\partial_{t}\Phi
=\biggl(-\frac{\hbar^{2}\nabla^{2}}{2m}+V_{ext} $$
\begin{equation}\label{aSRIff NLSE spin 1 TOIR}
+(6\pi^{2})^{\frac{2}{3}}\frac{\hbar^{2}}{2m}\mid\Phi\mid^{4/3}
+\frac{4}{5}g_{2}(6\pi^{2})^{\frac{2}{3}}\mid\Phi\mid^{10/3}\biggr)\Phi.\end{equation}

This non-linear Schr\"{o}dinger equation contains two non-liner terms.
Both of them have fractional nonlinearity.
One of them is the third term on the right-hand side
which is caused by the Fermi pressure.
The second nonlinear term is related to the interaction between fermions of the same spin projection.
It is presented by the last term in equation (\ref{aSRIff NLSE spin 1 TOIR}).

Equation (\ref{aSRIff NLSE spin 1 TOIR}) shows some resemblance to the energy density functional presented
by eq. 5 in Ref. \cite{Parker PRA 12} and eq. 12 in Ref. \cite{Roth PRA 02}.
It is based on an effective many-particle-look Hamiltonian containing the quasi-potential
including p-wave term containing nonzero momentum difference of the interacting particles.
If momentums are the operators of momentum it means higher derivatives of the wave function just like in our expansion.

Fermions are studied by a set of different hydrodynamic models.
For instance, there is the Thomas--Fermi–-von Weizs\"{a}cker hydrodynamic theory \cite{Zyl PRA 14, Zyl PRA 13}
applied for examination of a zero-temperature, spin-polarized, harmonically trapped, dipolar Fermi gas.

If equation of state is considered,
present the macroscopic wave function $\Phi$
as $\Phi=\Phi_{0}e^{-\imath\mu t/\hbar}$,
where $\mu$ is the chemical potential
and $\Phi_{0}$ is a constant at zero external field $V_{ext}=0$.
Consequently, find the derivation of the chemical potential from the Fermi energy caused by the interaction between fermions of the same spin projection:
\begin{equation}\label{aSRIff chem pot TOIR}
\mu=(6\pi^{2})^{\frac{2}{3}}\frac{\hbar^{2}}{2m}n^{2/3}
+\frac{4}{5}g_{2}(6\pi^{2})^{\frac{2}{3}}n^{5/3},\end{equation}
where $n=\mid\Phi_{0}\mid^{2}$.
The change of the chemical potential is found for the zero temperature.
The second term in equation (\ref{aSRIff chem pot TOIR}) has stronger dependence on the concentration of fermions.
Since, chemical potential (\ref{aSRIff chem pot TOIR}) is found in the limit of small interaction.
Hence, the second term in (\ref{aSRIff chem pot TOIR}) can be applied for the relatively small concentrations,
which, nevertheless, are suitable for the experiments with ultracold fermions.

\subsection{Partial spin polarization and separate spin evolution}

Separate spin evolution presented by equations (\ref{aSRIff Euler FOIR spin up}) and (\ref{aSRIff Euler FOIR spin down})
can be studied in more details at the account of the third order by the interaction radius:
$$mn_{\uparrow}(\partial_{t} +\textbf{v}_{\uparrow}\cdot\nabla)v^{\alpha}_{\uparrow} +n_{\uparrow}\partial^{\alpha}V_{ext}
-\frac{\hbar^{2}}{2m}n_{\uparrow}\partial^{\alpha}\frac{\triangle\sqrt{n_{\uparrow}}}{\sqrt{n_{\uparrow}}}$$
$$+\frac{(6\pi^{2})^{\frac{2}{3}}\hbar^{2}}{3m}n_{\uparrow}^{\frac{2}{3}}\partial^{\alpha}n_{\uparrow} =-g_{2}\frac{4}{3}(6\pi^{2})^{\frac{2}{3}}n_{\uparrow}^{\frac{5}{3}}\partial^{\alpha}n_{\uparrow}$$
\begin{equation}\label{aSRIff Euler TOIR SSE spin up} -g_{\uparrow\downarrow} n_{\uparrow}\partial^{\alpha}n_{\downarrow}
-\frac{g_{2,\uparrow\downarrow}}{2}n_{\uparrow}\partial^{\alpha}\triangle n_{\downarrow}
,\end{equation}
and
$$mn_{\downarrow}(\partial_{t} +\textbf{v}_{\downarrow}\cdot\nabla)v^{\alpha}_{\downarrow} +n_{\downarrow}\partial^{\alpha}V_{ext}
-\frac{\hbar^{2}}{2m}n_{\downarrow}\partial^{\alpha}\frac{\triangle\sqrt{n_{\downarrow}}}{\sqrt{n_{\downarrow}}}$$
$$+\frac{(6\pi^{2})^{\frac{2}{3}}\hbar^{2}}{3m}n_{\downarrow}^{\frac{2}{3}}\partial^{\alpha}n_{\downarrow}
=-g_{2}\frac{4}{3}(6\pi^{2})^{\frac{2}{3}}n_{\downarrow}^{\frac{5}{3}}\partial^{\alpha}n_{\downarrow}$$
\begin{equation}\label{aSRIff Euler TOIR SSE spin down} -g_{\uparrow\downarrow} n_{\downarrow}\partial^{\alpha}n_{\uparrow}
-\frac{g_{2,\uparrow\downarrow}}{2}n_{\downarrow}\partial^{\alpha}\triangle n_{\uparrow}
,\end{equation}
where $g_{2}\equiv g_{2,\uparrow\uparrow}=g_{2,\downarrow\downarrow}$,
and the right-hand sides of the Euler equations include the interspecies interaction in accordance with equation (\ref{aSRIff F up to TOIR via n1 n2}).

Equations (\ref{aSRIff Euler TOIR SSE spin up}) and (\ref{aSRIff Euler TOIR SSE spin down}) have similar to each other structure.
Hence, describe structure of one of them.
The first term on the left-hand side of equation (\ref{aSRIff Euler TOIR SSE spin up}) is the substantional derivative of the velocity field.
the second term is the action of the external field on the particles.
The third (the fourth) term is the quantum Bohm potential (the gradient of partial Fermi pressure).
The right-hand side contains the interparticle interaction.
The first term on the right-hand side of equation (\ref{aSRIff Euler TOIR SSE spin up}) presents
the interaction between spin-up fermions
which appears in the third order by the interaction radius.
Other terms on the right-hand side describe action of the spin-down fermions on the spin-up fermions
(in the first and third orders by the interaction radius, correspondingly).

Equations (\ref{aSRIff Euler TOIR SSE spin up}) and (\ref{aSRIff Euler TOIR SSE spin down})
together with the continuity equations (\ref{aSRIff cont eq via vel spin up}) and (\ref{aSRIff cont eq via vel spin down})
can be rewritten via the non-linear Pauli equation:
$$\imath\hbar\partial_{t}\Phi
=\Biggl[-\frac{\hbar^{2}\nabla^{2}}{2m}
+V_{ext}
+(6\pi^{2})^{\frac{2}{3}}\frac{\hbar^{2}}{2m}\left(
                                           \begin{array}{cc}
                                             n_{\uparrow}^{\frac{2}{3}} & 0 \\
                                             0 & n_{\downarrow}^{\frac{2}{3}} \\
                                           \end{array}
                                         \right)$$
$$+g_{\uparrow\downarrow} \left(
                                           \begin{array}{cc}
                                             n_{\downarrow} & 0 \\
                                             0 & n_{\uparrow} \\
                                           \end{array}
                                         \right)
+\frac{1}{2}g_{2,\uparrow\downarrow} \left(
                                           \begin{array}{cc}
                                             \triangle n_{\downarrow} & 0 \\
                                             0 & \triangle n_{\uparrow} \\
                                           \end{array}
                                         \right)$$
\begin{equation}\label{aSRIff NLSE Pauli TOIR}
+\frac{4}{5}g_{2}(6\pi^{2})^{\frac{2}{3}}\left(
                                           \begin{array}{cc}
                                             n^{\frac{5}{3}}_{\uparrow} & 0 \\
                                             0 & n^{\frac{5}{3}}_{\downarrow} \\
                                           \end{array}
                                         \right)\Biggr]\Phi.\end{equation}
Equation (\ref{aSRIff NLSE Pauli TOIR}) is a generalization of equations
(\ref{aSRIff NLSE Pauli first appearence}), (\ref{aSRIff NLSE first appearence after Pauli}), and (\ref{aSRIff NLSE spin 1 TOIR}).
Equation (\ref{aSRIff NLSE Pauli TOIR}) contains four nonlinear terms.
The second term on the right-hand side is caused by the Fermi pressure.
Three other terms are related to the interparticle interaction.
The fourth and fifth terms on the right-hand side describe interaction between fermions
with different spin projections in the first and third orders by the interaction radius, correspondingly.
The fifth term includes the nonlocal nonlinearity since it contains the second order derivatives of the particle concentration.
The last term in equation (\ref{aSRIff NLSE Pauli TOIR}) describes the interaction between fermions with the same spin projection.
Equation (\ref{aSRIff NLSE Pauli TOIR}) allows to make a reduction to the single fluid form
similarly to the reduction of equation (\ref{aSRIff NLSE Pauli first appearence}) to equation (\ref{aSRIff NLSE first appearence after Pauli}).

Consider the single fluid regime at the partial spin polarization appearing from equations
(\ref{aSRIff Euler TOIR SSE spin up}) and (\ref{aSRIff Euler TOIR SSE spin down})
$$ mn(\partial_{t} +\textbf{v}\cdot\nabla)v^{\alpha}
-\frac{\hbar^{2}}{2m}n\partial^{\alpha}\frac{\triangle\sqrt{n}}{\sqrt{n}}$$
$$+n\partial^{\alpha}V_{ext}
+\vartheta\frac{(3\pi^{2})^{\frac{2}{3}}\hbar^{2}}{3m}n^{\frac{2}{3}}\partial^{\alpha}n$$
$$=-g_{2}\frac{(3\pi^{2})^{\frac{2}{3}}}{3}\biggl((1+\eta)^{\frac{8}{3}}+(1-\eta)^{\frac{8}{3}}\biggr)n^{\frac{5}{3}}\partial^{\alpha}n$$
\begin{equation}\label{aSRIff Euler TOIR partial spin pol Single Fluid}
-g_{\uparrow\downarrow}\frac{1-\eta^{2}}{4} n\partial^{\alpha}n
-\frac{g_{2,\uparrow\downarrow}}{2}\frac{1-\eta^{2}}{4} n\partial^{\alpha}\triangle n,
\end{equation}
where $\vartheta$ is given by equation (\ref{aSRIff vartheta def}).

The Euler equation (\ref{aSRIff Euler TOIR partial spin pol Single Fluid}) is the single fluid reduction of
equations (\ref{aSRIff Euler TOIR SSE spin up}) and (\ref{aSRIff Euler TOIR SSE spin down}).
Therefore, the physical meaning of different terms is similar to the corresponding terms in
equations (\ref{aSRIff Euler TOIR SSE spin up}) and (\ref{aSRIff Euler TOIR SSE spin down}).

\section{Separate spin evolution hydrodynamic model with the pressure evolution}

Present the full set of the separate spin evolution quantum hydrodynamic equations including the pressure tensor evolution equation.

Start with the continuity equations;
\begin{equation}\label{aSRIff cont eq via vel spin s with set with p}
\partial_{t}n_{s}+\nabla\cdot (n_{s}\textbf{v}_{s})=0. \end{equation}

Next, the Euler equations are shown:
$$mn_{s}(\partial_{t} +\textbf{v}_{s}\cdot\nabla)v^{\alpha}_{s} +n_{s}\partial^{\alpha}V_{ext}
-\frac{\hbar^{2}}{2m}n_{s}\partial^{\alpha}\frac{\triangle\sqrt{n_{s}}}{\sqrt{n_{s}}}$$
$$+m\partial^{\beta}p_{s}^{\alpha\beta} =-g_{2}\frac{m^{2}}{2\hbar^{2}}I_{0}^{\alpha\beta\gamma\delta}\partial^{\beta}(n_{s}p^{\gamma\delta}_{s})$$
\begin{equation}\label{aSRIff Euler TOIR SSE spin s with set with p} -g_{\uparrow\downarrow} n_{s}\partial^{\alpha}n_{s'}
-\frac{g_{2,\uparrow\downarrow}}{2}n_{s}\partial^{\alpha}\triangle n_{s'},\end{equation}
where $s=\uparrow,\downarrow$ and $s'\neq s$, so $s'$ presents the different spin projection.
In equation (\ref{aSRIff Euler TOIR SSE spin s with set with p}) there is a difference
with equations (\ref{aSRIff Euler TOIR SSE spin up}) and (\ref{aSRIff Euler TOIR SSE spin down})
since no equation of state is used for the pressure tensor.

The pressure tensor evolution equation for fermions with a chosen spin projection has the following form:
$$\partial_{t}p_{s}^{\alpha\beta} +v_{s}^{\gamma}\partial_{\gamma}p_{s}^{\alpha\beta} +p_{s}^{\alpha\gamma}\partial_{\gamma}v_{s}^{\beta} +p_{s}^{\beta\gamma}\partial_{\gamma}v_{s}^{\alpha}$$
$$+p_{s}^{\alpha\beta}\partial_{\gamma}v_{s}^{\gamma} +\partial_{\gamma}T_{s}^{\alpha\beta\gamma} +\frac{\hbar^{2}}{4m^{2}}\biggl[\partial_{\alpha}\partial_{\beta}\partial_{\gamma}(n_{s}v_{s}^{\gamma})$$
$$-v_{s}^{\gamma}\partial_{\alpha}\partial_{\beta}\partial_{\gamma}n_{s}
-\frac{1}{n_{s}}(\partial_{\gamma}v_{s}^{\gamma})\partial_{\alpha}n_{s}\cdot\partial_{\beta}n_{s}$$
$$-\frac{\partial_{\beta}n_{s}}{n_{s}}\cdot \partial_{\gamma}(n_{s}\cdot\partial_{\alpha}v_{s}^{\gamma})
-\frac{\partial_{\alpha}n_{s}}{n_{s}}\cdot \partial_{\gamma}(n_{s}\cdot\partial_{\beta}v_{s}^{\gamma})\biggr]$$
$$=-\frac{m}{8\hbar^{2}}g_{2} \{I_{0}^{\alpha\gamma\delta\mu}
[3 n_{s}^{2} v_{s}^{\beta}v_{s}^{\delta}\partial^{\gamma}v_{s}^{\mu}
+2n_{s}p_{s}^{\mu\delta}(\partial^{\gamma}v_{s}^{\beta}-\partial^{\beta}v_{s}^{\gamma})]$$
\begin{equation} \label{aSRIff eq for p alpha beta II with set spin s}
+I_{0}^{\beta\gamma\delta\mu}
[3 n_{s}^{2} v_{s}^{\alpha}v_{s}^{\delta}\partial^{\gamma}v_{s}^{\mu}
+2n_{s}p_{s}^{\mu\delta}(\partial^{\gamma}v_{s}^{\alpha}-\partial^{\alpha}v_{s}^{\gamma})]\}, \end{equation}
where $s=\uparrow,\downarrow$, and tensor $T_{s}^{\alpha\beta\gamma}$ is given by equation (\ref{aSRIff T alpha beta gamma explicit}).

As it is demonstrated in Sec. V, the interspecies interaction does not enter equation for the pressure tensor evolution.

\section{Collective excitations}

\subsection{Minimal coupling model up to the third order by the interaction radius}

Consider the small amplitude perturbations of the equilibrium state
and focus on the linear properties.

\subsubsection{Full spin polarization}

Considering single fluid model we have two functions $n$ and $\textbf{v}$.
Consider the uniform equilibrium concentration $n_{0}$ and zero velocity field.
After account of the perturbations functions have the following structure $n=n_{0}+\delta n$ and $\textbf{v}=\delta\textbf{v}$.
Perturbations are considered as the plane wave
$\delta n=N e^{-\imath\omega t+\imath k x}$
and $\delta\textbf{v}=\textbf{U} e^{-\imath\omega t+\imath k x}$.

Spectrum of the collective excitations propagating as the plane wave in the infinite uniform degenerate fermions
(an analog of the Bogoliubov spectrum in BEC) is
\begin{equation}\label{aSRIff spectrum Full spin}
\omega^{2}= \frac{(6\pi^{2}n_{0})^{\frac{2}{3}}\hbar^{2}}{3m^{2}}k^{2}
+\frac{g_{2}n_{0}k^{2}}{m}\frac{4}{3}(6\pi^{2}n_{0})^{\frac{2}{3}}+\frac{\hbar^{2}k^{4}}{4m^{2}}. \end{equation}
Spectrum (\ref{aSRIff spectrum Full spin}) corresponds to the result of Ref. \cite{Andreev PRA08} (see equation 64),
but coefficients caused by pressure are different.

Spectrum of collective excitations in the regime of full spin polarization is discussed in more details in \cite{Andreev 1912}.

\subsubsection{Partial and zero spin polarizations in the single fluid approach}

If the partially spin polarized fermions are described as the single fluid
we obtain one wave solution
similar to the previous case.
However, the interaction between fermions of different spin projections gives extra contribution in the spectrum.
The partial spin contribution modifies almost all coefficients
$$\omega^{2}= \vartheta\frac{(3\pi^{2}n_{0})^{\frac{2}{3}}\hbar^{2}}{3m^{2}}k^{2}+\frac{\hbar^{2}k^{4}}{4m^{2}}$$
$$+\frac{g_{2}n_{0}k^{2}}{m}\frac{4}{3}\biggl((1+\eta)^{\frac{8}{3}}+(1-\eta)^{\frac{8}{3}}\biggr)(6\pi^{2}n_{0})^{\frac{2}{3}} $$
\begin{equation}\label{aSRIff spectrum partial spin SF}
+\frac{g_{\uparrow\downarrow}}{m}\frac{1-\eta^{2}}{4}n_{0}k^{2}
-\frac{g_{2\uparrow\downarrow}}{2m}\frac{1-\eta^{2}}{4}n_{0}k^{4}. \end{equation}
Coefficients $\vartheta$ (see (\ref{aSRIff vartheta def}))
and $(1+\eta)^{\frac{8}{3}}+(1-\eta)^{\frac{8}{3}}$ in the first and third terms on the right-hand side show that
the increase of the spin polarization increases contribution of these terms via the increase of the pressure.
However, the third term is negative (for the repulsive interaction $g_{2}>0$). Hence, its increase decreases the frequency $\omega$.
The third term has higher dependence on the spin polarization.
Consequently, the increase of the spin polarization decreases the frequency.
Moreover, two last terms decreases with the grough of the spin polarization.
Their combined contribution is negative (for the repulsive interaction $g_{2}>0$).
Hence, it gives a mechanism for the frequency increase at the increase of the spin polarization.
Therefore, there is a competition between different terms
and a way of change of the frequency as the function of the spin polarization depends on the relation between parameters of the system.

Dropping terms caused by the TOIR and quantum Bohm potential find two terms:
the pressure caused term and the FOIR term ($s$-scattering).
These terms show stability of the spectrum at the attractive interaction between fermions
with opposite spin projections.
This conclusion is in conflict with the well-known phenomenon:
the formation of Cooper pairs and formation BCS-state.
It corresponds to instability of fermions as system described as a composition of independent fermions.
Hence, the developed fluid model fails to described this effect.
Therefore, this model is applicable for the repulsive interaction between fermions of different spin polarization
as it is mention in the beginning of this paper.

\subsubsection{SSE at the partial spin polarization}

Here we study spectrum of waves appearing in the two fluid description of the partially spin polarized spin-1/2 fermions.
Our analysis is based on equations (\ref{aSRIff Euler TOIR SSE spin up}) and (\ref{aSRIff Euler TOIR SSE spin down})
which are a generalizations of equations (\ref{aSRIff Euler FOIR spin up}) and (\ref{aSRIff Euler FOIR spin down}).

In this regime the system is described by two concentrations $n_{\uparrow}$ and $n_{\downarrow}$ and two velocity fields $\textbf{v}_{\uparrow}$ and $\textbf{v}_{\downarrow}$.
Equilibrium state is described by nonzero concentrations $n_{0\uparrow}$ and $n_{0\downarrow}$ and zero velocity fields.
Hence, the considering functions have the following structure $n_{s}=n_{0s}+\delta n_{s}$ and $\textbf{v}_{s}=\delta\textbf{v}_{s}$
with the following structure of perturbations
$\delta n_{s}=N_{s} e^{-\imath\omega t+\imath k x}$
and $\delta\textbf{v}_{s}=\textbf{U}_{s} e^{-\imath\omega t+\imath k x}$.

As the result obtain spectrum
$$\omega^{2}=\frac{1}{2}k^{2}\Biggl\{\frac{\hbar^{2}k^{2}}{2m^{2}} +\frac{1}{3}\frac{\hbar^{2}}{m^{2}}(6\pi^{2})^{\frac{2}{3}}(n_{0\uparrow}^{\frac{2}{3}}+n_{0\downarrow}^{\frac{2}{3}}) $$
$$+\frac{g_{2}}{m}\frac{4}{3}(6\pi^{2})^{\frac{2}{3}}(n_{0\uparrow}^{\frac{5}{3}}+n_{0\downarrow}^{\frac{5}{3}})
\pm\Biggl[\biggl[
\frac{1}{3}\frac{\hbar^{2}}{m^{2}}(6\pi^{2})^{\frac{2}{3}}(n_{0\uparrow}^{\frac{2}{3}}-n_{0\downarrow}^{\frac{2}{3}})$$
\begin{equation}\label{aSRIff spectrum partial spin SSE}
+\frac{g_{2}}{m}\frac{4}{3}(6\pi^{2})^{\frac{2}{3}}(n_{0\uparrow}^{\frac{5}{3}}-n_{0\downarrow}^{\frac{5}{3}})\biggr]^{2}
+4\frac{n_{0\uparrow}n_{0\downarrow}}{m^{2}}
\biggl(g-\frac{g_{2,\uparrow\downarrow}}{2}k^{2}\biggr)^{2}\Biggr]^{\frac{1}{2}}\Biggr\}\end{equation}
which consist of two acoustic waves
while single fluid approaches given by equations (\ref{aSRIff spectrum Full spin}) and (\ref{aSRIff spectrum partial spin SF}) show the single acoustic wave.

\subsection{Extended hydrodynamic model up to the third order by the interaction radius}

Spectra for the single fluid and two fluid models of spin-1/2 fermions are given above for the regime of minimal coupling
which does not include the pressure evolution.
Let us use the extended hydrodynamic model containing the pressure evolution equation to find generalizations of obtained spectra.

In this regime include nonzero equilibrium values of the partial concentrations $n_{0u}\neq n_{0d}$ and partial diagonal pressures $p_{0u}\neq p_{0d}$
while nondiagonal equilibrium elements of the pressure tensor and the equilibrium velocity field are equal to zero.
Considering functions present in the following form $n_{s}=n_{0s}+\delta n_{s}$, $\textbf{v}_{s}=\delta\textbf{v}_{s}$, and $p_{s}^{\alpha\gamma}=p_{0s}^{\alpha\gamma}+\delta p_{s}^{\alpha\gamma}$.
Perturbations are considered as the plane wave
$\delta n_{s}=N_{s} e^{-\imath\omega t+\imath k x}$,
$\delta\textbf{v}_{s}=\textbf{U}_{s} e^{-\imath\omega t+\imath k x}$,
and $\delta p_{s}^{\alpha\beta}=P_{s}^{\alpha\beta} e^{-\imath\omega t+\imath k x}$.

Present the set of linearized hydrodynamic equations
(\ref{aSRIff cont eq via vel spin s with set with p})-(\ref{aSRIff eq for p alpha beta II with set spin s}):
\begin{equation}\label{aSRIff cont lin arbitrary k} -\imath\omega\delta n_{s} +n_{0s}\imath \textbf{k}\delta \textbf{v}_{s}=0, \end{equation}
$$-\imath\omega mn_{0s}\delta v_{s}^{\alpha} +\frac{\hbar^{2}}{4m}\imath k^{\alpha}k^{2}\delta n_{s} +\imath m k^{\beta}\delta p_{s}^{\alpha\beta}$$
$$=-\imath k^{\beta}n_{0s}g_{2}\frac{m^{2}}{2\hbar^{2}}I_{0}^{\alpha\beta\gamma\delta}\delta p_{s}^{\gamma\delta}
-5\imath k^{\alpha} p_{0s}g_{2}\frac{m^{2}}{2\hbar^{2}}\delta n_{s}$$
\begin{equation}\label{aSRIff Euler lin arbitrary k}
-g_{\uparrow\downarrow}n_{0s}\imath k^{\alpha}\delta n_{s'}
+\frac{1}{2}g_{2\uparrow\downarrow}n_{0s}\imath k^{\alpha}k^{2}\delta n_{s'}, \end{equation}
and
$$-\imath\omega\delta p_{s}^{\alpha\beta}
+\imath k^{\alpha}p_{0s}\delta v_{s}^{\beta}
+\imath k^{\beta}p_{0s}\delta v_{s}^{\alpha}
+\imath k^{\gamma}p_{0s} \delta^{\alpha\beta}\delta v_{s}^{\gamma}$$
\begin{equation}\label{aSRIff pr evol lin arbitrary k}
-\imath\frac{\hbar^{2}}{4m^{2}}n_{0s}
(k^{\alpha}k^{2}\delta v_{s}^{\beta} +k^{\beta}k^{2}\delta v_{s}^{\alpha}-2k^{\alpha}k^{\beta}k^{\gamma}\delta v_{s}^{\gamma})=0, \end{equation}
where it has been used that $p_{0s}^{\alpha\beta}=p_{0s} \delta^{\alpha\beta}$, and $I_{0}^{\alpha\beta\gamma\gamma}=5\delta^{\alpha\beta}$.

Linearized equations (\ref{aSRIff cont lin arbitrary k})-(\ref{aSRIff pr evol lin arbitrary k}) are presented for the arbitrary wave vector.
Simplify these equations for the chosen regime.
As the result find the following set of equations:
\begin{equation}\label{aSRIff} \omega\delta n_{s}=n_{0s}k\delta v_{sx}, \end{equation}
and
$$n_{0s}\omega\delta v_{sx}-\frac{\hbar^{2}k^{3}}{4m^{2}}\delta n_{s}
=kn_{0s}\frac{g_{2}m}{2\hbar^{2}}(3\delta p_{s}^{xx}+\delta p_{s}^{yy}+\delta p_{s}^{zz}) $$
\begin{equation}\label{aSRIff}+k\delta p_{s}^{xx}
+kp_{0s}g_{2}\frac{5m}{2\hbar^{2}}\delta n_{s}
+\frac{g_{\uparrow\downarrow}}{m}n_{0s}k\delta n_{s'}-\frac{g_{2\uparrow\downarrow}}{2m}n_{0s}k^{3}\delta n_{s}, \end{equation}
where $I_{0}^{xx\gamma\delta}=\delta^{\gamma\delta}+2\delta^{x\gamma}\delta^{x\delta}$.

Also it includes
\begin{equation}\label{aSRIff} \omega\delta p_{s}^{xx}=3p_{0s}k\delta v_{sx}, \end{equation}
and
\begin{equation}\label{aSRIff} \omega\delta p_{s}^{yy}=\omega\delta p_{s}^{zz}=p_{0s}k\delta v_{sx}. \end{equation}

If there is no interspecies interaction find the single acoustic wave with the following spectrum:
\begin{equation}\label{aSRIff spectrum SF with pressure} \omega^{2}=\frac{3p_{0}}{n_{0}}k^{2}\biggl[1+8g_{2}\frac{mn_{0}}{3\hbar^{2}}\biggr]
+\frac{\hbar^{2}k^{4}}{4m^{2}}, \end{equation}
where equilibrium pressure can be used in the standard form
$p_{0}=(3\pi^{2})^{\frac{2}{3}}\hbar^{2}n_{0}^{\frac{5}{3}}/5m^{2}$ for the zero spin polarization
and $p_{0}=(6\pi^{2})^{\frac{2}{3}}\hbar^{2}n_{0}^{\frac{5}{3}}/5m^{2}$ for the full spin polarization.

Let us repeat that in this paper the pressure includes
an extra multiplier $1/m$ so its physical dimension differs from traditional.
As it is mentioned above,
the modification of the pressure physics dimension is made to give symmetric form of equations.

Equation (\ref{aSRIff spectrum SF with pressure}) gives the spectrum of excitations in uniform medium for isotropic Fermi surface.
However, the trap influence and the anisotropy of the Fermi surface are discussed in Ref. \cite{Andreev 1912}.

For the nontrivial two species evolution with the interspecies interaction included up to the third order by the interaction radius,
find the following spectrum consisting of two acoustic waves:
$$\omega^{2}=\frac{\hbar^{2}k^{4}}{4m^{2}}+\frac{3}{2}k^2 \biggl(\frac{p_{0u}}{n_{0u}}+\frac{p_{0d}}{n_{0d}}\biggr)
-g_{2}4k^{2}\frac{m}{\hbar^{2}}[p_{0u} +p_{0d}]$$
$$\pm\Biggl\{\Biggl[\frac{3}{2}k^2 \biggl(\frac{p_{0u}}{n_{0u}}-\frac{p_{0d}}{n_{0d}}\biggr)
-g_{2}4k^{2}\frac{m}{\hbar^{2}}[p_{0u}-p_{0d}]\Biggr]^{2}$$
\begin{equation}\label{aSRIff spectrum TwoF with pressure}
+\frac{n_{0u}n_{0d}}{m^2}k^{4}\biggl(g_{\uparrow\downarrow}-\frac{1}{2}g_{2,\uparrow\downarrow}k^{2}\biggr)^{2}\Biggr\}^{\frac{1}{2}}. \end{equation}

Equation (\ref{aSRIff spectrum SF with pressure}) shows that
the account of the pressure evolution changes coefficients in the first and second terms of equation (\ref{aSRIff spectrum Full spin})
which defines the speed of sound.
The last term can be modified either.
However, it requires further extend of the set of hydrodynamic equations.
The obtained model gives a correct description of terms up to the second order on the wave vector in the expression for the frequency square.
Hence, we have a model suitable for the sound wave modeling in the main order.
The sound wave spectrum for the two fluid description of spin-1/2 fermions (\ref{aSRIff spectrum TwoF with pressure})
contains similar generalization of equation (\ref{aSRIff spectrum partial spin SSE}) caused by the pressure evolution.

Spectra (\ref{aSRIff spectrum SF with pressure}) and (\ref{aSRIff spectrum TwoF with pressure})
gives an illustration of the pressure evolution contribution.
It gives considerable contribution in main order in the sound wave spectrum.
Further analysis of these effects including the spectra of collective excitations will be addressed elsewhere.
Thus, the pressure evolution account is the necessary step to obtain a proper hydrodynamic model of repulsing ultracold fermions.

\section{Conclusion}

A set of hydrodynamic models for the degenerate fermions with the short-range interaction has been developed.
The short-range interaction has been calculated up to the third order by the interaction radius.
It has been found that the hydrodynamic model based on the continuity and the Euler equation gives a rough,
but qualitatively good description of fermions.
This approximation allows to introduce the non-linear Schr\"{o}dinger and nonlinear Pauli equations for the macroscopic wave function at the eddy-free motion.

A more realistic hydrodynamic model of fermions requires the account of the pressure tensor evolution.
This approximate allows better description of kinetic effects
in degenerate fermions.
The short-range interaction is calculated in the weakly interacting limit.
Both, the interaction between fermions of the same spin projection and the interaction between fermions with different spin projections are included.
Ultracold fermions is a complex system.
If interaction between spin-up and spin-down fermions is repulsive there is a normal degenerate system.
While at the attractive interaction system undergoes the transition to be formed Cooper pairs.

Presented derivation is started from the concentration of fermions.
So, the basic definition does not include the information about Cooper pairs formation.
Hence, the described derivation is unsuitable for
the attractive interaction between fermions with different spin projections.
Consistent account of tensors of higher dimensions
$v^{\alpha}$, $\Pi^{\alpha\beta}$, $Q^{\alpha\beta\gamma}$, etc,
gives better description of the kinetic effects in terms of hydrodynamic model.
Hence, two kinds of truncations are presented.
One includes the concentration $n$ and the velocity field $v^{\alpha}$ evolution and
uses an equation of state for the pressure.
It provides a traditional form of hydrodynamic equations consisting of continuity and Euler equations.
The Euler equation includes interaction between fermions of different spin projections
(it has been considered up to the third order by the interaction radius)
and interaction between fermions of the same spin projection
(it has nontrivial contribution in the third order by the interaction radius
while zero contribution appears in the first order by the interaction radius).
The terms appearing in the third order by the interaction radius show resemblance to the p-wave scattering terms.

This approximation based on the continuity and Euler equations is called the minimal coupling model.
It is presented in two regimes.

The first regime presents all fermions as one system (as a single fluid).
The spin polarization enters equations as a parameter.
A nonlinear Schr\"{o}dinger equation is derived for fermions for the eddy-free motion.
The single fluid limit leads to single bulk wave solution
which is the sound wave supported by the Fermi pressure and affected by the interaction.

The second regime for the minimal coupling model presents evolution of spin-1/2 fermions as dynamics of two fluids.
Each of them is associated with fermions with chosen spin projection.
Consequently, two continuity equations and two Euler equations are found.
Corresponding nonlinear Pauli equation is derived for the eddy-free motion of each fluid.
Analysis of collective dynamics of this regime gives two sound waves.

Next, an extended set of hydrodynamic equation has been presented.
It includes the evolution of the pressure tensor.
The results are presented for the two fluid approach,
but it can be straightforwardly reduced to the single fluid by methods described in the paper.
The pressure evolution equation does not contain the external field.
Moreover,
the interaction between fermions having different spin projections does not contribute in the pressure evolution equation either.
The interaction between fermions having same spin projection enters the pressure evolution equation,
but it has nontrivial contribution in the third order by the interaction radius.
The first order contribution is equal to zero.
Same as it is for the Euler equation.
It is related to the antisymmetry of the wave function of fermions.
Using analogy described above,
it can be described as the pressure evolution equation,
where the interaction between fermions is considered in p-wave scattering limit.

The extended hydrodynamic model is applied to study the bulk collective excitations.

The spectra are considered in the single fluid and two fluid limits.

The single fluid limit gives one longitudinal sound wave.
Its spectrum is a generalization of corresponding spectrum found in the single fluid minimal coupling model.
The speed of sound is changed in this limit from $v_{Fe}/\sqrt{3}$ to $\sqrt{3/5}v_{Fe}$
(this is an illustration for the zero spin polarization).
Moreover, generalized spectrum gives modified dependence on the interaction constant $g_{2}$.

Obviously, the two-fluid extended hydrodynamic model demonstrates two longitudinal waves
and gives generalization of spectra found from the minimal coupling model.

A hydrodynamic model covering wide range of phenomena in degenerate repulsive fermions is derived from the microscopic model.
Fundamental collective excitations of fermions can be studied in terms of derived model.

\section{Appendix A: Expansion of the derivatives of the wave functions}

Calculation of $g^{\alpha\beta}(\textbf{r},\textbf{r}',t)$ includes the expansion of the derivatives of the many-particle wave functions
which are demonstrated here:
$$\partial^{\gamma}_{1}\Psi^{*}(R'',t)\cdot\partial^{\delta}_{1}\Psi(R'',t)=\frac{1}{2}\sum_f \sum_{f'\neq f}
\frac{n_f}{N} \frac{n_{f'}}{N-1}$$
$$\left(\:
\partial_{\gamma}\langle \textbf{r},t | f\rangle \: \cdot\langle\textbf{r}',t | f'\rangle-
\langle\textbf{r}',t | f\rangle \: \partial_{\gamma} \langle\textbf{r},t | f'\rangle \:\right)\times$$
\begin{equation}\label{aSRIff} \times\left(\:
\partial_{\delta}\langle f| \textbf{r},t\rangle \: \cdot\langle f'|\textbf{r}',t \rangle-
\langle f|\textbf{r}',t \rangle \: \partial_{\delta}\langle f'|\textbf{r},t \rangle \:\right),\end{equation}

$$\partial^{\gamma}_{2}\Psi^{*}(R'',t)\cdot\partial^{\delta}_{2}\Psi(R'',t)=\frac{1}{2}\sum_f \sum_{f'\neq f}
\frac{n_f}{N} \frac{n_{f'}}{N-1}$$
$$\left(\:
\langle \textbf{r},t | f\rangle \: \partial_{\gamma}' \langle\textbf{r}',t | f'\rangle-
\partial_{\gamma}' \langle\textbf{r}',t | f\rangle \: \cdot\langle\textbf{r},t | f'\rangle \:\right)\times$$
\begin{equation}\label{aSRIff} \times\left(\:
\langle f| \textbf{r},t\rangle \: \partial_{\delta}' \langle f'|\textbf{r}',t \rangle-
\partial_{\delta}' \langle f|\textbf{r}',t \rangle \: \cdot\langle f'|\textbf{r},t \rangle \:\right),\end{equation}

$$\partial^{\gamma}_{1}\Psi^{*}(R'',t)\cdot\partial^{\delta}_{2}\Psi(R'',t)=\frac{1}{2}\sum_f \sum_{f'\neq f}
\frac{n_f}{N} \frac{n_{f'}}{N-1}$$
$$\left(\:
\partial_{\gamma}\langle \textbf{r},t | f\rangle \: \langle\textbf{r}',t | f'\rangle-
\langle\textbf{r}',t | f\rangle \: \partial_{\gamma}\langle\textbf{r},t | f'\rangle \:\right)\times$$
\begin{equation}\label{aSRIff} \times\left(\:
\langle f| \textbf{r},t\rangle \: \partial_{\delta}' \langle f'|\textbf{r}',t \rangle-
\partial_{\delta}' \langle f|\textbf{r}',t \rangle \: \langle f'|\textbf{r},t \rangle \:\right),\end{equation}
and
$$\partial^{\gamma}_{2}\Psi^{*}(R'',t)\cdot\partial^{\delta}_{1}\Psi(R'',t)=\frac{1}{2}\sum_f \sum_{f'\neq f}
\frac{n_f}{N} \frac{n_{f'}}{N-1}$$
$$\left(\:
\langle \textbf{r},t | f\rangle \: \partial_{\gamma}' \langle\textbf{r}',t | f'\rangle-
\partial_{\gamma}' \langle\textbf{r}',t | f\rangle \: \langle\textbf{r},t | f'\rangle \:\right)\times$$
\begin{equation}\label{aSRIff} \times\left(\:
\partial_{\delta}\langle f| \textbf{r},t\rangle \: \langle f'|\textbf{r}',t \rangle-
\langle f|\textbf{r}',t \rangle \: \partial_{\delta}\langle f'|\textbf{r},t \rangle \:\right),\end{equation}
where $\partial_{\gamma}'$ is the derivative on gamma projection of the vector $\textbf{r}'$.

\section{Appendix B: Expressions of tensors $B^{\alpha\beta\gamma}$ and $D^{\alpha,\beta\gamma\delta}$ via hydrodynamic functions}

Expressions of for real and imaginary parts of tensor $B^{\alpha\beta\gamma}$ and for the imaginary part of tensor $D^{\alpha,\beta\gamma\delta}$ written in terms of the hydrodynamic functions are given below.
The real part of tensor $D^{\alpha,\beta\gamma\delta}$ is not required for derivation presented in the paper.
\begin{widetext}
$$Re B^{\alpha\beta\gamma}=
m^{3}\Biggl[Q^{\alpha\beta\gamma}+v^{\alpha}p^{\beta\gamma}+v^{\beta}p^{\alpha\gamma}+v^{\gamma}p^{\alpha\beta}
+nv^{\alpha}v^{\beta}v^{\gamma}
-\frac{\hbar^{2}}{m^{2}}\Biggl(
\frac{1}{3}n(\partial^{\alpha}\partial^{\beta} v^{\gamma} +\partial^{\beta}\partial^{\gamma} v^{\alpha} +\partial^{\alpha}\partial^{\gamma} v^{\beta})
+\frac{1}{4}[\partial^{\alpha}n (\partial^{\beta}v^{\gamma}+\partial^{\gamma}v^{\beta})$$
$$+\partial^{\beta}n (\partial^{\alpha}v^{\gamma}+\partial^{\gamma}v^{\alpha})
+\partial^{\gamma}n (\partial^{\alpha}v^{\beta}+\partial^{\beta}v^{\alpha})]
+(v^{\alpha}\sum_{f}n_{f}a_{f}\partial^{\beta}\partial^{\gamma}a_{f}
+v^{\beta}\sum_{f}n_{f}a_{f}\partial^{\alpha}\partial^{\gamma}a_{f}
+v^{\gamma}\sum_{f}n_{f}a_{f}\partial^{\alpha}\partial^{\beta}a_{f})$$
$$+\frac{1}{3}\sum_{f}n_{f}a_{f}^{2}(\partial^{\alpha}\partial^{\beta} u_{f}^{\gamma}
+\partial^{\beta}\partial^{\gamma} u_{f}^{\alpha}
+\partial^{\alpha}\partial^{\gamma} u_{f}^{\beta})
+\sum_{f}n_{f}a_{f}(u_{f}^{\alpha}\partial^{\beta}\partial^{\gamma}a_{f}
+u_{f}^{\beta}\partial^{\gamma}\partial^{\alpha}a_{f}
+u_{f}^{\gamma}\partial^{\alpha}\partial^{\beta}a_{f})$$
\begin{equation} \label{aSRIff}
+\frac{1}{2}\sum_{f}n_{f}a_{f}(\partial^{\alpha}a_{f}\cdot(\partial^{\beta}u_{f}^{\gamma}+\partial^{\gamma}u_{f}^{\beta})
+\partial^{\beta}a_{f}\cdot(\partial^{\gamma}u_{f}^{\alpha}+\partial^{\alpha}u_{f}^{\gamma})
+\partial^{\gamma}a_{f}\cdot(\partial^{\alpha}u_{f}^{\beta}+\partial^{\beta}u_{f}^{\alpha}))\Biggl)\Biggr],\end{equation}
\begin{equation} \label{aSRIff}
Im B^{\alpha\beta\gamma}=-\frac{1}{2}m^{2}\hbar\biggl(\partial^{\alpha}(nv^{\beta}v^{\gamma}+p^{\beta\gamma})
+\partial^{\beta}(nv^{\alpha}v^{\gamma}+p^{\alpha\gamma})
+\partial^{\gamma}(nv^{\alpha}v^{\beta}+p^{\alpha\beta})
-\frac{\hbar^{2}}{m^{2}}\sum_{f}n_{f}a_{f}\partial^{\alpha}\partial^{\beta}\partial^{\gamma}a_{f}\biggr) ,\end{equation}
and
$$Im D^{\alpha,\beta\gamma\delta}=
-m^{3}\hbar\Biggl\{
\frac{1}{2}(\partial^{\beta}n\cdot v^{\alpha}v^{\gamma}v^{\delta}
+\partial^{\gamma}n\cdot v^{\alpha}v^{\beta}v^{\delta}
+\partial^{\delta}n\cdot v^{\alpha}v^{\beta}v^{\gamma}
-\partial^{\alpha}n\cdot v^{\beta}v^{\gamma}v^{\delta})
+\frac{1}{2}nv^{\alpha}[v^{\beta}(\partial^{\gamma}v^{\delta}+\partial^{\delta}v^{\gamma})$$
$$+v^{\gamma}(\partial^{\beta}v^{\delta}+\partial^{\delta}v^{\beta})
+v^{\delta}(\partial^{\beta}v^{\gamma}+\partial^{\gamma}v^{\beta})]
+\frac{1}{2}\biggl( p^{\alpha\beta}(\partial^{\gamma}v^{\delta}+\partial^{\delta}v^{\gamma})
+p^{\alpha\gamma}(\partial^{\beta}v^{\delta}+\partial^{\delta}v^{\beta})
+p^{\alpha\delta}(\partial^{\beta}v^{\gamma}+\partial^{\gamma}v^{\beta}) \biggr)$$
$$+\frac{1}{2}v^{\alpha}\biggl(v^{\beta}\sum_{f}n_{f}(u_{f}^{\gamma}\partial^{\delta}a_{f}^{2}+u_{f}^{\delta}\partial^{\gamma}a_{f}^{2})
+v^{\gamma}\sum_{f}n_{f}(u_{f}^{\beta}\partial^{\delta}a_{f}^{2}+u_{f}^{\delta}\partial^{\beta}a_{f}^{2})
+v^{\delta}\sum_{f}n_{f}(u_{f}^{\beta}\partial^{\gamma}a_{f}^{2}+u_{f}^{\gamma}\partial^{\beta}a_{f}^{2})\biggr)$$
$$+\frac{1}{2}v^{\alpha}\biggl( \sum_{f}n_{f}\partial^{\beta}a_{f}^{2}\cdot u_{f}^{\gamma}u_{f}^{\delta}
+\sum_{f}n_{f}\partial^{\gamma}a_{f}^{2}\cdot u_{f}^{\beta}u_{f}^{\delta}
+\sum_{f}n_{f}\partial^{\delta}a_{f}^{2}\cdot u_{f}^{\beta}u_{f}^{\gamma}\biggr)$$
$$-\frac{1}{2}\biggl(v^{\beta}v^{\gamma}\sum_{f}n_{f}\partial^{\alpha}a_{f}^{2}\cdot u_{f}^{\delta}
+v^{\gamma}v^{\delta}\sum_{f}n_{f}\partial^{\alpha}a_{f}^{2}\cdot u_{f}^{\beta}
+v^{\beta}v^{\delta}\sum_{f}n_{f}\partial^{\alpha}a_{f}^{2}\cdot u_{f}^{\gamma}\biggr)$$
$$-\frac{1}{2}\biggl( v^{\beta}\sum_{f}n_{f}\partial^{\alpha}a_{f}^{2}\cdot u_{f}^{\gamma}u_{f}^{\delta}
+v^{\gamma}\sum_{f}n_{f}\partial^{\alpha}a_{f}^{2}\cdot u_{f}^{\beta}u_{f}^{\delta}
+v^{\delta}\sum_{f}n_{f}\partial^{\alpha}a_{f}^{2}\cdot u_{f}^{\beta}u_{f}^{\gamma}\biggr)
-\frac{1}{2}\sum_{f}n_{f}\partial^{\alpha}a_{f}^{2}\cdot u_{f}^{\beta}u_{f}^{\gamma}u_{f}^{\delta}$$
$$+\frac{1}{2}\biggl(v^{\beta}v^{\gamma} \sum_{f}n_{f}\partial^{\delta}a_{f}^{2}\cdot u_{f}^{\alpha}
+v^{\gamma}v^{\delta} \sum_{f}n_{f}\partial^{\beta}a_{f}^{2}\cdot u_{f}^{\alpha}
+v^{\beta}v^{\delta} \sum_{f}n_{f}\partial^{\gamma}a_{f}^{2}\cdot u_{f}^{\alpha}\biggr)
+\frac{1}{2}\biggl(v^{\beta} (\sum_{f}n_{f}\partial^{\gamma}a_{f}^{2}\cdot u_{f}^{\delta}u_{f}^{\alpha}
+\sum_{f}n_{f}\partial^{\delta}a_{f}^{2}\cdot u_{f}^{\gamma}u_{f}^{\alpha})$$
$$+v^{\gamma} (\sum_{f}n_{f}\partial^{\beta}a_{f}^{2}\cdot u_{f}^{\delta}u_{f}^{\alpha}
+\sum_{f}n_{f}\partial^{\delta}a_{f}^{2}\cdot u_{f}^{\beta}u_{f}^{\alpha})
+v^{\delta} (\sum_{f}n_{f}\partial^{\beta}a_{f}^{2}\cdot u_{f}^{\gamma}u_{f}^{\alpha}
+\sum_{f}n_{f}\partial^{\gamma}a_{f}^{2}\cdot u_{f}^{\beta}u_{f}^{\alpha})\biggr)
+\frac{1}{2}\biggl( \sum_{f}n_{f}\partial^{\beta}a_{f}^{2}\cdot u_{f}^{\gamma}u_{f}^{\delta}u_{f}^{\alpha}$$
$$+\sum_{f}n_{f}\partial^{\gamma}a_{f}^{2}\cdot u_{f}^{\beta}u_{f}^{\delta}u_{f}^{\alpha}
+\sum_{f}n_{f}\partial^{\delta}a_{f}^{2}\cdot u_{f}^{\beta}u_{f}^{\gamma}u_{f}^{\alpha}\biggr)
+\frac{1}{2}\biggl( v^{\alpha}v^{\beta}\sum_{f}n_{f}a_{f}^{2}(\partial^{\gamma}u_{f}^{\delta}+\partial^{\delta}u_{f}^{\gamma})
+v^{\alpha}v^{\gamma}\sum_{f}n_{f}a_{f}^{2}(\partial^{\beta}u_{f}^{\delta}+\partial^{\delta}u_{f}^{\beta})$$
$$+v^{\alpha}v^{\delta}\sum_{f}n_{f}a_{f}^{2}(\partial^{\beta}u_{f}^{\gamma}+\partial^{\gamma}u_{f}^{\beta})\biggr)
+\frac{1}{2}v^{\alpha}\biggl(\sum_{f}n_{f}a_{f}^{2}\partial^{\beta}(u_{f}^{\gamma}u_{f}^{\delta})
+\sum_{f}n_{f}a_{f}^{2}\partial^{\gamma}(u_{f}^{\beta}u_{f}^{\delta})
+\sum_{f}n_{f}a_{f}^{2}\partial^{\delta}(u_{f}^{\beta}u_{f}^{\gamma})\biggr)$$
$$+\frac{1}{2}\biggl(v^{\beta}\sum_{f}n_{f}a_{f}^{2}u_{f}^{\alpha}(\partial^{\gamma}u_{f}^{\delta}+\partial^{\delta}u_{f}^{\gamma})
+v^{\gamma}\sum_{f}n_{f}a_{f}^{2}u_{f}^{\alpha}(\partial^{\beta}u_{f}^{\delta}+\partial^{\delta}u_{f}^{\beta})
+v^{\delta}\sum_{f}n_{f}a_{f}^{2}u_{f}^{\alpha}(\partial^{\beta}u_{f}^{\gamma}+\partial^{\gamma}u_{f}^{\beta})\biggr)$$
$$+\frac{1}{2}\biggl(\sum_{f}n_{f}a_{f}^{2}u_{f}^{\alpha}
(\partial^{\beta}(u_{f}^{\gamma}u_{f}^{\delta})+\partial^{\gamma}(u_{f}^{\beta}u_{f}^{\delta})+\partial^{\delta}(u_{f}^{\beta}u_{f}^{\gamma}))\biggr)
+\frac{\hbar^{2}}{m^{2}}\Biggl[
\frac{1}{6}\partial^{\alpha}n\cdot(\partial^{\beta}\partial^{\gamma}v^{\delta}
+\partial^{\gamma}\partial^{\delta}v^{\beta}
+\partial^{\beta}\partial^{\delta}v^{\gamma})
+v^{\beta}\sum_{f}n_{f}\partial^{\alpha}a_{f}\cdot\partial^{\gamma}\partial^{\delta}a_{f}$$

$$+v^{\gamma}\sum_{f}n_{f}\partial^{\alpha}a_{f}\cdot\partial^{\beta}\partial^{\delta}a_{f}
+v^{\delta}\sum_{f}n_{f}\partial^{\alpha}a_{f}\cdot\partial^{\beta}\partial^{\gamma}a_{f}
-v^{\alpha}\sum_{f}n_{f}a_{f}\partial^{\beta}\partial^{\gamma}\partial^{\delta}a_{f}
-\sum_{f}n_{f}u_{f}^{\alpha}a_{f}\partial^{\beta}\partial^{\gamma}\partial^{\delta}a_{f}$$
$$+\sum_{f}n_{f}(u_{f}^{\beta}\partial^{\alpha}a_{f}\cdot \partial^{\gamma}\partial^{\delta}a_{f}
+u_{f}^{\gamma}\partial^{\alpha}a_{f}\cdot \partial^{\beta}\partial^{\delta}a_{f}
+u_{f}^{\delta}\partial^{\alpha}a_{f}\cdot \partial^{\beta}\partial^{\gamma}a_{f})
+\frac{1}{6}\sum_{f}n_{f}\partial^{\alpha}a_{f}\cdot
(\partial^{\beta}\partial^{\gamma}u_{f}^{\delta}+\partial^{\gamma}\partial^{\delta}u_{f}^{\beta}+\partial^{\beta}\partial^{\delta}u_{f}^{\gamma})$$

$$+\frac{1}{2}\biggl( (\partial^{\beta}v^{\gamma}+\partial^{\gamma}v^{\beta})\cdot\sum_{f}n_{f}\partial^{\delta}a_{f}\cdot\partial^{\alpha}a_{f}
+(\partial^{\beta}v^{\delta}+\partial^{\delta}v^{\beta})\cdot\sum_{f}n_{f}\partial^{\gamma}a_{f}\cdot\partial^{\alpha}a_{f}
+(\partial^{\gamma}v^{\delta}+\partial^{\delta}v^{\gamma})\cdot\sum_{f}n_{f}\partial^{\beta}a_{f}\cdot\partial^{\alpha}a_{f}\biggr)$$
\begin{equation} \label{aSRIff}
+\frac{1}{2}\biggl(\sum_{f}n_{f}(\partial^{\gamma}u_{f}^{\delta}+\partial^{\delta}u_{f}^{\gamma})\cdot\partial^{\beta}a_{f}\cdot\partial^{\alpha}a_{f}
+\sum_{f}n_{f}(\partial^{\beta}u_{f}^{\delta}+\partial^{\delta}u_{f}^{\beta})\cdot\partial^{\gamma}a_{f}\cdot\partial^{\alpha}a_{f}
+\sum_{f}n_{f}(\partial^{\beta}u_{f}^{\gamma}+\partial^{\gamma}u_{f}^{\beta})\cdot\partial^{\delta}a_{f}\cdot\partial^{\alpha}a_{f}\biggr)\Biggr]\Biggr\}.\end{equation}

\end{widetext}

\section{Acknowledgements}
Work is supported by the Russian Foundation for Basic Research (grant no. 20-02-00476).

\end{document}